\documentclass[fleqn,usenatbib]{mnras}

\usepackage{newtxtext,newtxmath}

\usepackage[T1]{fontenc}

\DeclareRobustCommand{\VAN}[3]{#2}
\let\VANthebibliography\thebibliography
\def\thebibliography{\DeclareRobustCommand{\VAN}[3]{##3}\VANthebibliography}

\usepackage{graphicx}	
\usepackage{bm}
\usepackage{subfig }
\usepackage{amsmath,mathtools}
\usepackage{enumitem}
\hypersetup{
    colorlinks=true,
    linkcolor=blue,
    filecolor=blue,      
    urlcolor=blue,
    citecolor=blue
}

\DeclarePairedDelimiter{\Ev}{\langle}{\rangle}

\renewcommand{\Re}{\operatorname{Re}}

\title[]{An analytical model for non-local stochastic field-level galaxy bias on the sphere}

\author[M. von Wietersheim-Kramsta et al.]{Maximilian von Wietersheim-Kramsta,$^{1,2}$\thanks{E-mail: maximilian.von-wietersheim-kramsta@durham.ac.uk}
Nicolas Tessore,$^{3}$
Benjamin Joachimi,$^{4}$
Qianjun Hang,$^{4}$
\newauthor Niall Jeffrey,$^{5,4}$
Joop Schaye$^{6}$
\\
$^{1}$Department of Physics, Institute for Computational  Cosmology, Durham University, South Road, Durham DH1 3LE, UK\\
$^{2}$Department of Physics, Centre for Extragalactic Astronomy, Durham University, South Road, Durham DH1 3LE, UK\\
$^{3}$Mullard Space Science Laboratory, University College London, Holmbury St Mary, Dorking, Surrey RH5 6NT, UK\\
$^{4}$Department of Physics and Astronomy, University College London, Gower Street, London WC1E 6BT, UK\\
$^{5}$Department of Physics \& King’s Institute for Artificial Intelligence, King’s College London, Strand, London WC2R 2LS, UK\\
$^{6}$Leiden Observatory, Leiden University, PO Box 9513, 2300 RA Leiden, the Netherlands
}

\date{Accepted XXX. Received YYY; in original form ZZZ}

\pubyear{\the\year{}}

\begin{document}
\label{firstpage}
\pagerange{\pageref{firstpage}--\pageref{lastpage}}
\maketitle

\begin{abstract}
    We present the `Bias of Overdensities in Large-scale Tracers on the Sphere' (BOLTS) model, a novel analytic galaxy bias model for projected fields on the sphere that incorporates stochasticity and non-locality already at linear order while enforcing physicality and statistical isotropy. Rather than as a deterministic transfer function, we model galaxy bias as a statistically isotropic random field, allowing stochasticity to enter through fluctuations in the bias itself. The construction yields an explicit decomposition into a linear, non-local stochastic and deterministic components in harmonic space. We fit the model parameters to the projected power spectra measured in large-volume hydrodynamical simulations (FLAMINGO), enabling sampling of galaxy populations from simulated matter fields while conditioning on galaxy properties such as halo mass or redshift. By construction, the  model reproduces the two-point statistics down to the simulation resolution, and without being calibrated on it, BOLTS recovers the equilateral bispectrum within the sample variance of the simulation down to comoving scales of $\mathcal{O}(10 \, \mathrm{Mpc})$ for halo masses $ M_{\mathrm{h}} > 10^{11.5}\,\mathrm{M}_{\odot} $ over redshifts $0 < z < 3$. Relative to local bias prescriptions, it more accurately captures amplitude-phase correlations and mode coupling in the measured wavelet phase harmonics wherever the stochasticity of the tracers exceeds their discreteness, including low redshifts and lower halo masses $M_{\mathrm{h}} = 10^{11.5}\,\mathrm{M}_{\odot}$ where non-linearities become important. This model enables efficient field-level forward modelling for simulation-based inference and model comparison with large datasets from forthcoming galaxy surveys including \textit{Euclid}, Rubin LSST, and \textit{Roman}, while improving the accuracy of galaxy bias modelling with respect to standard linear deterministic models.
\end{abstract}




\section{Introduction}

The precision cosmology era, marked by large-scale structure (LSS) surveys such as the Kilo-Degree Survey (KiDS; e.g. \citealt{wright2025kids}), Dark Energy Survey (DES; e.g. \citealt{des2026dark}), Hyper-Suprime Cam survey (HSC; e.g. \citealt{dalal2023hsc}), and the Dark Energy Spectroscopic Instrument survey (DESI; e.g. \citealt{desi2025results}), has solidified most aspects of the Lambda Cold Dark Matter ($\Lambda$CDM) paradigm. In the near future, Stage IV surveys such as the \textit{Euclid} mission \citep{laureijs2011euclid, euclid2025mission}, the Vera C. Rubin Observatory (LSST; \citealt{lsst2009lsst}), and the \textit{Roman} mission \citep{spergel2015wide} are expected to deliver sub-percent constraints on the dark energy equation-of-state, $w$, the sum of neutrino masses, $\sum m_{\nu}$, the dark matter density, $\Omega_{\mathrm{CDM}}$, and the growth of matter density fluctuations, $\sigma_8 (z)$, from cosmic shear, galaxy clustering, and their cross-correlations. Achieving this unprecedented precision requires theoretical modelling of cosmological observables that matches or exceeds the accuracy of the data, especially concerning the highly complex relationship between the observed galaxy field ($\delta^{\mathrm{g}}$) and the underlying dark matter density field ($\delta$). Accurately modelling and measuring these quantities at the field-level enables the extraction of more cosmological and astrophysical information than with standard summary statistics.

This requirement is sharpened by the rapid adoption of field-level forward modelling and simulation-based inference (SBI) in weak lensing and galaxy clustering: e.g. full-field
deep-learning analysis of KiDS-1000 \citep{fluri2022full}, neural SBI of DES Y3 weak lensing maps \citep{jeffrey2025dark, gatti2024des, thomsen2026des}, neural SBI of KiDS-1000 and KiDS-Legacy cosmic shear \citep{lin2023simulation, vonwietersheimkramsta2024kidssbi, saoulis2026field}, and field-level SBI of spectroscopic large-scale structure \citep{lemos2023simbig}. These analyses require a forward model of the observed tracer field that is accurate beyond two-point statistics and, for observables projected along the line-of-sight, respects the spherical geometry of the sky. To be practical for SBI, galaxy bias must be sampled cheaply at every forward-model evaluation, without necessarily having access to three-dimensional halo catalogues.

Since galaxies form within dark matter halos through baryonic physics \citep{kaiser1984on}, at small enough scales, they are non-linear, non-local, and stochastic tracers of the matter field (see e.g. \citealt{McDonald2009clustering, Desjacques2018large, wechsler2018connection, Kokron2022priors, Zennaro2022priors}), meaning that the observed galaxy overdensity $\delta^{\mathrm{g}}$ is not simply proportional to the dark matter overdensity $\delta^{\mathrm{m}}$. Inadequate modelling of this bias translates directly into systematic uncertainties in cosmological parameter estimation (e.g. \citealt{Zentner2014galaxy, paviot2024impact, Pandey2025cosmology}). Consequently, for precision observational cosmology, a model of galaxy bias should be robust over the widest possible range of cosmological scales.

A variety of analytical approaches have been developed to model galaxy bias at the level of the matter density field, $\delta^{\mathrm{m}}$. On very large scales, the simplest approach assumes a constant linear bias factor, $b$, such that $\delta^{\mathrm{g}} = b \, \delta^{\mathrm{m}}$. While observationally useful, this assumption fails once non-linear gravitational evolution becomes significant. In the quasi-linear regime, the frameworks of Perturbation Theory \citep{Bernardeau2002large} and Effective Field Theory (EFT; \citealt{Carrasco2012effective}) provide a robust high-dimensional parametrisation of a tracer of the dark matter density on small scales. A common simplification in this regime is the Local-In-Matter-Density (LIMD) assumption where the number density of tracers is assumed to only depend on an expansion of the local matter density field \citep{fry1993biasing}. This expansion can be generalised by incorporating other local gravitational observables such as the tidal field and higher order derivatives of the matter density field \citep{mirbabayi2015biased, lazeyras2018beyond, lazeyras2019robust}. Additionally, the stochastic nature of the galaxy-halo connection, driven by complex baryonic processes, introduces noise typically via scale-dependent stochasticity terms \citep{dekel1999stochastic, taruya1999stochastic, matsubara1999stochasticity, rubira2025non}. Although analytically tractable, the convergence of the expansion is only guaranteed in the quasi-linear regime. Even in this regime, the large number of degrees-of-freedom in EFT may also be impractical for the purposes of some large-scale cosmological analyses.

For the non-linear regime, parametric galaxy bias models have been developed that do not rely on the assumptions of perturbation theory, but they usually still contain the LIMD assumption (e.g. \citealt{szalay1988constraints, matsubara2011nonlinear, Frusciante2012lagrangian, Neyrinck2014Transformationally, Ata2015Bayesian}) or rely on the assumption that the tracer is only a function of the smoothed matter density field (e.g. \citealt{schmoldt1999density, erdogdu2004the, kitaura2009cosmic, jasche2010bayesian, jasche2015past, modi2018cosmological,  ramanah2019cosmological}). However, as noted in \citet{bartlett2024bye}, the LIMD assumption significantly biases the modelled halo field across all scales (linear, quasi-linear, and non-linear) and halo masses. Therefore, to achieve sub-percent level accuracy, any field-level biasing model cannot rely on the Local-In-Matter-Density assumption, even in the linear regime. Effective descriptions that inject non-locality through the cosmic web have been applied in three dimensions: classifying the tidal environment hierarchically and assigning each web class its own deterministic-plus-stochastic bias reproduces the halo power spectrum to 2\% at $k \sim 0.8 \, h \mathrm{Mpc}^{-1}$ \citep{coloma2024hierarchical}.

To address the failure of Eulerian LIMD, recent field-level inference frameworks have increasingly adopted Lagrangian Perturbation Theory (LPT) and Hybrid Effective Field Theory (HEFT) approaches (e.g. \citealt{modi2020simulations, kokron2021cosmology, pellejero2022modelling}). Unlike Eulerian models, these formalisms define bias relationships in the initial Lagrangian space and subsequently advect the biased field to the late-time Universe using particle displacements. This advection naturally incorporates the bulk of non-local gravitational evolution, allowing these models to capture the matter-galaxy connection down to non-linear scales without relying on expensive full N-body gravity solvers.

However, the LPT and HEFT frameworks remain fundamentally three-dimensional: inferring cosmological parameters from projected surveys requires generating full lightcones and projecting them onto the sphere at every likelihood evaluation, with significant computational overhead and geometric complexity once window functions and redshift uncertainties are included. Furthermore, their stochasticity is often treated phenomenologically or through ad-hoc noise fields that are difficult to marginalise over analytically (see how this can be addressed in \citealt{friedrich2022pdf}).

Alternatively, one can avoid the LIMD assumption through the incorporation of non-locality through N-body simulations and then populating halos with galaxies according to a halo occupation distribution (HOD; \citealt{peacock2000halo, Berlind2002halo, Zheng2005theoretical, zheng2007galaxy}) or a sub-halo abundance matching (SHAM) model \citep{conroy2006modeling, vale2006non, Reddick2013connection}. The standard HOD framework assumes that all galaxies are hosted by a dark matter halo \citep{seljak2000analytic, cooray2002halo} and the probability of a halo hosting a galaxy is solely determined by the mass of the host halo. The SHAM model instead assumes a direct mapping from the peak peculiar velocities of dark matter sub-halos to stellar masses of the hosted galaxies. Both the SHAM model and extensions of the HOD formalism can incorporate non-locality (e.g. \citealt{tinker2005on}), stochasticity, and the dependence on other halo properties in the halo bias, i.e. halo assembly bias \citep{gao2007assembly, dalal2008halo}, and in the galaxy bias, i.e. galaxy assembly bias \citep{croton2007halo, hearin2016introducing, artale2018impact, zehavi2018impact, bose2019revealing, Hadzhiyska2020limitations, Hadzhiyska2023millennium, Shiferaw2025how}. While these frameworks accurately populate dark matter simulations with galaxies, they require the three-dimensional halo field, whereas many observations and simulations for imaging surveys (e.g. GLASS; \citealt{tessore2023glass}) are projected along the line-of-sight, without a well-defined halo mass, scale radius, or substructure hierarchy. An effective bias model in projection must therefore mimic the statistical properties of the three-dimensional HOD or SHAM without reconstructing halos explicitly.

In this work, we present a novel analytical bias model that linearly maps an underlying density field, $\delta$, to a biased tracer field, $\delta^{\mathrm{b}}$, in projection on the sphere, incorporating non-locality and stochasticity while preserving the homogeneity and realness of the resulting field. Working in spherical harmonics, the Bias of Overdensities in Large-scale Tracers on the Sphere (BOLTS) model is computationally efficient and adopts the geometry natural to astronomical observations.

Sections~\ref{galaxy_bias} and~\ref{derivation} construct the simplest model in spherical projection that produces a galaxy field consistent with the cosmological principle while incorporating non-locality and stochasticity, describe the calibration of its parameters, and show how they can be conditioned on halo or galaxy properties. Section~\ref{flamingo} describes the hydrodynamical simulations used to calibrate and test the model in Section~\ref{calibration_testing}. Section~\ref{forward_model} presents a semi-empirical extension that interpolates the scale dependence of the model for efficient forward modelling and for parametrising galaxy clustering and galaxy-galaxy lensing likelihoods. We conclude in Section~\ref{conclusion}, and additional derivations and tests are in the appendices.
 
\section{The galaxy-halo connection: galaxy bias}\label{galaxy_bias}

Both halos and galaxies can be thought of as tracers of an underlying continuous density contrast field, $\delta(\bm{x}, \tau)$, given by
\begin{equation}
    \delta (\bm{x}, \tau) = \frac{\rho(\bm{x}, \tau) - \overline{\rho}(\tau)}{\overline{\rho}(\tau)},
    \label{eqn:matter_density_field}
\end{equation}
\noindent where $\rho$ is the matter density field as a function of the comoving spatial position, $\bm{x}$, and proper time, $\tau$. The overline indicates the mean over all space. Similarly, one can construct the overdensity field of a tracer, $\delta^{\mathrm{b}}$ as follows
\begin{equation}
    \delta^{\mathrm{b}}(\bm{x}, \tau) = \frac{N(\bm{x}, \tau) - \overline{N}(\tau)}{\overline{N}(\tau)},
    \label{eqn:bias_density_field}
\end{equation}
\noindent where $N$ is the discrete number count of tracers. If the discrete number count field is conditional on another field like the matter overdensity field, $\delta$, we can denote $\delta^{\mathrm{b}}$ as a `biased' tracer of $\delta$.

When observing tracers like galaxies, the field is projected onto the curved sky such that lookback time and comoving distance have a unique mapping. In this case, we can describe the fields in terms of the angular two-dimensional coordinate on the sky, $\bm{\theta}$, and the redshift, $z$, such that $\delta \rightarrow \delta(\bm{\theta}, z)$. Hence, a given redshift slice of this projected field is given by
\begin{equation}
    \delta(\bm{\theta}) = \int_{-\infty}^{+\infty} \mathrm{d}z \, W(z) \,\delta(\bm{\theta}, z),
    \label{eqn:projection}
\end{equation}
\noindent where $W(z)$ is the weighting function for the projection. Throughout this work, $W(z)$ is always a rectangular function over various redshift ranges. This two-dimensional scalar field, $\delta(\bm{\theta})$, can then be expanded into spherical harmonic space as follows
\begin{equation}
    \delta_{\ell m} = \int \mathrm{d}^2 \bm{\theta} \, \delta(\bm{\theta})\, _0Y_{\ell m}^{*}(\bm{\theta}),
\end{equation}
\noindent where $_0Y_{\ell m}$ are the spin-zero spherical harmonics for angular number, $\ell$, and order, $m$, while $\delta_{\ell m}$ gives the angular coefficients of the field such that
\begin{equation}
    \delta(\bm{\theta}) = \sum_{\ell m} \delta_{\ell m} \, _0Y_{\ell m}(\bm{\theta}).
    \label{eqn:spherical_harmonics_exp}
\end{equation}

\section{BOLTS Model: Bias of Overdensities in Large-scale Tracers on the Sphere}\label{derivation}

\subsection{Derivation of the model}

The standard linear galaxy bias in real (angular) space is typically defined as
\begin{equation}
    \langle \delta^{\mathrm{b}} (\bm{\theta}, z) \rangle = b(z) \, \delta(\bm{\theta}, z) ,
    \label{eqn:standard_bias}
\end{equation}
\noindent where $\langle \delta^{\mathrm{b}} (\bm{\theta}, z) \rangle$ is the expectation of the overdensity field of biased tracers and $b(z)$ is the constant linear bias as a function of redshift. This standard bias prescription acknowledges that galaxy bias is stochastic through the fact that the equality in Equation~\ref{eqn:standard_bias} only holds in expectation. However, this model does not make any predictions for the higher-order moments of the distribution of $ \delta^{\mathrm{b}} (\bm{\theta}, z)$ beyond its mean. Additionally, the linear bias term, $b(z)$, is typically taken to be scale-independent, i.e. local. To fully incorporate the non-locality and stochasticity of galaxy bias in the linear regime, we consider a broader ansatz.

If we assume the mapping from the underlying matter overdensity field, $\delta$, to the tracer field, $\delta^{\mathrm{b}}$, is linear in harmonic space, the most general bias model can be defined as follows
\begin{equation}
    \delta^{\mathrm{b}}_{\ell m}=\sum_{\ell'm'}B_{\ell m,\ell'm'}\delta_{\ell'm'},
\end{equation}
\noindent where the $B_{\ell m,\ell'm'}$ is a tensor that defines the coupling across all angular scales, $\ell$, and modes, $m$. For the purposes of this work, we assume that there is not any coupling between the angular scales and modes, $\ell-m$, while also assuming that, at a given $\ell$ and $m$, the bias does not cause any scale or mode mixing such that
\begin{equation}
    \delta_{\ell m}^{\mathrm{b}} = b_{\ell m} \, \delta_{\ell m},
\label{eq:general_bias_model}
\end{equation}
\noindent where $\delta_{\ell m}^{\mathrm{b}}$ and $b_{\ell m}$ are the angular coefficients of the tracer field and the biasing field, respectively. This bias model is inherently non-local, as there is an explicit scale dependence in the bias field, $b$, as a function of angular scale, $\ell$.

A deterministic linear map on the sphere commutes with rotations if and only if its harmonic representation is diagonal with $m$-independent eigenvalues. Each $\ell$-subspace transforms irreducibly under rotations, so a covariant map must act on it as $b_{\ell}$ times the identity, i.e. as an isotropic convolution. The diagonal ansatz of Equation~(\ref{eq:general_bias_model}) is therefore written in a chosen coordinate frame, and once $b_{\ell m}$ depends on $m$ (as every realisation of a stochastic bias field does), the realised map singles out that frame. Rotating the input field and applying the model is not equivalent to applying the model and then rotating. Statistical isotropy is instead only recovered in expectation, and only order by order. The biased field is exactly homogeneous and isotropic at the two-point level by construction, while at third order it is only approximately isotropic, with corrections confined to coincident-mode configurations whose weight in isotropically averaged estimators is $\mathcal{O}((2\ell+1)^{-1})$. For the same reason, the split of a realised $b_{\ell m}$ into real and imaginary parts depends on the frame, while the parameters of the model do not, as in expectation they must obey isotropy (Section~\ref{derivation}).

To satisfy the homogeneity and isotropy assumptions set out in the Friedmann-Lemaître-Robertson-Walker (FLRW) metric of the standard model of cosmology, the biased field must be a homogeneous random field such that
\begin{equation}
    \Ev*{ \delta_{\ell m}^{\mathrm{b}} \, (\delta_{\ell' m'}^{\mathrm{b}})^{*} } = \delta_{\ell \ell'}^{\mathrm{K}} \, \delta_{m m'}^{\mathrm{K}} \, C_{\ell}^{\mathrm{bb}},
    \label{eqn:bb_spectrum}
\end{equation}
\noindent where~$\Ev{\:\cdot\:}$ denotes the expectation value, $\delta_{\ell \ell'}^{\mathrm{K}}$ and $\delta_{mm'}^{\mathrm{K}}$ are Kronecker deltas for their respective indices, while $C_\ell^{\mathrm{bb}}$ is the angular power spectrum of the biased field. Assuming that the $b_{\ell m}$ and $\delta_{\ell m}$ are independent, we recompute this relation while substituting Equation~(\ref{eq:general_bias_model}), we obtain that
\begin{equation}
    \begin{aligned}
        \Ev*{ \delta_{\ell m}^{\mathrm{b}} \, (\delta_{\ell' m'}^{\mathrm{b}})^{*} } & = \Ev*{ b_{\ell m} \, b_{\ell' m'}^{*}} \, \Ev*{ \delta_{\ell m} \, \delta_{\ell' m'}^{*}} \\
        & = \delta_{\ell \ell'}^{\mathrm{K}} \, \delta_{m m'}^{\mathrm{K}} \, \Ev*{ |b_{\ell m}|^2 } \, C_{\ell},
    \end{aligned} 
    \label{eqn:galaxy_galaxy_c_ell}
\end{equation}
\noindent where $C_{\ell}$ is the angular power spectrum of the matter overdensity field. To satisfy homogeneity, the expectation of $|b_{\ell m}|^2$ must be independent of $m$, i.e. $\Ev*{|b_{\ell m}|^2} = \Ev*{|b_{\ell}|^2}\, \forall\,  m$.

The matter field, $\delta$, and the biased field, $\delta^{\mathrm{b}}$, also have to jointly satisfy homogeneity such that
\begin{equation}
    \begin{aligned}
        \Ev*{ \delta_{\ell m} \, (\delta_{\ell' m'}^{\mathrm{b}})^{*} } & = \Ev*{ \delta^{\mathrm{b}}_{\ell m} \, \delta_{\ell' m'}^{*} } =  \delta_{\ell \ell'} \, \delta_{m m'} \,  C_{\ell}^{\mathrm{b}} \\
        & = \delta_{\ell \ell'}^{\mathrm{K}} \, \delta_{m m'}^{\mathrm{K}} \, \Ev*{ b_{\ell} } \,  C_{\ell},
    \end{aligned}
    \label{eqn:bdensity_spectrum}
\end{equation}
\noindent where $C_{\ell}^{\mathrm{b}}$ is the cross angular power spectrum between the matter overdensity field, $\delta$, and the biased field, $\delta^{\mathrm{b}}$.

\begin{figure*}
    \centering
    \includegraphics[trim={0cm 1.05cm 0cm 0cm},clip,width=0.9\linewidth]{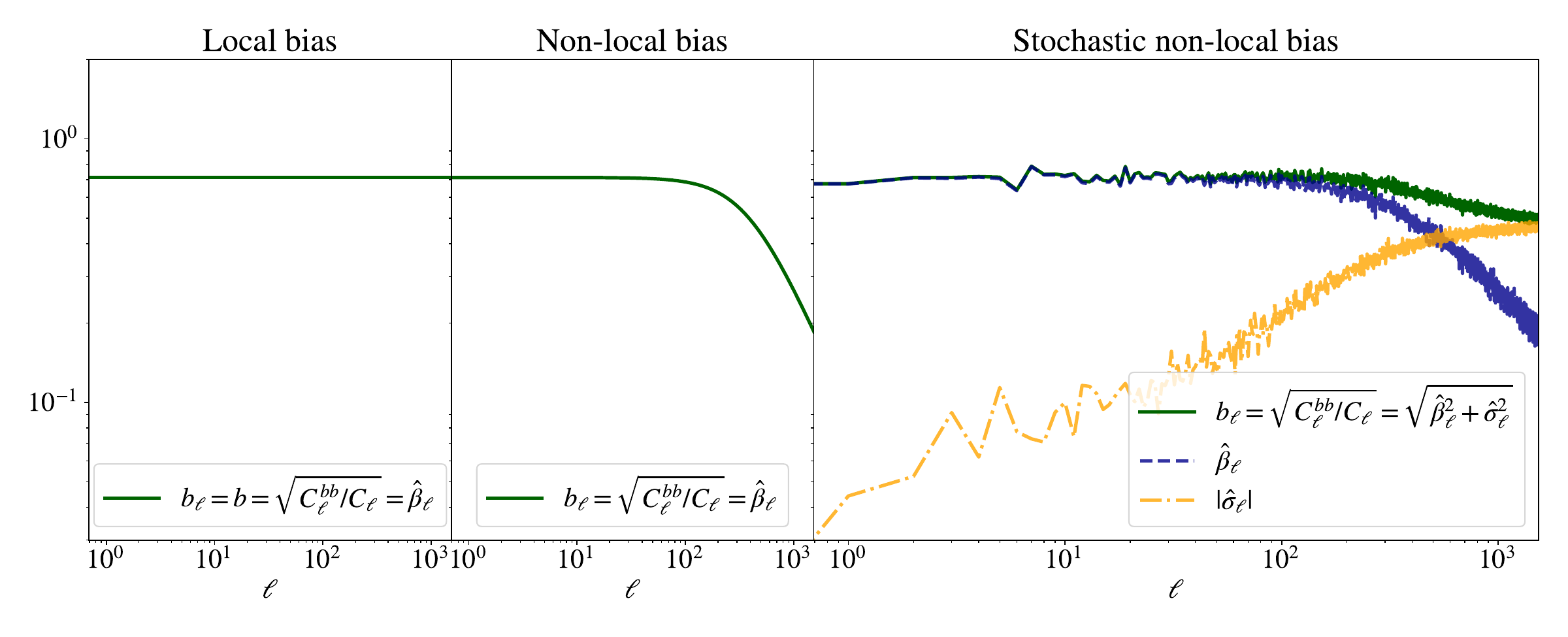}
    \caption{Square root of the angular power spectrum of the bias, $b_{\ell}$, over angular scales $\ell$ for three different models of linear galaxy bias (green solid line). The left panel shows the bias for a local linear bias, i.e. without any scale dependence. The middle panel shows a model where there is a scale dependence, i.e. non-locality. The right panel shows a realisation of the non-local and stochastic bias model presented in this work. The blue dashed line shows the deterministic contribution, $\beta_{\ell}$, and the dot-dashed orange line shows the stochastic contribution, $\sigma_{\ell}$. The values of  $\beta_{\ell}$ and $\sigma_{\ell}$ are chosen to be in approximate agreement with the measurements from the \texttt{FLAMINGO} simulations over the redshift range from $z=0.50$ to $z=0.55$ (see Figure~\ref{fig:calibration}).}
    \label{fig:model_interpretation}
\end{figure*}

The requirement of joint homogeneity together with the requirement that both $\delta$ and $\delta^{\mathrm{b}}$ must be real-valued fields implies that $\Ev*{ \delta_{\ell m} \, (\delta_{\ell' m'}^{\mathrm{b}})^{*} } \in \mathbb{R}$ and $C_{\ell} \in \mathbb{R}$, such that $\Ev*{ b_{\ell} }\in \mathbb{R}$. For a real-valued field the condition $(\delta_{lm}^{\rm b})^* = (-1)^m \, \delta_{l-m}^{\rm b}$ implies
\begin{equation}
    b_{lm}^* = b_{l-m} \,.
\end{equation}
Note that the coefficients~$b_{lm}$ do not have the symmetry of a real-valued field: the right-hand side is missing a factor of~$(-1)^m$ here.
However, similar to a real-valued field, it follows that the $m=0$ coefficients $b_{l0}^* = b_{l0}$ must be real-valued. We note that realness itself correlates the coefficients: it fixes the $m < 0$ coefficients through $b_{\ell m}^{*} = b_{\ell, -m}$ and makes $b_{\ell 0}$ real. Beyond this pairing, joint homogeneity places no constraint at the two-point level on correlations among the independent coefficients $\{b_{\ell m},\, m \ge 0\}$, so we choose the simplest assumption: they are mutually uncorrelated.

Therefore, starting from the most general non-local linear biasing model, we have shown that, to satisfy homogeneity and realness of all fields, the expectation of the angular coefficients of the biasing field, $b$, must be constant for a given value of $\ell$ and real-valued. 

We know from simulations and observations (see e.g. \citealt{dekel1999stochastic, taruya1999stochastic, matsubara1999stochasticity, McDonald2009clustering, Desjacques2018large, Kokron2022priors, Zennaro2022priors}) that galaxy formation and non-linear structure formation induce stochasticity to perturbations of the galaxy density over a broad range of scales. The initial conditions of the matter and galaxy density fields at small scales are not correlated between each other over large distances. This implies that the process of mapping the matter density field $\delta$ to the galaxy density field $\delta^{\mathrm{b}}$ has an inherently stochastic component. Consequently, a deterministic linear non-local biasing model, $b_{\ell}$, may not fully capture the stochastic bias. 

Additionally, when sampling galaxies from the $\delta^{\mathrm{b}}$ field through, for example, a Poissonian process, one can induce an additional layer of stochasticity to the resulting galaxy count field $N$. However, as we show in Appendix~\ref{appendix:stochasticity_in_n}, in the context of a linear galaxy bias model, this sampling stochasticity cannot account for any bias besides $b=1$. Therefore, the number count stochasticity is insufficient to be the source of a stochastic bias in a general linear bias model. An additional stochastic contribution at the mapping $\delta \rightarrow \delta^{\mathrm{b}}$ is required.

Given that the assumptions made only require the biasing field, $b_{\ell m}$, to be real-valued and independent of $m$ in the expectation, we have the freedom to add the contribution from a random variable, $\gamma_{\ell m}$, to the biasing field. By incorporating this additional term in the bias, BOLTS extends standard linear bias as follows
\begin{equation}
    b_{\ell m} = \beta_{\ell} + \gamma_{\ell m},
\end{equation}
\noindent where $\beta_{\ell}$ is the deterministic real-valued bias contribution. As the spherical harmonic coefficients of a field are in general complex, without loss of generality, we allow $\gamma_{\ell m} = \epsilon_{\ell m} + \mathrm{i} \eta_{\ell m} \in \mathbb{C}$ with  $\epsilon_{\ell m}, \eta_{\ell m} \in \mathbb{R}$ for $m>0$.

To avoid violating homogeneity, $\Ev*{ \gamma_{\ell m}}_{m} = 0 \, \forall \, \ell$ in any distribution from which $\gamma_{\ell m}$ is drawn. In principle, these fields could be drawn from a general normalisable distribution. If the distributions are given by a Gaussian, for a given $\ell$, the distributions of $\epsilon_{\ell m}$ and $ \eta_{\ell m}$ are given by
\begin{equation}
    \begin{pmatrix}
        \epsilon_{\ell m}\
        \eta_{\ell m}
    \end{pmatrix} 
    \sim \mathcal{N} \Bigg( \bm{\mu_{\ell}} = 
    \begin{pmatrix}
        0\\
        0
    \end{pmatrix},
    \bm{\Sigma}_{\ell} =
    \begin{pmatrix}
        \sigma_{1, \ell}^2 & \rho \, \sigma_{1, \ell} \, \sigma_{2, \ell}\\
        \rho \, \sigma_{1, \ell} \, \sigma_{2, \ell} & \sigma_{2, \ell}^2
    \end{pmatrix}
    \Bigg),
\end{equation}
\noindent where $\mathcal{N}$ is a two-dimensional multivariate Gaussian or normal distribution with mean $\bm{\mu}_{\ell}$ and a covariance matrix $\bm{\Sigma_{\ell}}$, with $\sigma_{1, \ell}$ and $\sigma_{2, \ell}$ as the standard deviations of the real and imaginary parts of $\gamma_{\ell m}$, respectively. The distribution accounts for a correlation between the real and imaginary parts through the $\rho$ correlation factor. Alternatively, this model can also be expressed as follows
\begin{equation}
    b_{\ell m} \sim \mathcal{CN}(\beta_{\ell} + \mathrm{i} \, 0,
   \bm{\Sigma}_{\ell}) \, \mathrm{for} \, 1 \le m \le \ell,
\end{equation}
\noindent where $\mathcal{CN}$ is a complex Gaussian or normal distribution. Note that $\beta_{\ell} \in \mathbb{R}$ throughout. The $m = 0$ mode requires separate treatment. Realness makes $b_{\ell 0}$ real ($\eta_{\ell 0} = 0$), while homogeneity requires $\Ev*{|b_{\ell m}|^2}$ to be independent of $m$ (Equation~\ref{eqn:galaxy_galaxy_c_ell}). Since this expectation equals $\beta_{\ell}^2 + \sigma_{1,\ell}^2 + \sigma_{2,\ell}^2$ for the complex $m > 0$ modes, the real $m = 0$ coefficient must carry the full stochastic variance, \begin{equation}   
    b_{\ell 0} \sim \mathcal{N}\big(\beta_{\ell},\,   \sigma_{1,\ell}^2 + \sigma_{2,\ell}^2\big), 
\end{equation} 
\noindent rather than the per-component variance. This mirrors the standard convention for the spherical harmonic coefficients of a real isotropic Gaussian field, whose $m = 0$ coefficients are real with variance $C_{\ell}$ while the real and imaginary parts of each $m > 0$ coefficient carry $C_{\ell}/2$. The $m < 0$ coefficients are fixed by conjugation, $b_{\ell, -m} = b_{\ell m}^{*}$.

From simulations, as shown in Appendix~\ref{appendix:two-point}, we find that the distribution of spherical harmonic coefficients, $a_{\ell m}$, per angular scale, $\ell$, is approximately Gaussian and broadly consistent with this model.

Given the deterministic and stochastic contributions to the bias model, using Equation~(\ref{eqn:galaxy_galaxy_c_ell}), we can derive each contribution to the angular power spectrum of the biased field as follows
\begin{equation}
    \begin{aligned}
        \Ev*{ |b_{\ell m}|^2} & = \Ev*{ (\beta_{\ell} + \epsilon_{\ell})^2 + \eta_{\ell}^2 } \\
        & = \beta_{\ell}^2 +\Ev*{ \epsilon_{\ell}^2 } + \Ev*{ \eta_{\ell}^2 } \\
        & = \beta_{\ell}^2 + \sigma_{1, \ell}^2 + \sigma_{2, \ell}^2.
    \end{aligned}
    \label{eqn:bias_two_point_contribution}
\end{equation}
Hence, the stochasticity in the mapping from $\delta$ to $\delta^{\mathrm{b}}$ boosts the angular auto power spectrum of the biased field according to the variance of $b_{\ell m}$ for a given $\ell$, i.e. $\sigma_{\ell}^2 \equiv \sigma_{1, \ell}^2 + \sigma_{2, \ell}^2$.

We can also derive the predicted third-order moments. Because $b$ and $\delta$ are independent, $\Ev*{\delta^{\mathrm{b}}_{\ell_1 m_1}\delta^{\mathrm{b}}_{\ell_2 m_2}\delta^{\mathrm{b}}_{\ell_3 m_3}} = \Ev*{b_{\ell_1 m_1} b_{\ell_2 m_2} b_{\ell_3 m_3}}\,
\Ev*{\delta_{\ell_1 m_1}\delta_{\ell_2 m_2}\delta_{\ell_3 m_3}}$, and for a Gaussian $\gamma_{\ell m}$ the third moment of the bias field is
\begin{equation}
    \begin{aligned}
        \Ev*{b_1 b_2 b_3} = \beta_{\ell_1}\beta_{\ell_2}\beta_{\ell_3} 
        + \sum_{\{i,j\}} \delta^{\mathrm{K}}_{\ell_i \ell_j} \beta_{\ell}\Big[ \sigma_{\ell_i}^{2} \delta^{\mathrm{K}}_{m_i, -m_j} \\
        +\, Q_{\ell_i}(m_i) \delta^{\mathrm{K}}_{m_i m_j} \big(1 - \delta^{\mathrm{K}}_{m_i 0}\big) \Big],
    \end{aligned}
    \label{eqn:bispectrum_b}
\end{equation}
\noindent where the sum runs over the three unordered index pairs $\{i,j\} \subset \{1,2,3\}$, with $\sigma_\ell^2 \equiv \sigma_{1,\ell}^2 + \sigma_{2,\ell}^2$, and where $Q_{\ell_i}(m_{i})$ is given by
\begin{equation}
    Q_{\ell}(m) = \begin{cases}
     \sigma_{1,\ell}^{2} - \sigma_{2,\ell}^{2}
       + 2 \mathrm{i} \rho \sigma_{1,\ell} \sigma_{2,\ell} , & m > 0 ,\\
     \sigma_{1,\ell}^{2} - \sigma_{2,\ell}^{2}
       - 2 \mathrm{i} \rho \sigma_{1,\ell} \sigma_{2,\ell} , & m < 0 ,
    \end{cases}
\end{equation}
\noindent which is the pseudo-variance $\Ev*{\gamma_{\ell m}^2}$ of the stochastic component (where its conjugation for $m < 0$ follows from $\gamma_{\ell, -m} = \gamma_{\ell m}^{*}$). 

The stochastic corrections are confined to configurations in which two coefficients coincide and are not proportional to the Wigner-3$j$ structure of an isotropic bispectrum (leading to the approximate third-order isotropy). In an isotropically averaged bispectrum estimator, which is real for any real-valued field, the imaginary part $\propto \rho$ cancels between $\pm m$. Realness therefore does not constrain $\rho$, but isotropy fixes $\rho = 0$ (see Section~\ref{derivation}). For the equilateral configuration,
\begin{equation}
  \frac{B^{\mathrm{bbb}}_{\ell\ell\ell}}{B_{\ell\ell\ell}}
  = \hat{\beta}_{\ell}^3
  + \frac{3\hat{\beta}_{\ell}\hat{\sigma}_{\ell}^2}{2\ell+1}
  + 3\hat{\beta}_{\ell}\big(\hat{\sigma}_{1,\ell}^2 -
    \hat{\sigma}_{2,\ell}^2\big)\, W_2(\ell),
  \label{eqn:analytical_blll}
\end{equation}
where $W_2(\ell) = \sum_{m \neq 0} \big(\begin{smallmatrix} \ell & \ell & \ell \\ m & m & -2m \end{smallmatrix}\big)^2 \approx 0.6/(2\ell+1)$. Both stochastic terms are suppressed by $\sim (2\ell+1)^{-1}$, so at $\ell \gtrsim 100$, they are sub-per-cent corrections to $\hat{\beta}^3_{\ell}$ even where $\sigma_{\ell} \sim \beta_{\ell}$. We verified this numerically on non-Gaussian mock fields, and found these relations to hold.

The equal split of the variance and the vanishing correlation between the real and imaginary parts follow from statistical isotropy. For the spherical-harmonic coefficients of any real isotropic random field with finite second moments, the real and imaginary parts are uncorrelated with equal variance for every $m \geq 1$ \citep{baldi2007characterizations}. Applied to the stochastic residual field $\delta^{\mathrm{g}} - K_{\beta} \ast \delta^{\mathrm{c}}$, whose coefficients are $\gamma_{\ell m}\,\delta^{\mathrm{c}}_{\ell m}$, this fixes $\rho = 0$ and $\sigma_{1,\ell} = \sigma_{2,\ell}$ exactly at the field level. For the multiplicative coefficient $\gamma_{\ell m}$ itself, the rotation by $\pi$ about an axis in the equatorial plane maps every per-mode ratio to its complex conjugate, $b_{\ell m} \to b^{*}_{\ell m}$, so isotropy still forces $\rho = 0$. Equal variances hold exactly when the residuals are Gaussian or its phase is independent of that of $\delta^{\mathrm{c}}_{\ell m}$, and otherwise up to coincident-mode corrections of the same $\mathcal{O}((2\ell+1)^{-1})$ as the third-order terms previously discussed. We measure the asymmetry $(\sigma_{1,\ell}^{2} - \sigma_{2,\ell}^{2})/(\sigma_{1,\ell}^{2} + \sigma_{2,\ell}^{2})$ in \texttt{FLAMINGO} and find it to be below $1\%$ at $\ell \geq 256$ for all halo-mass bins and redshifts (Appendix~\ref{appendix:two-point}). The one-point distributions of the coefficients are also consistent with equal variances, and it does not bias the two-point statistics of the bias field nor significantly the higher-order statistics of Section~\ref{calibration_testing}.

Finally, we can conclude that a general linear bias model in harmonic space, $b_{\ell m}$, which conserves the homogeneity, isotropy, and realness of the field:
\begin{itemize}[leftmargin=15pt]
    \item Must be independent of $m$ in its expectation: $\Ev*{ b_{\ell m} } = \beta_{\ell}$.
    \item Must be real-valued in its expectation: $\beta_{\ell} \in \mathbb{R}$.
    \item Must have a zero-mean complex random contribution, $\gamma_{\ell m}$, to incorporate stochasticity, with uncorrelated real and imaginary parts ($\rho = 0$) and, for the complex $m > 0$ modes, equal variances $\sigma_{1,\ell} = \sigma_{2,\ell}$. The single real $m = 0$ mode carries the total variance $\sigma_{\ell}^2$.
\end{itemize}

For a fixed redshift bin or shell $j$, defined by a weighting function $W^{(j)}(z)$ in Equation~(\ref{eqn:projection}), we define a redshift-dependent linear non-local stochastic bias BOLTS model as follows
\begin{equation}
    b_{\ell m}^{(j)} \sim \mathcal{CN} \left( \beta_{\ell}^{(j)} + \mathrm{i} \, 0, \, 
    \bm{\Sigma}_{\ell}^{(j)} =
    \begin{pmatrix}
        \frac{1}{2}\Big[\sigma_{\ell}^{(j)}\Big]^2 & 0\\
        0 & \frac{1}{2}\Big[\sigma_{\ell}^{(j)}\Big]^2
    \end{pmatrix}
    \right),
    \label{eqn:final_model}
\end{equation}
\noindent for $1 \le m \le \ell$, while $b_{\ell 0}^{(j)} \sim \mathcal{N}\big(\beta_{\ell}^{(j)}, [\sigma_{\ell}^{(j)}]^2\big)$.

For a given redshift bin $j$ and a given angular scale $\ell$, the model has three free parameters which can be fit from simulations and/or observations. As measured in simulations (see Section~\ref{calibration_testing}), in expectation, $\beta_{\ell}$ and $\sigma_{\ell}$ should vary smoothly as a function of $\ell$. The number of free parameters can be drastically reduced further through fitting a semi-empirical function of $\beta_{\ell}$ and $\sigma_{\ell}$ as a function of $\ell$ (see Section~\ref{forward_model}). Additionally, if available in the simulation, the model can also be fit on bins of or conditional on galaxy types, star-formation rate, stellar mass, baryonic feedback models, cosmological parameters, or any other property that may affect galaxy bias. These dependencies are beyond the scope of this paper and will be explored in future work.

\subsection{Calibration of the model parameters}\label{calibration}
As the model is linear and analytic, we can analytically derive the mean square error, $\chi^2$, as follows
\begin{equation}
    \chi^2 = \sum_{\ell m} |\delta^{\mathrm{b}}_{\ell m} - \beta_{\ell} \delta_{\ell m} |^2.
\end{equation}
Defining the minimum mean square error estimator of $\beta_{\ell}$ as follows
\begin{equation}
    \frac{\partial \chi^2}{\partial \beta_{\ell}} \Bigg|_{\beta_{\ell} = \hat{\beta}_{\ell}} = 0 \Rightarrow \hat{\beta}_{\ell} = \frac{\sum_{\ell m } \delta^{\mathrm{b}}_{\ell m} \delta_{\ell m}^{*}}{\sum_{\ell m}|\delta_{\ell m}|^2} = \frac{C_{\ell}^{\mathrm{b}}}{C_{\ell}},
    \label{eqn:beta_ml}
\end{equation}
\noindent where $\hat{\cdot}$ indicates the best estimate of the respective parameter. From Equations~\ref{eqn:galaxy_galaxy_c_ell} and~\ref{eqn:bias_two_point_contribution}, we know that
\begin{equation}
    \hat{b}_{\ell}^2 = \frac{C_{\ell}^{\mathrm{bb}}}{C_{\ell}} = \hat{\beta}_{\ell}^2 + \hat{\sigma}_{\ell}^2.
\end{equation}
From this, one finds that the best estimator for the variance of the stochastic bias is given by
\begin{equation}
    \hat{\sigma}_{\ell}^2 = \frac{C_{\ell}^{\mathrm{bb}}}{C_{\ell}} - \hat{\beta}_{\ell}^2.
    \label{eqn:sigma_ml}
\end{equation}
Note that from the power spectra alone one can only measure the total variance from the stochastic contribution, $\hat{\sigma}_{\ell}^2$. As shown in Appendix~\ref{appendix:model_independent}, only the total variance of a field is measurable at the two-point level, since Gaussian random fields are fully characterised by their mean and power spectra alone.

\subsection{Interpretation of the model}

The non-local stochastic galaxy bias model defined in Section~\ref{derivation} incorporates the known phenomena of non-locality and stochasticity into a description of linear galaxy bias in a spherical projection of a density field. As opposed to standard linear bias, this naturally incorporates and separates various contributions to the galaxy bias: deterministic bias and stochastic bias (from the uncoupled small-scale initial conditions and galaxy formation processes). Notably, the model can also separate any stochastic contribution from shot noise, i.e. random sampling of tracers from the biased density field, so when fitting the model, any shot noise bias is accounted for and when using the model as a simulator, shot noise-free fields can be generated when the shot-noise-subtracted calibration is used. In practice, we do not subtract a shot-noise term when measuring the galaxy bias from simulations. The stochastic amplitude $\hat{\sigma}_{\ell}^2$ then absorbs the sampling noise of the discrete galaxy field, which for Poisson sampling contributes $(4\pi/N_{\mathrm{g}})/C_{\ell}^{\mathrm{cc}}$. This choice is deliberate and self-consistent for the field-level comparisons of Section~\ref{calibration_testing}. The simulated $\delta^{\mathrm{g}}$ maps contain their sampling noise and the forward-modelled mocks reproduce it through $\sigma_{\ell}$ without sampling discrete counts a second time. It also keeps $\hat{\sigma}^2_{\ell} \ge 0$ at every multipole whereas subtracting an estimated shot term drives $\hat{\sigma}^2_{\ell}$ negative on hundreds of multipoles for the sparse high-mass bin. If required, the model can separate the intrinsic stochasticity from the Poisson noise (see Appendix~\ref{appendix:order_of_ops}). If the true sampling noise departs from being Poissonian \citep{casas-miranda2002on, baldauf2013halo, Ginzburg2017shot}, the residuals are absorbed into $\hat{\sigma}^2_{\ell}$.

As shown in Figure~\ref{fig:model_interpretation}, a standard scale-independent linear bias model, i.e. $\delta^{\mathrm{b}}_{\ell m} = b \,\delta_{\ell m}$, produces scale-independent angular power spectrum of the bias field, $|b_{\ell}|^2$. We know this is unphysical; both from the known violation of local-in-matter-density galaxy bias as well as from the fact that, at small scales, baryonic physics, exclusion effects, etc., modulate the galaxy bias. The model shown in the middle panel of Figure~\ref{fig:model_interpretation} addresses this by allowing for a scale-dependence in the bias model, i.e. $\delta^{\mathrm{b}}_{\ell m} = b_{\ell} \,\delta_{\ell m}$. This additional freedom in the model allows for a physical scale dependence, but it does not allow for stochastic bias as discussed in Section~\ref{derivation}. The right panel of Figure~\ref{fig:model_interpretation} shows a realistic $b_{\ell}$ derived from the model defined in Equation~(\ref{eqn:final_model}). The added stochasticity in the bias has two effects: it adds variance to the realised $b_{\ell}$ and $\beta_{\ell}$, and this variance in turn adds additional power to the biasing field quantified by $\sigma_{\ell}^2$. This flexibility in the model is necessary to explain the galaxy density field from current state-of-the-art cosmological simulations, such as \texttt{FLAMINGO}. Without assuming any model, when measuring $C^\mathrm{b}_{\ell}/C_{\ell}$ and $C^{\mathrm{bb}}_{\ell}/C_{\ell}$, if galaxy bias was purely deterministic, $(C^\mathrm{b}_{\ell}/C_{\ell})^2 = C^{\mathrm{bb}}_{\ell}/C_{\ell}$ for a homogeneous random field as set out in Equations~\ref{eqn:bb_spectrum},~\ref{eqn:galaxy_galaxy_c_ell}, and~\ref{eqn:bdensity_spectrum}. As is discussed in detail in Section~\ref{calibration}, when measuring these quantities from simulations, one consistently finds $(C^\mathrm{b}_{\ell}/C_{\ell})^2 < C^{\mathrm{bb}}_{\ell}/C_{\ell}$ over most scales, implying that there is an additional contribution in power to the bias field, i.e. $\Ev{b_{\ell}} < \sqrt{\Ev{b_{\ell}^2}}$. This, combined with the observed variance in $b_{\ell}$ and the known stochasticity in galaxy bias, implies that the simplest model to explain linear galaxy bias at the level of two-point statistics must be non-local and stochastic as set out in Section~\ref{derivation}.

As the spherical harmonics of the stochastic term can affect the entire sky, one could reach the conclusion that this term could violate causality on top of locality. However, at the two-point level it is statistically equivalent to adding an independent isotropic random field with angular spectrum $\sigma_{\ell}^2 C_{\ell}^{\mathrm{cc}}$, whose correlation function is short-ranged because $\sigma_{\ell}$ rises towards small scales (the fraction of the stochastic variance at $\ell < 20$ is below $10^{-3}$ for every shell and mass bin we measure).

\begin{figure}
    \centering
    \includegraphics[width=\columnwidth]{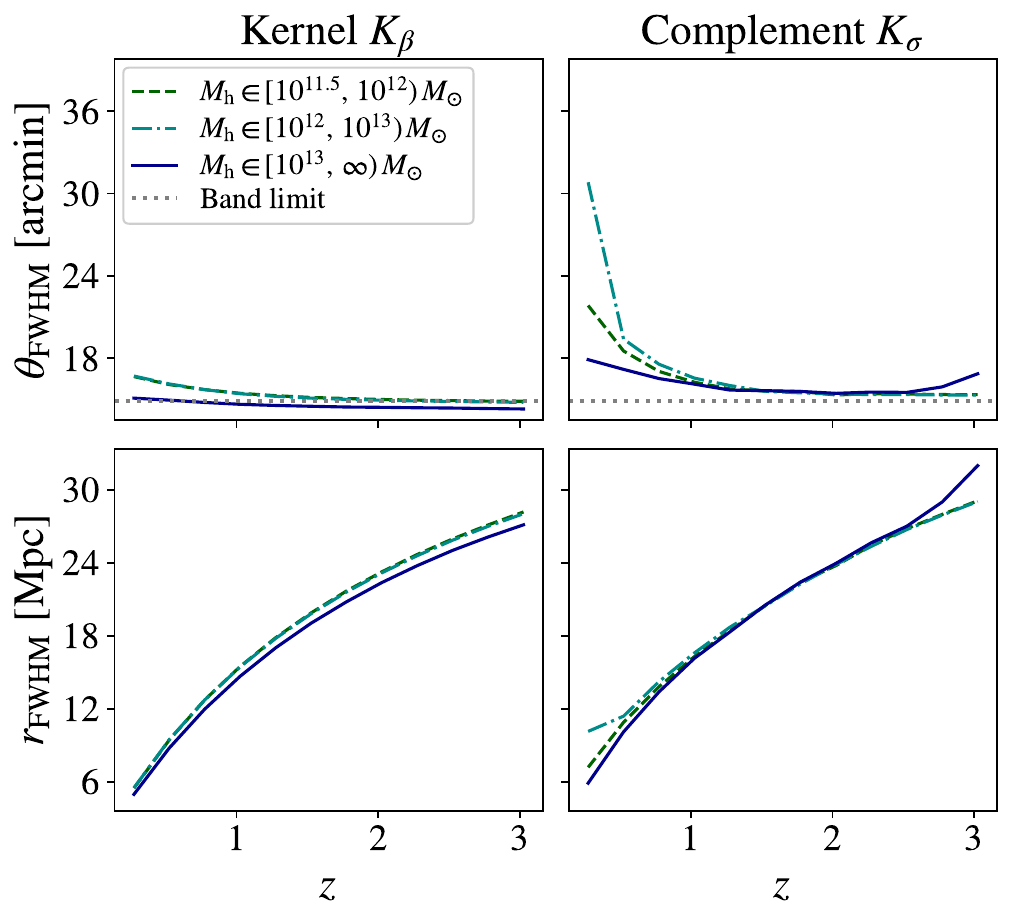}
    \caption{Real-space widths of the BOLTS kernels as a function of redshift for the three halo-mass bins consistent with the \texttt{FLAMINGO} simulations. Top row: the FWHM of the deterministic convolution kernel $K_{\beta}(\theta)$ (left) and of the low-pass complement of the stochastic amplitude, $K_{\sigma}(\theta)$ (right), the grey dotted line marks the band limit, i.e. the FWHM of a flat (perfectly local) spectrum over $\ell \le 2 N_{\mathrm{side}} = 1024$. Bottom row: the same widths as comoving lengths $r_{\mathrm{FWHM}} = \chi(z) \theta_{\mathrm{FWHM}}$.}
    \label{fig:kernel_width}
\end{figure}

The real-space widths of the deterministic and stochastic kernels are quantified in Figure~\ref{fig:kernel_width}. The Full width at half maximum (FWHM) of the deterministic convolution kernel $K_{\beta}(\theta) = \sum_{\ell} \frac{2\ell+1}{4\pi}\beta_{\ell}P_{\ell}(\cos\theta)$ and of the stochastic convolution kernel $K_{\sigma}(\theta)$ as the low-pass complement of the stochastic amplitude, whose width is the coherence scale below which stochasticity reaches full strength. As can be seen in Figure~\ref{fig:kernel_width}, the deterministic kernel width is small. Its half-width exceeds the band-limit floor of the calibration ($\ell \le 1024$, $\theta \sim 14.9'$) by at most 12\% at $z \simeq 0.3$ and by under 2\% beyond $z \sim 1.5$, so at map resolution $\beta_{\ell}$ acts as a quasi-local, arcminute-scale smoothing where non-locality only comes in through the small-scale variation. The scale at which the stochastic term begins to pick up is $22'$-$31'$, i.e. $r \sim 7$-$10\,\mathrm{Mpc}$ comoving at $z \approx 0.3$. This scale is an order of magnitude larger than the virial radii of the host halos ($R_{200} \sim 0.3$-$1.5\,\mathrm{Mpc}$ comoving for $10^{12}$-$10^{13.5}\,\mathrm{M}_{\odot}$). Instead, the scale is more consistent with the non-linear clustering length of the tracers. In this regime, the stochasticity is driven by the scatter in the galaxy-matter connection dependent on the halo environment. One can interpret this as an assembly-bias-like dependence on the large-scale tidal surroundings and from the line-of-sight arrangement of correlated structure within the $\Delta z = 0.05$ shell. At higher redshifts both kernel widths converge to the resolution limit as the scales where non-locality and stochasticity kick in become smaller than the resolution of the maps.

\section{The \texttt{FLAMINGO} simulations}\label{flamingo}
To test the model  presented in this work, we utilise the \texttt{FLAMINGO} (Full-hydro Large-scale structure simulations with All-sky Mapping for the Interpretation of Next Generation Observations) suite of simulations \citep{schaye2023flamingo, kugel2023flamingo}. \texttt{FLAMINGO} is a set of large-volume cosmological hydrodynamical simulations run using the \textsc{Swift} code \citep{schaller2024swift} with the \textsc{Sphenix} smoothed particle hydrodynamics (SPH) implementation \citep{borrow2022sphenix}. The simulations include a comprehensive model for baryonic physics, incorporating stellar mass loss \citep{wiersma2009chemical}, radiative cooling \citep{ploeckinger2020radiative}, star formation \citep{schaye2008relation}, and feedback from both supernovae \citep{chaikin2023thermal} and active galactic nuclei (AGN; \citealt{booth2009cosmological}), the latter two of which are calibrated to reproduce the observed $z=0$ galaxy stellar mass function and cluster gas fractions \citep{kugel2023flamingo}.

\begin{figure*}
    \centering
    \includegraphics[trim={0cm 0.4cm 0cm 0},clip,width=0.353\linewidth]{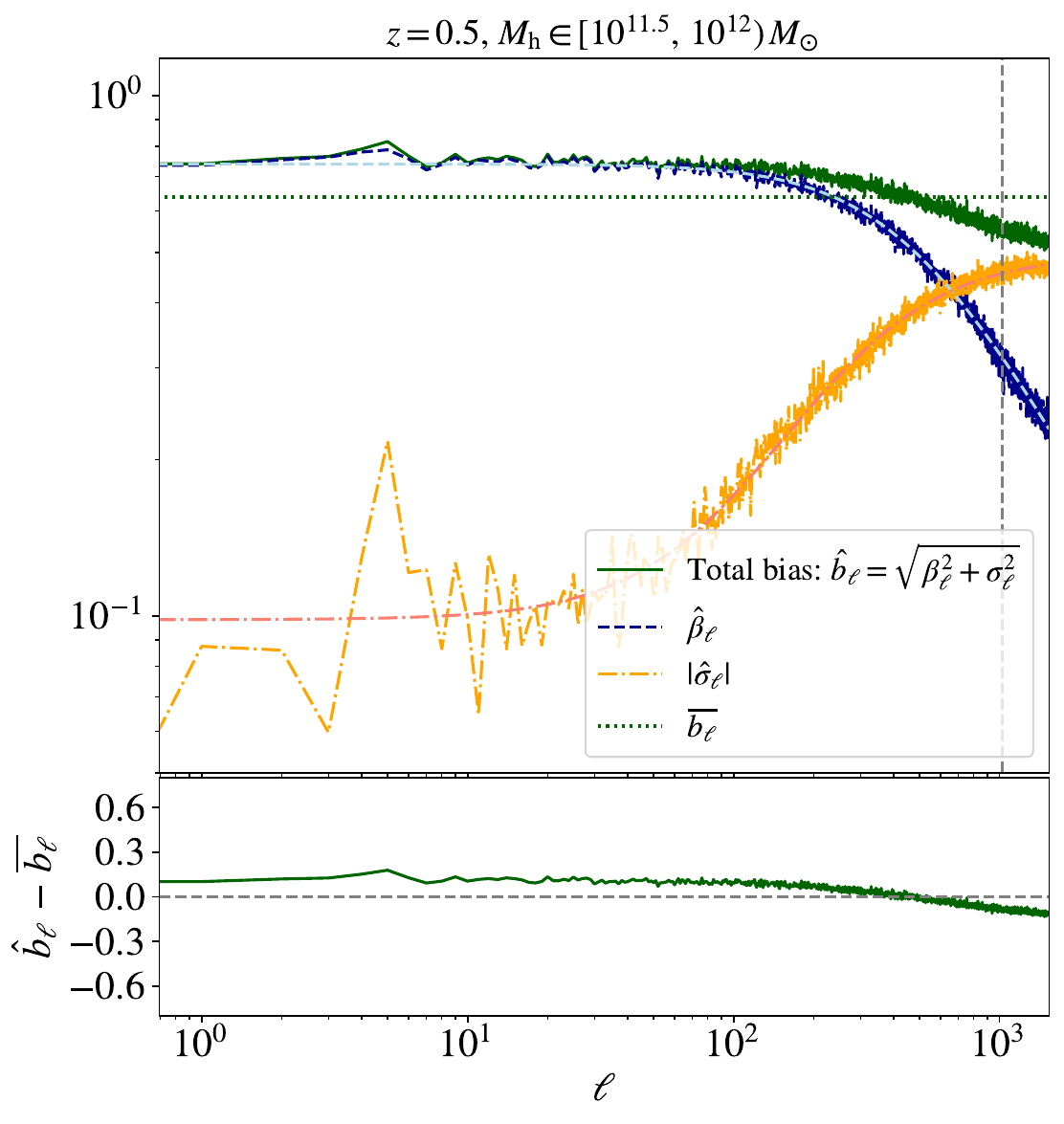}
    \includegraphics[trim={2.7cm 0.4cm 0.1cm 0},clip,width=0.3\linewidth]{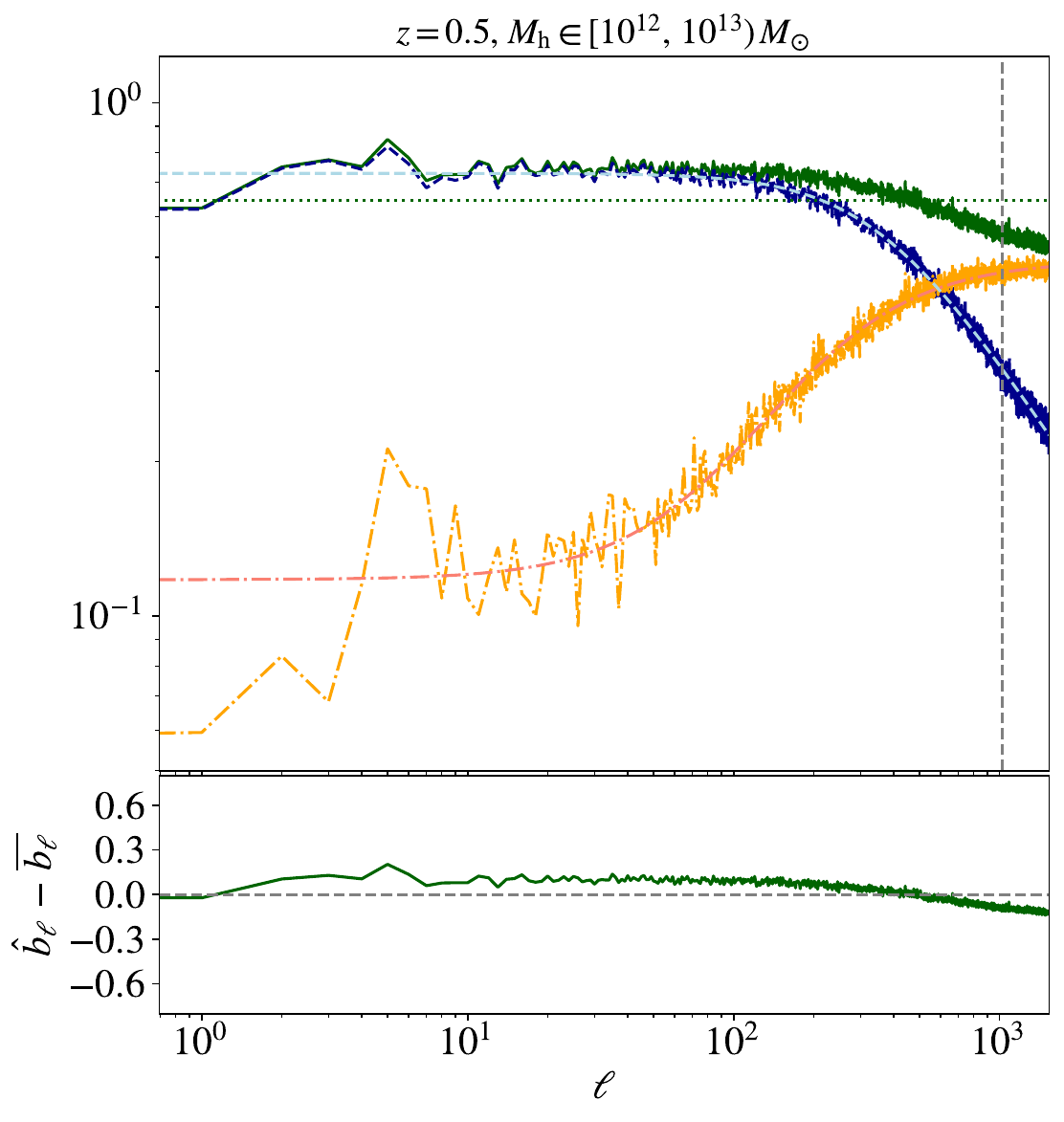}
    \includegraphics[trim={2.2cm 0.4cm 0.1cm 0},clip,width=0.3\linewidth]{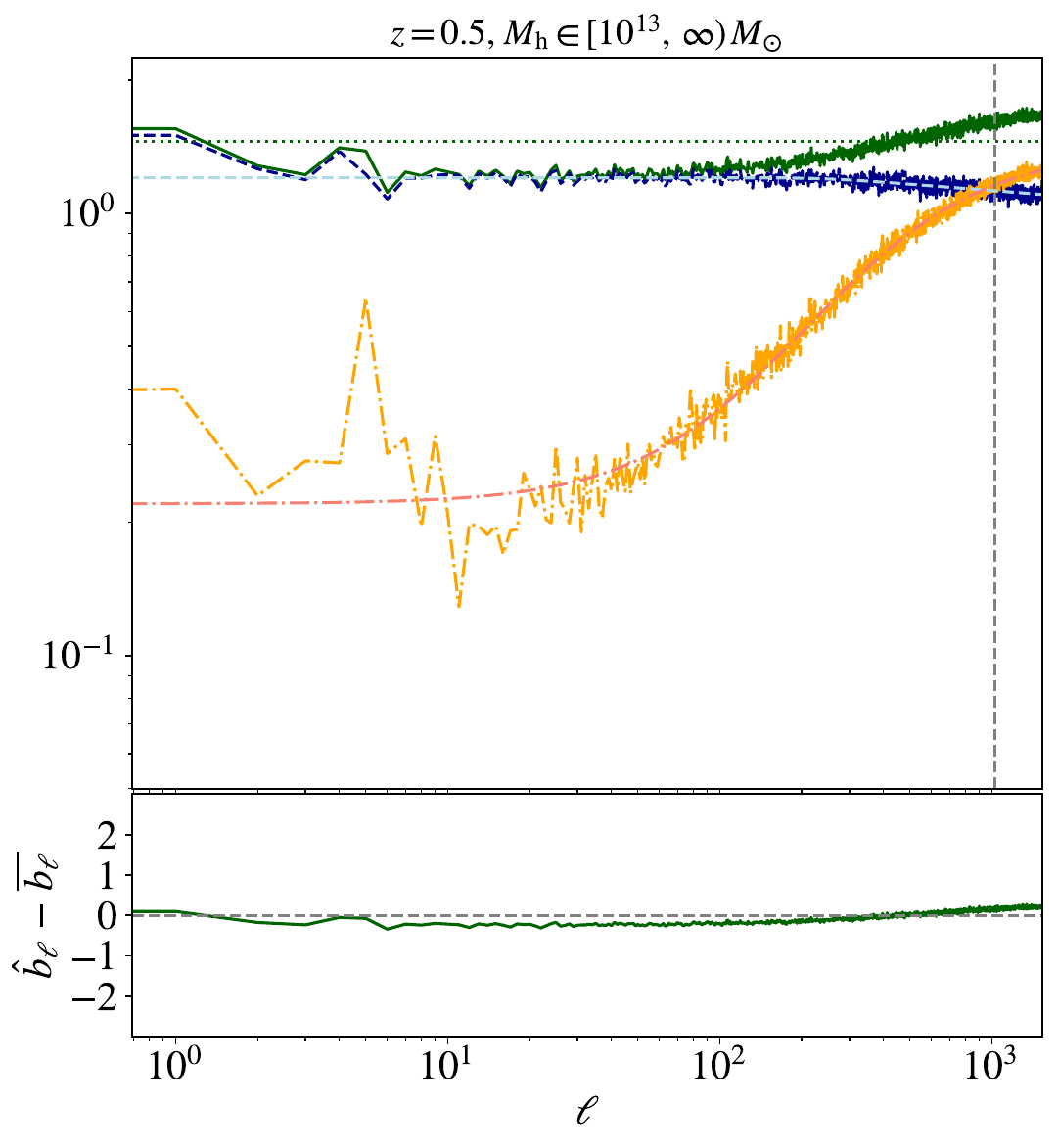}
    \caption{Galaxy bias in harmonic space as a function of angular scale, $\ell$, over the redshift range from $z=0.50$ to $z=0.55$ as measured from the `L2p8\_m9' \texttt{FLAMINGO} simulation. The dark blue dashed line shows the measured deterministic bias term as defined in Equation~\ref{eqn:beta_ml}, while the light blue dashed line shows its best fit line for the function outlined in Equation~\ref{eqn:beta_l_fit}. The orange dot-dashed line shows the contribution from the stochastic term as laid out in Equation~\ref{eqn:sigma_ml} with the best fit line according to Equation~\ref{eqn:sigma_l_fit} in the salmon-coloured dot-dashed line. The green solid line shows total galaxy bias as the sum in quadrature of the deterministic contribution and the stochastic contribution. The green dotted line shows the mean $b_{\ell}$ over all scales, while the grey vertical dashed line marks the \textsc{HealPix} pixel scale. From left to right, the panels show increasing ranges of halo mass, $M_{\mathrm{h}}$, of the galaxies in the $\delta^{\mathrm{g}}$ field. The bottom panels show the residuals between the total scale-dependent galaxy bias, $b_{\ell}$, and an equivalent local galaxy bias set to the mean, i.e. $b_{\ell} = \overline{b_{\ell}}$.}
    \label{fig:calibration}
\end{figure*}

Specifically, we employ the `L2p8\_m9'  hydrodynamical run and its corresponding dark-matter-only (DMO) counterpart within a periodic box of side length. The hydrodynamical run contains $5040^3$ baryonic particles and $5040^3$ cold dark matter (CDM) particles, alongside $2800^3$ massive neutrino particles modelled using the $\delta f$ method to suppress shot noise \citep{elbers2021optimal}. The initial particle mass is $m_{\mathrm{b}} = 1.07 \times 10^9 \, \mathrm{M}_{\odot}$ for baryons and $m_{\mathrm{CDM}} = 5.65 \times 10^9 \, \mathrm{M}_{\odot}$ for dark matter. The DMO simulation, `L2p8\_m9\_DMO', shares the same initial phases, amplitudes, and box size but treats all matter as collisionless, with the CDM particle mass increased to account for the cosmic baryon fraction, ensuring consistent total matter density $\Omega_{\mathrm{m}}$.

Note that in the linear regime the matter density fields of `L2p8\_m9' and `L2p8\_m9\_DMO' are virtually identical, as structure formation is gravity-dominated at these scales. At non-linear scales baryonic feedback suppresses small-scale structure and the halo mass function relative to the DMO run \citep{vandaalen2011effects, velliscig2014impact}, leaving galactic- and group-scale halos ($M_{\mathrm{h}} \sim 10^{11} - 10^{13} \, \mathrm{M}_{\odot}$) on average less abundant and with lower density profiles, while halo positions match given the same initial conditions \citep{vandaalen2014impact}.

Both simulations assume the `D3A' cosmology, which is the fiducial model for the \texttt{FLAMINGO} suite. The cosmological parameters are based on the \cite{planck2020results} constraints, with specific values: dimensionless Hubble constant $h = 0.681$, total matter density $\Omega_{\mathrm{m}} = 0.306$, baryonic density $\Omega_{\mathrm{b}} = 0.0486$, dark energy density $\Omega_{\Lambda} = 0.694$, spectral index $n_{\mathrm{s}} = 0.967$, and an amplitude of density fluctuations $\sigma_8 = 0.807$. The model includes a minimal sum of neutrino masses $\sum m_\nu = 0.06 \, \mathrm{eV}/\mathrm{c}^2$.

In addition to the standard snapshot outputs, the \texttt{FLAMINGO} simulations generate full-sky lightcone data on-the-fly. The particles are recorded as they traverse the backward lightcone of a virtual observer, and for the `L2p8\_m9' run lightcones are constructed for multiple observer positions. The outputs include raw particle catalogues and projected maps pixelated with the \textsc{Healpix} scheme \citep{gorski2005healpix}, integrated into concentric shells of thickness $\Delta z = 0.05$ from $z = 0$ to $3$.

For the purposes of this study, we define halos and subhalos using \textsc{HBT-Herons} (Hierarchical Bound-Tracing-Hydro-Enabled Retrieval of Objects in Numerical Simulations; \citealt{han2018hbt, forouhar2025assessing}), the fiducial subhalo finder for the \texttt{FLAMINGO} halo catalogues. It is a history-based finder. Subhalos are identified when they form in relative isolation and their particle memberships are tracked forward in time, which recovers substructure in dense environments more robustly than configuration- or phase-space methods \citep{forouhar2025assessing}. Subhalo centres are placed on the most bound particle which, unlike a centre-of-mass estimate, is not easily biased by massive substructures. This minimises spurious scatter in the measured centres of halos and galaxies, and consequently avoids biasing the measured stochasticity in the galaxy-halo connection.

Despite the efficiencies of halo finders such as \textsc{HBT-Herons}, running any subhalo finder on dark matter simulations may not always be computationally feasible when attempting to forward model many realisations of halo and galaxy catalogues. While we rely on \textsc{HBT-Herons} halo catalogues to test the validity of the BOLTS model as outlined in Section~\ref{derivation}, we outline practical considerations for the implementation of BOLTS within an efficient forward model in Section~\ref{forward_model}.

For the purposes of calibration and validation, we define `galaxies' as objects in the halo catalogue with stellar mass $M_{*} \ge 2\times10^{10}\,\mathrm{M}_{\odot}$ ($\sim20$ stellar particles). This selection is complete for $M_{\mathrm{h}} \gtrsim 10^{12}\,\mathrm{M}_{\odot}$ but not below. In the lowest bin that we consider, $[10^{11.5}, 10^{12}) \, M_{\odot}$, only 12–44\% of halos (depending on redshift) host a selected galaxy, 69–76\% of which are centrals, and the median halo mass of the selected sample is $M_{\mathrm{h}} = 10^{11.8}$–$10^{11.9} \, \mathrm{M}_{\odot}$, close to the upper bin edge, as expected when a stellar-mass floor selects within a steep mass function. We quote the median (16th–84th percentile) halo masses of the selected samples, $\log_{10} (M_{\mathrm{h}}/\mathrm{M}_{\odot}) = 11.8\,[11.6, 11.9]$, $12.2\,[12.1, 12.4]$, and $13.2\,[13.0, 13.4]$ for the three bins at $z \simeq 0.5$. We do not consider the halo mass range below $10^{11.5}\, \mathrm{M}_{\odot}$, where the same selection retains only $\lesssim 5$\% of halos, almost all of them satellites. We compute the biased tracer density, $\delta^{\mathrm{b}}$, as defined in Equation~\ref{eqn:bias_density_field}, from the number counts for a given redshift shell, halo mass range, and pixelated angular position. 

For each shell, we also compute the projected density field of gravitationally collapsed dark matter, $\rho_{\mathrm{c}}$, for a given shell. We define the gravitationally collapsed mass as the sum of all halo and subhalo dark matter-only masses at a given $\bm{\theta}$ and in a given redshift shell. We effectively treat the halos and subhalos as point masses in this process which at the pixel sizes and redshift shell widths considered in this work is a good approximation. As in Equation~\ref{eqn:matter_density_field}, we define the collapsed matter density contrast, $\delta^{\mathrm{c}}(\bm{\theta})$, from the density field. This field can be thought of as a continuous tracer of the halo count contrast, $\delta^{\mathrm{h}}$ and matter density contrast field, $\delta^{\mathrm{m}}$, simultaneously. In the linear regime and at large enough scales where exclusion effects can be ignored \citep{mo1996analytic}, the following holds
\begin{equation}
    \delta^{\mathrm{c}}(\bm{x}, \tau) \approx b_{\mathrm{h,M}}(\tau) \, \delta^{\mathrm{m}}(\bm{x}, \tau),
\end{equation}
\noindent where $b_{\mathrm{h, \,M}}$ is the mass-weighted halo bias which is defined as follows
\begin{equation}
    b_{\mathrm{h, M}}(\tau) = \frac{1}{\overline{\rho}_{\mathrm{h}}(\tau)} \int_{M_{\mathrm{min}}}^{M_{\mathrm{max}}} \mathrm{d}M \, M \, n(M, \tau) \, b(M, \tau),
\end{equation}
\noindent where $\overline{\rho}_{\mathrm{h}}(\tau)$ is the mean comoving mass density of the halo population at proper time $\tau$, $n(M, \tau)$ is the number of halos per unit volume with masses between $M$ and $M + \mathrm{d}M$, and $b(M, \tau)$ is the linear halo bias as a function of halo mass and time given by $\delta^{\mathrm{h}}(M, \tau) \approx b(M, \tau) \, \delta^{\mathrm{m}}(\tau)$. Note that in line-of-sight projection $\tau \rightarrow z$ and $\bm{x} \rightarrow \bm{\theta}$.

Alternatively, the collapsed matter density contrast, $\delta^{\mathrm{c}}$, can also be interpreted as a tracer of the halo count contrast field, $\delta^{\mathrm{h}}$, \citep{cooray2002halo} as follows
\begin{equation}
    \delta^{\mathrm{c}}(\bm{x}, \tau) \approx \frac{b_{\mathrm{h, M}}(\tau)}{b_{h, \, n}(\tau)} \, \delta^{\mathrm{h}}(\bm{x}, \tau),
\end{equation}
\noindent where $b_{h, n}(\tau)$ is the number-weighted halo bias which is defined as
\begin{equation}
    b_{\mathrm{h, n}}(\tau) = \frac{1}{\overline{n}_{\mathrm{h}}(\tau)} \int_{M_{\mathrm{min}}}^{M_{\mathrm{max}}} \mathrm{d}M \, n(M, \tau) \, b(M, \tau),
\end{equation}
\noindent where $\overline{n}_{\mathrm{h}}(\tau)$ is total number of halos per unit comoving volume at time $\tau$. For the calibration sample these integrals, evaluated with the \citet{tinker2010toward} bias over the measured halo mass function, give $b_{\mathrm{h,M}} = 1.35$–$2.35$ and $b_{\mathrm{h,n}} = 0.77$–$1.83$ between $z = 0.28$ and $2.5$. They characterise the collapsed field itself and should not be confused with the scalar bias $b$ of the local mock models of Section~\ref{calibration}, which is measured from the galaxy spectra. As a consistency check, the peak-background-split prediction for the large-scale BOLTS amplitude, $A \approx \langle b \rangle_{\mathrm{g}}/b_{\mathrm{h,M}}$, matches the amplitude of the best fit $\beta_{\ell}$ within a few per cent for the two largest halo mass bins. In the lowest halo mass bin, they deviate by 12–25\%, which is consistent with its selection incompleteness.

Understanding the mapping from $\delta^{\mathrm{c}} \rightarrow \delta^{\mathrm{g}}$ allows one to isolate the galaxy formation processes underpinning the galaxy-halo connection, while reducing the sensitivity to non-linear gravitational collapse processes. Hence, any linear non-locality and stochasticity in the galaxy-halo connection can be captured by the model. Any detected stochasticity can subsequently be largely attributed to randomness in the galaxy formation within halos (which is typically captured with HOD scatter; see e.g. \citealt{Zheng2005theoretical}), while capturing any non-locality in the stochasticity itself.

\section{Calibration and testing with hydrodynamical simulations}\label{calibration_testing}

\begin{figure}
    \centering
    \includegraphics[width=\linewidth]{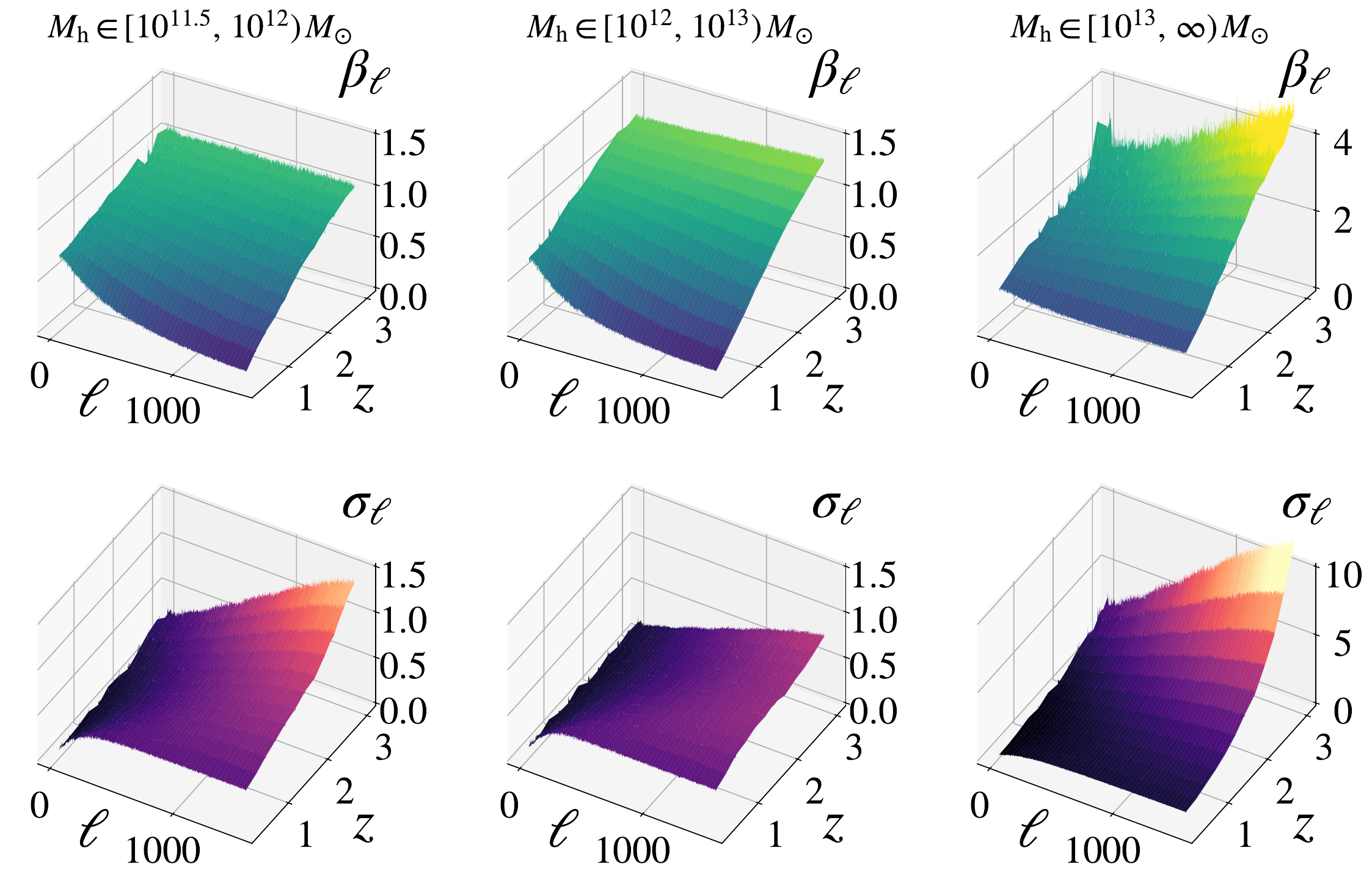}
    \caption{Measured galaxy bias in harmonic space as a function of angular scale, $\ell$, and mean redshift, $z$, of the respective shell as measured from `L2p8\_m9' \texttt{FLAMINGO} simulation. The top panels show the deterministic contribution to the galaxy, $\beta_{\ell}(z)$, while the bottom panels show the stochastic contribution, $\sigma_{\ell}(z)$. The panels are ordered in increasing halo mass, $M_{\mathrm{h}}$, ranges from left to right. The colours of the three-dimensional plots are coloured according to the $\beta_{\ell}$ and $\sigma_{\ell}$ values. Figure~\ref{fig:calibration} shows a cross-section of this three-dimensional plot.}
    \label{fig:beta_sigma_ell_z}
\end{figure}

To test whether the BOLTS model can capture the complexities of the galaxy-halo connection when compared to hydrodynamical simulations, we construct the following full-sky galaxy count contrast fields, $\delta^{\mathrm{g}}$, for a given redshift shell and halo mass, $M_{\mathrm{h}}$, range: 
\begin{itemize}[leftmargin=15pt]
    \item A `\textit{True}' $\delta^{\mathrm{g}}$ field as measured from `L2p8\_m9' \texttt{FLAMINGO} simulation. This model is the most accurate as it incorporates baryonic feedback processes when modelling the underlying matter density field, while also incorporating a full hydrodynamical treatment of galaxy formation.
    \item  A `\textit{BOLTS Mock}' of the $\delta^{\mathrm{g}}$ field forward modelled with the BOLTS model by biasing the collapsed matter density contrast, $\delta^{\mathrm{c}}$, field from the `L2p8\_m9' \texttt{FLAMINGO} simulation: $\delta^{\mathrm{m}} \rightarrow\delta^{\mathrm{c}} \rightarrow \delta^{\mathrm{c}}_{\ell m} \rightarrow b_{\ell m}\delta^{\mathrm{c}}_{\ell m} \rightarrow \delta^{\mathrm{g}}$. This model takes in the collapsed matter density contrast field from \texttt{FLAMINGO} and predicts the galaxy field, $\delta^{\mathrm{g}}$ based only on the galaxy angular power spectra as defined in Equations~\ref{eqn:beta_ml} and~\ref{eqn:sigma_ml}, as shown in the fits in Figures~\ref{fig:calibration} and~\ref{fig:beta_sigma_ell_z}. Over the scales where the BOLTS model is valid and for large halo masses ($M_{\mathrm{h}} > 10^{13} \, \mathrm{M}_{\odot}$), the predicted $\delta^{\mathrm{g}}$ by biasing $\delta^{\mathrm{c}}$ from the `L2p8\_m9' and the `L2p8\_m9\_DMO' simulations are virtually identical. Since the BOLTS model is stochastic in nature, we run 200 independent realisations of the mocks when drawing $b_{\ell m}^{(j)}$ as defined in Equation~\ref{eqn:final_model}. All subsequent summaries of the `\textit{BOLTS Mock}' field show the mean and standard deviation across all the realisations. Note that the mock uncertainties vary only the stochastic bias draws on a single realisation of the matter field. The realisations do not account for cosmic variance. The eight lightcones of `L2p8\_m9' can only partially supply it. Their observers are offset by half a box length ($1.4\,\mathrm{Gpc}$), so where the lightcones overlap they view the same structures from different directions and at different epochs, and the scatter between them would measure the sample variance on scales much smaller than the box. Nevertheless, they cannot capture the variance on the largest scales, as all lightcones share the same large-scale modes, the modes larger than the box are absent altogether, and beyond a box length in comoving distance every lightcone samples the same volume periodically. Sub-dividing a single lightcone into sky patches probes the same range of scales, and in the absence of independent realisations of the simulation, we adopt this simpler approach, estimating the sample variance of the `True' fields by dividing the sky into 48 patches.
    \item A `\textit{Local mock from $\delta^{\mathrm{c}}$}' $\delta^{\mathrm{g}}$ derived from a linear local deterministic bias (see Equation~\ref{eqn:standard_bias}) from $\delta^{\mathrm{c}}$ from the `L2p8\_m9' \texttt{FLAMINGO} simulation: $\delta^{\mathrm{m}} \rightarrow \delta^{\mathrm{c}} \rightarrow b\delta^{\mathrm{c}}\rightarrow \delta^{\mathrm{g}}$. The bias is set to the mean bias of the `True' $\delta^{\mathrm{g}}$ field for a given shell and $M_{\mathrm{h}}$ bin. We determine $b$ by measuring $\hat{b}_{\ell} = (C_{\ell}^{\mathrm{gg}}/C_{\ell}^{\mathrm{xx}})^{1/2}$ over large scales ($2 \leq \ell \leq 100$) where the angular power spectra are signal dominated. This model is equivalent to the BOLTS model if the non-locality and stochasticity vanish (i.e. $\beta_{\ell} = b \, \forall \, \ell$ and $\sigma_{\ell} = 0$). 
    \item A `\textit{Local mock from $\delta^{\mathrm{m}}$}' $\delta^{\mathrm{g}}$ derived from a linear local deterministic bias (see Equation~\ref{eqn:standard_bias}) from the $\delta^{\mathrm{m}}$ field from the `L2p8\_m9\_DMO' \texttt{FLAMINGO} simulation: $\delta^{\mathrm{m}} \rightarrow b\delta^{\mathrm{m}}\rightarrow \delta^{\mathrm{g}}$. The bias is set to the mean bias of the `True' $\delta^{\mathrm{g}}$ field for a given shell and $M_{\mathrm{h}}$ bin. This model not only neglects non-locality and stochasticity in the galaxy-halo connection, by using the $\delta^{\mathrm{m}}$ field directly, it conflates halo and galaxy formation into a single step. This means that it can also not account for any additional non-locality, stochasticity or non-linearity in the matter-halo connection. Nevertheless, we still choose to present this model here as a baseline, as the linear mapping from  $\delta^{\mathrm{m}}$ and  $\delta^{\mathrm{g}}$ is a common assumption for analyses of two-dimensional projected galaxy density fields in the linear regime.
\end{itemize}

As the free parameters in the BOLTS model only depend on the angular power spectra of the traced and tracer density fields, we test its accuracy by measuring higher-order summary statistics. As the construction of BOLTS does not put any inherent constraints to match the higher-order moments of the biasing field, any improvements in the accuracy in the measured higher-order statistics of the biased fields imply that the incorporation of non-locality and stochasticity when linearly biasing the fields in harmonic space allows for more accurate modelling of biased tracers at the field level. 

\subsection{Accuracy of three-point statistics: angular bispectrum}\label{calibration:bispectrum}

\begin{figure*}
    \centering
    \includegraphics[trim={0cm 1.5cm 0cm 1cm},clip,width=0.9\linewidth]{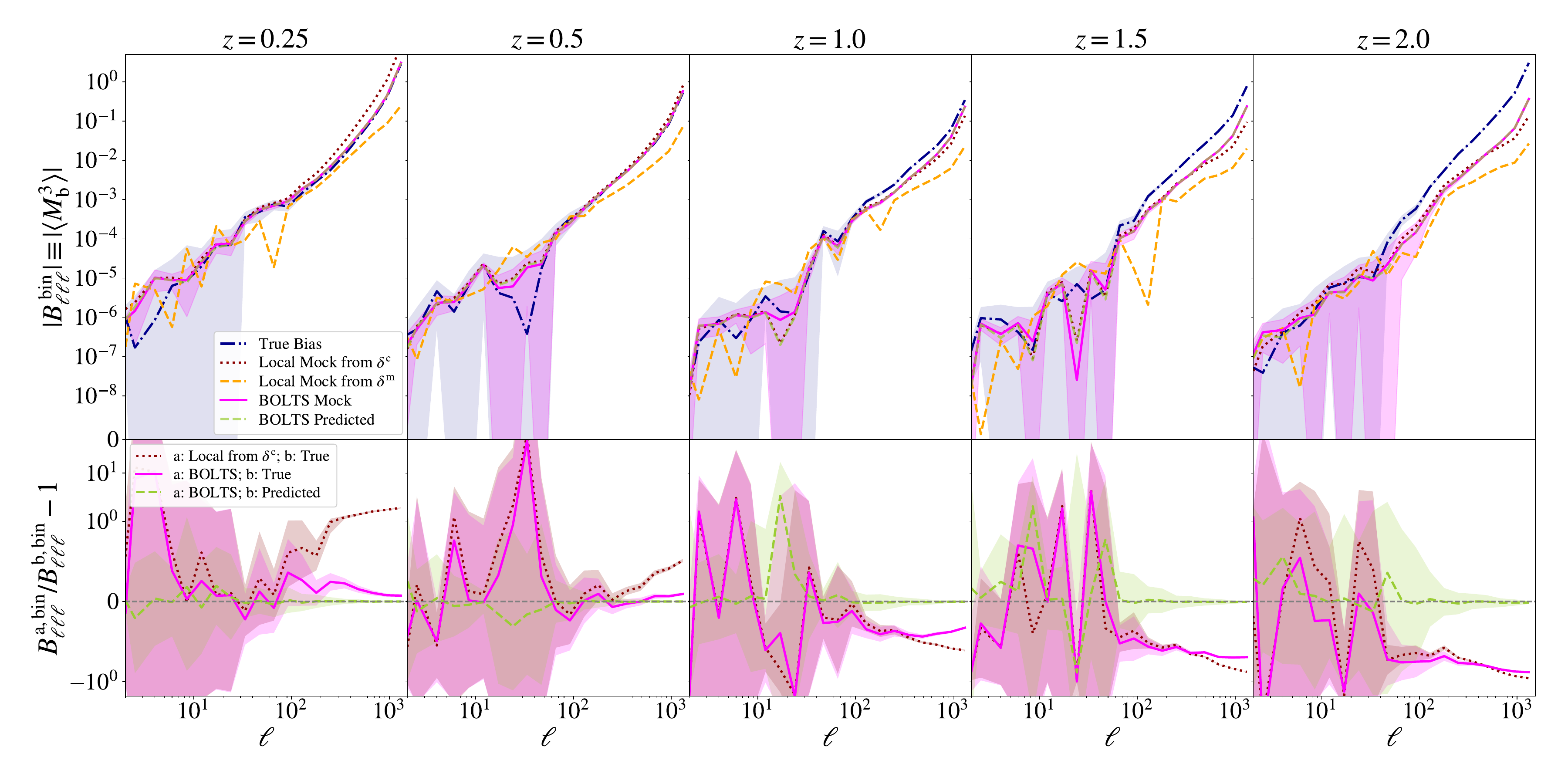}
    \caption{Equilateral terms of the full-sky angular bispectrum in over 20 log-spaced $\ell$-bins of the projected bias fields, $\delta^{\mathrm{b}}$, for objects with a halo mass, $M_{\mathrm{h}} > 10^{13} \, \mathrm{M}_{\odot}$. The panels from left to right show the bispectra within shells of increasing redshift from $z=0.25$ to $z=2.00$. The blue dot-dashed line shows the measurement from $\delta^{\mathrm{b}}$ in the hydrodynamical simulations (i.e. the `L2p8\_m9' \texttt{FLAMINGO} simulation, which we consider the `truth'). The grey band shows the `truth' $\pm$ its 48-patch sub-sample error. The mean measured bispectrum from 200 mock realisations from the BOLTS model is shown with the solid magenta line (with the standard deviation between them shown as the magenta shaded region; note this does not include cosmic variance as this is for a single realisation of the phases of the field). The green dashed line shows the analytical prediction for BOLTS based on Equation~\ref{eqn:analytical_blll} while assuming that $\sigma_{\ell, 1}^2 = \sigma_{\ell, 2}^2$. The dark red dotted line shows the bispectrum measured for a local linear bias applied to the $\delta^{\mathrm{c}}$ field, as is the case for the BOLTS model, with the same average amplitude over $2 \leq \ell \leq 100$ as all other models. The orange dashed line shows the prediction for a local linear bias applied to the $\delta^{\mathrm{m}}$ field. The bottom panel of each plot shows the residuals between different combinations of the bispectra in the top panels.}
    \label{fig:bispectrum}
\end{figure*}

As a first test beyond two-point statistics, we measure the equilateral angular bispectrum, $B_{\ell \ell \ell}^{\mathrm{bbb}}$, which probes the non-Gaussian structure of the biased density field. Since the BOLTS model parameters are calibrated exclusively on the angular power spectra (Equations~\ref{eqn:beta_ml} and~\ref{eqn:sigma_ml}), the bispectrum constitutes a genuinely independent validation. Figure~\ref{fig:bispectrum} presents the equilateral bispectrum measured from the four bias models described at the beginning of this section for galaxies residing in halos with $M_{\mathrm{h}} > 10^{13} \, \mathrm{M}_{\odot}$ across four redshift shells spanning $z = 0.25$ to $z = 2.00$.

As can be seen from the residual panel in Figure~\ref{fig:bispectrum}, the analytical prediction for the BOLTS bispectrum, derived from Equation~\ref{eqn:analytical_blll} under the assumption that $\sigma_{1,\ell}^2 = \sigma_{2,\ell}^2$, agrees exceptionally well with the measured bispectrum of the mocks at all scales. Given that this remains true at all halo masses and redshift, the agreement confirms the validity of the analytical prediction for the mock ensemble.

The bispectrum of the galaxies from the BOLTS mocks agree well with the true bispectrum measured from the hydrodynamical simulation at large scales, in particular, at low redshift. For $M_{\mathrm{h}} > 10^{13}\,\mathrm{M}_{\odot}$, as shown in Figure~\ref{fig:bispectrum}, the BOLTS mocks remain consistent with the truth within the combined mock scatter and sample variance down to $\ell_{\mathrm{break}} \sim 250$-$680$ at $z \le 0.5$, corresponding to comoving scales $1/k = \chi/\ell \sim 3$-$4.5\,\mathrm{Mpc}$, degrading to $\ell_{\mathrm{break}} \sim 130$ ($\chi/\ell \sim 27$-$35\,\mathrm{Mpc}$) at $z = 1$-$1.5$. At larger redshift, $2 \le z \le 3$, the bispectra only agree down to $\sim 50$-$90\,\mathrm{Mpc}$. In this regime, the model fits rely on fewer halos which may inflate errors which may lead to biases. As we do not propagate the BOLTS model fitting error or any errors arising from the halo selection to error bounds to the bispectrum measurements, the reported errors in Figure~\ref{fig:bispectrum} are likely underestimated, particularly, at high redshifts and large halo masses. In the future, any biases when calibrating the BOLTS model can be mitigated fitting the model on simulations with larger volumes or using more realisations of the same cosmology.

In the lower halo mass regime, with $ 10^{11.5}\,\mathrm{M}_{\odot} < M_{\mathrm{h}} <  10^{13}\,\mathrm{M}_{\odot}$, the measured bispectrum is accurate down to $\sim11$-$27\,\mathrm{Mpc}$ over $0.25 \le z \le 1.5$ and to $9$-$19\,\mathrm{Mpc}$ at $2 \le z \le 3$. Within this halo mass range, there are approximately $10^2$ times more halos in \texttt{FLAMINGO} than in the $M_{\mathrm{h}} > 10^{13}\,\mathrm{M}_{\odot}$ range. This confirms that when the calibration sample of galaxies is large enough, the BOLTS model can accurately recover higher-order moments of the galaxy field down to $\mathcal{O}(10\,\mathrm{Mpc})$ for various redshifts and halo masses. At the same time, despite the lower shot noise, the lowest mass bin has the largest uncertainties. This is driven by the larger stochastic-to-deterministic ratio $\sigma_{\ell}/\beta_{\ell}$.

In contrast, both local bias models deviate from \texttt{FLAMINGO} systematically. These residuals depend on how the constant bias is fixed. When measuring it over $2 \leq \ell \leq 100$, the local bias model have significant residuals at small scales. Any such discrepancies with the true galaxy field and the BOLTS galaxy field can be mostly attributed to the lack of scale dependence in the `local' models, which BOLTS captures through $\beta_{\ell}$.

As shown in in Figure~\ref{fig:beta_sigma_ell_z}, at higher redshifts, the deterministic bias of the high-mass bin rises with $\ell$. This is largely a discreteness effect of the collapsed matter field. Because $\delta^{\mathrm{c}}$ deposits every halo and subhalo of mass $m_i$ as a point source, its auto-spectrum carries, besides the clustering term, the white mass-weighted discreteness power of the form $4\pi \sum_i m_i^2 / (\sum_im_i)^2$ \citep{hamaus2010minimizing} where the sums run over all deposited objects in the shell. Every selected galaxy is itself one such deposit, so the cross-spectrum $C^{\mathrm{gc}}_{\ell}$ contains the self-pair term $4\pi \langle m_{\mathrm{h}} \rangle_{\mathrm{g}} / \sum_i m_i$, with $\langle m_{\mathrm{h}} \rangle_{\mathrm{g}}$ as the mean deposited (sub)halo mass of the galaxy sample. Where these terms dominate, the ratio tends to
\begin{equation}  
    \hat{\beta}_{\ell \to \infty} \longrightarrow  \frac{\langle m_{\mathrm{h}} \rangle_{\mathrm{g}} \sum_i m_i}       
    {\sum_i m_i^2}.
    \label{eqn:beta_l_limit}
\end{equation}

For galaxies selected in the most massive halos this ratio exceeds the large-scale amplitude and grows with redshift as massive halos become rare. This analytical limit tracks the measured $\hat{\beta}_{\ell \sim 10^3}$ within 8-17\% at $z \le 1$ (at higher redshifts Equation~\ref{eqn:beta_l_limit} overestimates the measurements, as clustering still contributes at $\ell \sim 10^3$ and the ratio is only the $\ell \to \infty$ limit). For the lower mass bins the same ratio lies far below the large-scale amplitude, which is why in these cases $\hat{\beta}_{\ell}$ decreases with $\ell$. The stochastic term of the high-mass bin is correspondingly dominated by its Poisson noise.

\subsection{Accuracy of n-point statistics: wavelet phase harmonic moments}\label{calibration:wph}

\begin{figure}
    \centering
    \includegraphics[trim={1.3cm 0 1.3cm 0},clip,width=0.39\linewidth]{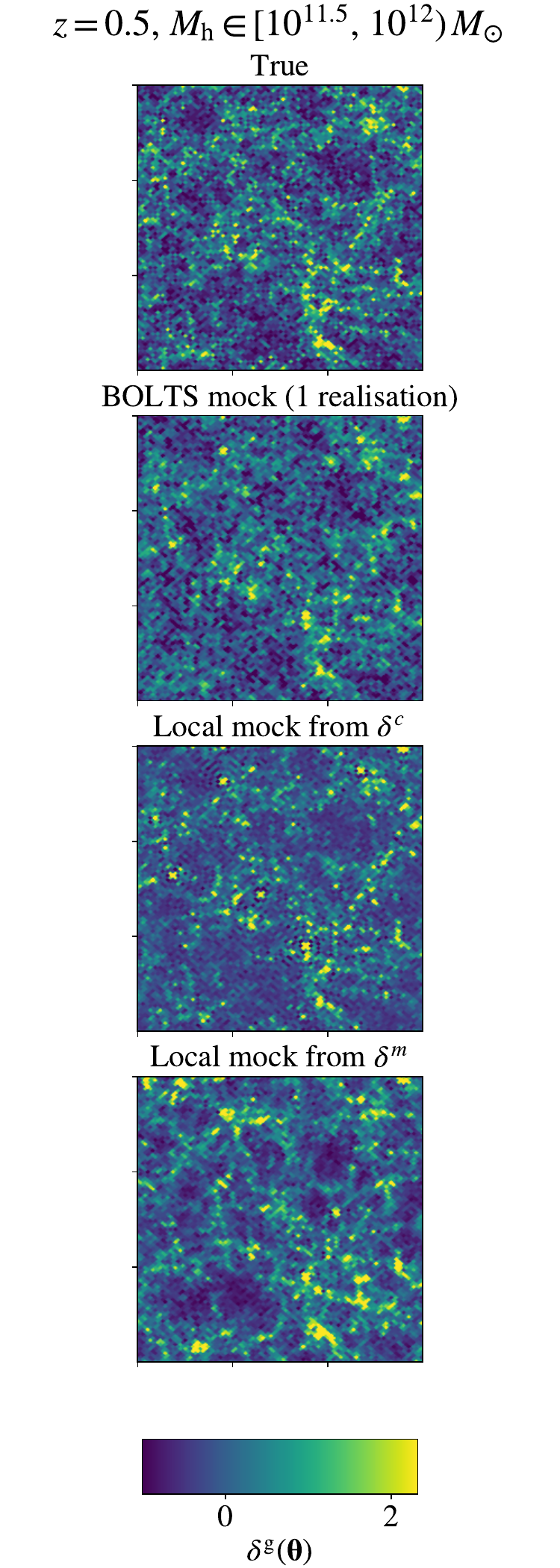}
    \includegraphics[trim={1.3cm 0 1.3cm 0},clip,width=0.39\linewidth]{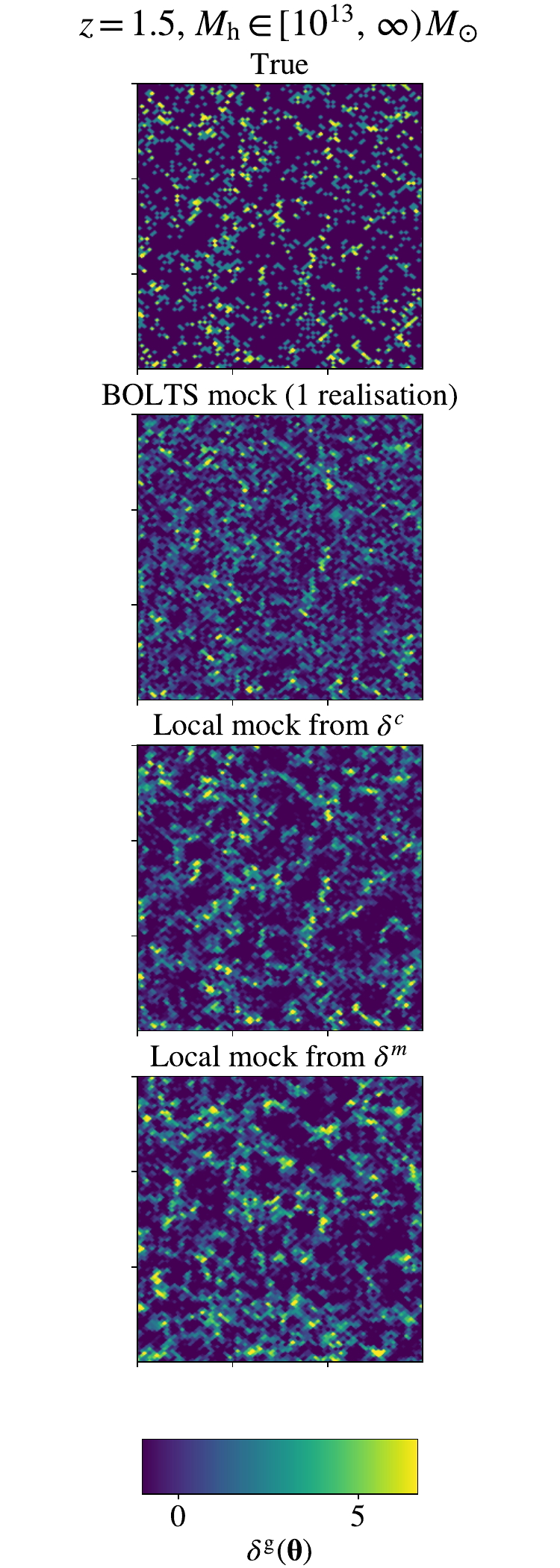}
    \caption{Cartesian projection of the $10 \times 10 \deg$ cutouts from the HealPix maps of the biased density field, $\delta^{\mathrm{b}}$. The left column shows a shell at relatively low redshift, $z\in[0.5, 0.55)$, with low mass halos, $M_{\mathrm{h}} \in [10^{11.5}, 10^{12}) \mathrm{M}_{\odot}$, while the right column show the other extreme: higher redshift, $z\in[1.50, 1.55)$, with higher-mass halos, $M_{\mathrm{h}} > 10^{13} \mathrm{M}_{\odot}$. The top panels show the `True' map from \texttt{FLAMINGO}, while the second row of panels shows the \texttt{FLAMINGO} $\delta^{\mathrm{c}}$ field as biased by the BOLTS model, the third row shows the collapsed matter density contrast field, $\delta^{\mathrm{c}}$, with a local deterministic linear bias, and the bottom row the matter density contrast field, $\delta^{\mathrm{m}}$, with a local deterministic linear bias model.}
    \label{fig:cartesian_maps}
\end{figure}

The Wavelet Phase Harmonic statistics (WPH; introduced by \citealt{mallat2019phase, zhang2019maximum}, first applied to large-scale structure by \citealt{allys2020new}) provide a summary of the non-Gaussian properties of a random field $X$. These statistics are constructed from the coefficients of a wavelet transform, using a non-linear operator known as the phase harmonic.

Let $\psi_{j,\theta}$ denote a complex-valued wavelet filter at scale $j$ and orientation $\theta$. The wavelet transform of the field $X$ is given by the convolution $X \otimes \psi_{j,\theta}$. The phase harmonic operator of order $p$, denoted $[\cdot]^p$, acts pointwise on a complex number $z = |z|e^{i\varphi}$ via
\begin{equation}
[z]^p = |z| \otimes e^{\mathrm{i} p \varphi}.
\end{equation}
For $p=1$, this is the identity operator, whilst for $p=0$, it extracts the modulus.

The WPH statistics go beyond two-point statistics by probing amplitude-phase and cross-scale couplings, making them well suited to discriminating between bias models that agree at the power spectrum level but differ in their non-Gaussian imprints. Their implementation in the \texttt{pywph} package \citep{regaldo2021new, regaldo2023generative} makes them computationally efficient to evaluate on flat two-dimensional fields. However, the WPH formalism is defined on regular Cartesian grids, requiring Cartesian re-projections from \textsc{HealPix} maps before the statistics can be computed. This limitation introduces a degree of distortion for wide-field analyses, although for the $10 \times 10 \, \mathrm{deg}^2$ cutouts used here the flat-sky approximation is well justified. We minimise the distortion by imposing a Cartesian window which uses  isotropic $1\,\mathrm{arcmin}$ pixels.

The WPH moments are defined as the covariance between two phase-harmonic transformed wavelet coefficients, potentially at different scales $\lambda=(j,\theta)$ and $\lambda'=(j',\theta')$, and separated by a spatial shift $\tau$,
\begin{equation}
C_{\lambda, p, \lambda', p'}(\tau) = \text{Cov}\left( [X \otimes \psi_{\lambda}(\mathbf{r})]^p , [X \otimes \psi_{\lambda'}(\mathbf{r} + \tau)]^{p'} \right).
\end{equation}
In practice, a subset of these coefficients is used to characterise the field efficiently.

The $S_{11}$ coefficients ($p=1, p'=1, \lambda=\lambda'$) are the auto-covariance of the wavelet coefficients at a single scale and orientation,
\begin{equation}
    S_{11}(\lambda) = \text{Cov}( X \otimes \psi_{\lambda} , X \otimes \psi_{\lambda} ) = \mathbb{E}[ |X \otimes \psi_{\lambda}|^2 ].
\end{equation}
These are equivalent to the wavelet power spectrum averaged over the spatial domain. They capture the field across scales and orientations and fully characterise the second-order (Gaussian) statistics of the field.

The $S_{01}$ coefficients ($p=0, p'=1, \lambda=\lambda'$) couple the modulus of the wavelet coefficient to the coefficient itself,
\begin{equation}
    S_{01}(\lambda) = \text{Cov}\left( |X \otimes \psi_{\lambda}|, \, X \otimes \psi_{\lambda} \right).
\end{equation}
A non-zero $S_{01}$ indicates a correlation between the local amplitude and phase of the wavelet coefficients, which is a signature of non-linear structure that vanishes for Gaussian fields.

The $C_{01}$ coefficients ($p=0, p'=1, \lambda \neq \lambda'$) are cross-scale moments that couple the modulus at one scale with the wavelet coefficient at another,
\begin{equation}
    C_{01}(\lambda, \lambda') = \text{Cov}\left( |X \otimes \psi_{\lambda}|, \, X \otimes \psi_{\lambda'} \right).
\end{equation}
They probe the non-linear mode coupling between different scales and are particularly sensitive to coherent extended structures such as filaments and sheets.

Finally, the $C_{\text{phase}}$ coefficients ($p=1, p'=2^{j'-j}, \lambda \neq \lambda'$) target the phase alignment across scales by matching frequency harmonics,
\begin{equation}
C_{\text{phase}}(\lambda, \lambda') = \text{Cov}\left( X \otimes \psi_{\lambda}, \, [X \otimes \psi_{\lambda'}]^{2^{j'-j}} \right).
\end{equation}
These moments distinguish coherent structures, where phases are locked across scales, from incoherent noise, capturing a hallmark of non-Gaussian morphology.

\begin{figure*}
    \centering
    \includegraphics[width=0.9\textwidth]{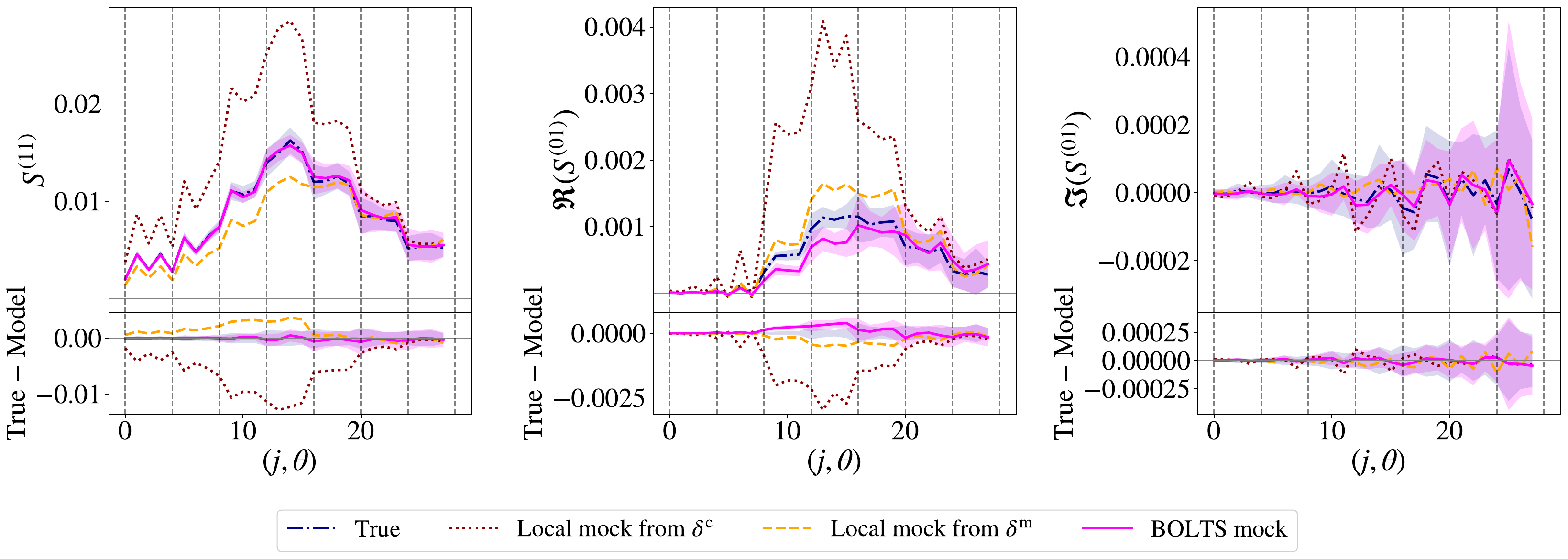}
    \caption{Wavelet phase harmonics measured on twelve $10 \times 10\,\deg$ flat cutouts of the band-limited ($\ell \leq 1535$) projected bias fields, $\delta^{\mathrm{b}}$, at $z\in[0.5, 0.55)$ with a halo mass, $M_{\mathrm{h}} \in [10^{11.5}, \, 10^{12}) \, \mathrm{M}_{\odot}$, as shown in Figure~\ref{fig:cartesian_maps}. The left panel shows the $S_{11}$ coefficients as a function of scale $j$ and orientation $\theta$ (the dashed vertical lines mark the boundaries of indices of constant scale $j$). The middle panel shows the real part of the $S_{01}$ coefficients over the same domain, while the right panel shows its imaginary part. The bottom panel of each plot shows the absolute residuals of each model with respect to the `truth' from \texttt{FLAMINGO}. Lines show the mean over the twelve windows. The shaded bands show the window-to-window standard deviation of the `True' field (blue) and of the BOLTS-mock mean over 20 realisations per window (magenta). The residual panels show the `True' band around zero. The local amplitudes are fitted over $2 \leq \ell \leq 100$. The models and styles are the same as in Figure~\ref{fig:bispectrum}.}
    \label{fig:wph_s01s11}
\end{figure*}

\begin{figure*}
    \centering
    \includegraphics[width=0.82\textwidth]{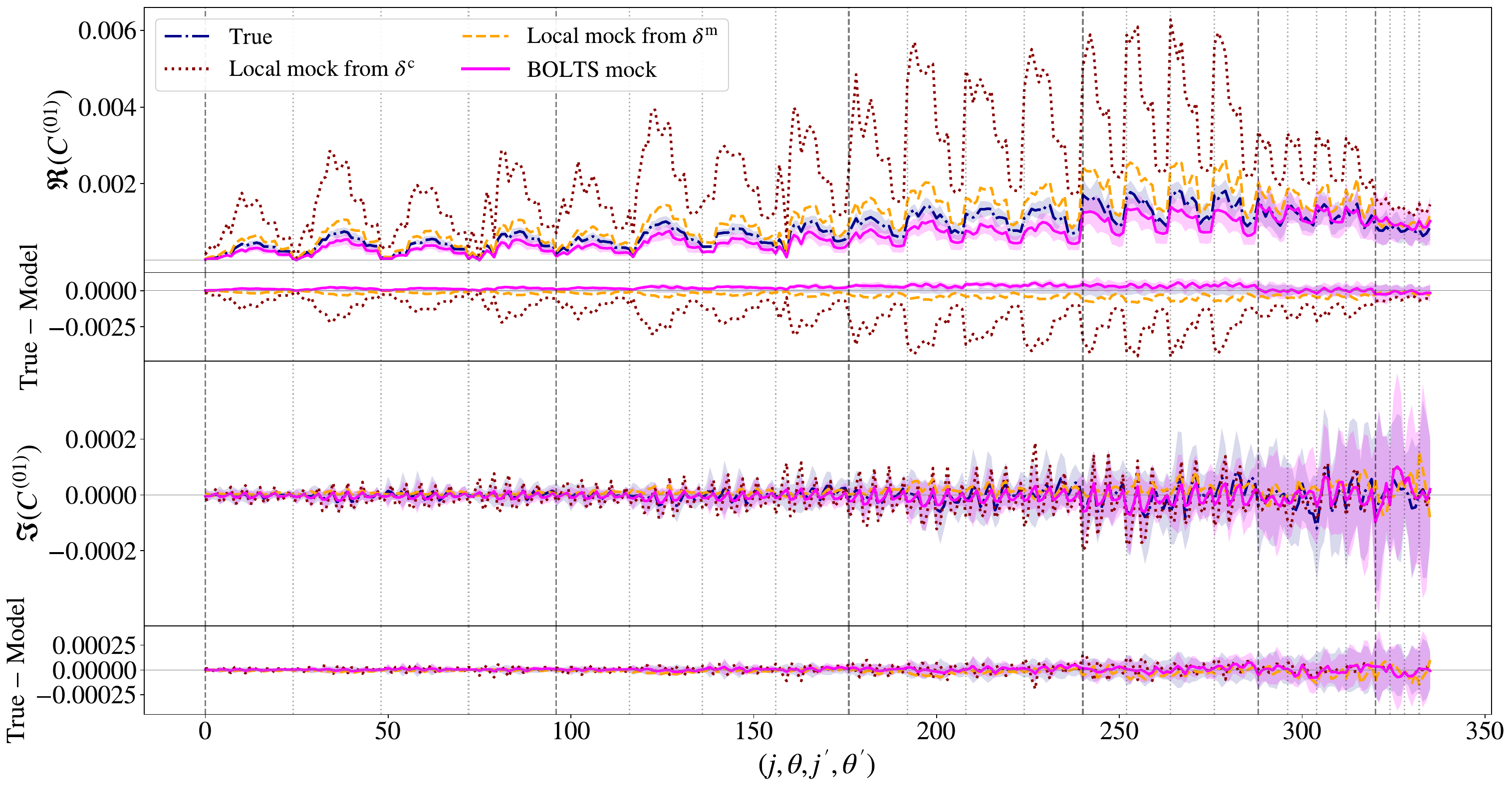}

    \caption{$C^{(01)}$ wavelet phase harmonic of the same bias field, $\delta^{\mathrm{b}}$, measured on the same twelve $10 \times 10\,\deg$ cutouts of the band-limited projected bias fields, $\delta^{\mathrm{b}}$, with $z\in[0.5, 0.55)$ for objects with a halo mass, $M_{\mathrm{h}} \in [10^{11.5}, \, 10^{12}) \, \mathrm{M}_{\odot}$, as shown in Figure~\ref{fig:cartesian_maps}. The top panels show the measurement and the respective residuals for the real part of $C^{(01)}$, while the bottom two panels show its imaginary part. All lines represent the same fields as in Figure~\ref{fig:wph_s01s11}.}
    \label{fig:wph_c01}
\end{figure*}

 We compute the WPH statistics on $10 \times 10\,\mathrm{deg}^2$ Cartesian cutouts re-projected from the \textsc{HealPix} maps onto isotropic $1'$ pixels. First, the BOLTS and local mocks are synthesised from spherical-harmonic coefficients and are therefore band-limited at $\ell_{\mathrm{max}} = 3 N_{\mathrm{side}} - 1 = 1535$, whereas the `True' count map carries aliased shot-noise power above $\ell_{\mathrm{max}}$ which a harmonic-space model cannnot contain. Therefore, we band-limit the `True' field in the same way before the cutout, so that all fields are compared within the calibrated band. Second, a single window gives no estimate of the sampling scatter of the statistics, so we measure every field on twelve non-overlapping windows spaced $30\,\deg$ apart along the equator and show the mean over windows with the window-to-window standard deviation as a band; for the BOLTS mock the band is the window scatter of the per-window mean over 20 realisations. Example cutouts are shown in Figure~\ref{fig:cartesian_maps} for two representative regimes: a low-redshift, low-mass bin ($z \in [0.5, 0.55)$, $M_{\mathrm{h}} \in [10^{11.5}, 10^{12}) \, \mathrm{M}_{\odot}$) and a high-redshift, high-mass bin ($z \in [1.50, 1.55)$, $M_{\mathrm{h}} > 10^{13} \, \mathrm{M}_{\odot}$). Visual inspection confirms that the BOLTS mocks reproduce the morphology and dynamic range of the true \texttt{FLAMINGO} galaxy field more faithfully than either local bias model, particularly in the low-mass regime where non-locality and stochasticity are most prominent. The local models, lacking stochastic scatter and scale-dependent bias, produce fields that are overly smooth and under-represent the fine-grained structure visible in the hydrodynamical simulation.

Figure~\ref{fig:wph_s01s11} presents the $S_{11}$ and $S_{01}$ WPH coefficients for the low-mass sample at $z \in [0.50, 0.55)$. The $S_{11}$ coefficients (left panel) quantify the energy distribution across wavelet scales and orientations; as these are equivalent to the wavelet power spectrum, all models that match the two-point statistics agree on $S_{11}$ by construction. BOLTS reproduces $S_{11}$ to within the window scatter at every scale and orientation (relative residual $2$ per cent, $0.2\sigma$).

The $S_{01}$ coefficients (middle and right panels) probe the coupling between the amplitude and phase of the wavelet coefficients, which is a diagnostic of non-Gaussianity induced by non-linear structure. The BOLTS mock tracks the true $S_{01}$ within the window scatter ($0.8\sigma$, with a relative residual of $0.23$), whereas the local mock from $\delta^{\mathrm{c}}$ over-estimates its real part by a factor of 2 to 3 at the scales where the coupling peaks ($6\sigma$, with a relative residual of 1.7), reflecting its failure to capture the scale-dependent suppression of the deterministic bias. The local mock from $\delta^{\mathrm{m}}$ lies between the two ($1.3\sigma$). The improvement from BOLTS arises because the non-local $\beta_{\ell}$ naturally modulates the amplitude-phase coupling as a function of scale, while the stochastic $\gamma_{\ell m}$ introduces additional scatter that decorrelates the amplitude and phase at small scales, consistent with the physical effect of stochastic galaxy formation. Quantitatively, the stochastic term suppresses the amplitude–phase coupling by 4-20\% per wavelet scale relative to a deterministic $\beta_{\ell}$ mapping (growing towards small scales). This is far less than the $r^3$ dilution an additive noise field of equal power would produce ($r_{\ell} = \beta_{\ell}/(\beta_{\ell}^2 +\sigma_{\ell}^2)^{1/2}$; $r^3 \approx 0.25$ at the smallest scales versus the measured $\approx 0.8$). Since the stochasticity is multiplicative in harmonic space, the mock inherits the input field's amplitude and phase pattern. The local models, in contrast, keep the full coupling and rescale it by $b^3$, over-predicting it when the true bias is stochastic.

\begin{figure*}
    \centering
    \includegraphics[width=0.82\textwidth]{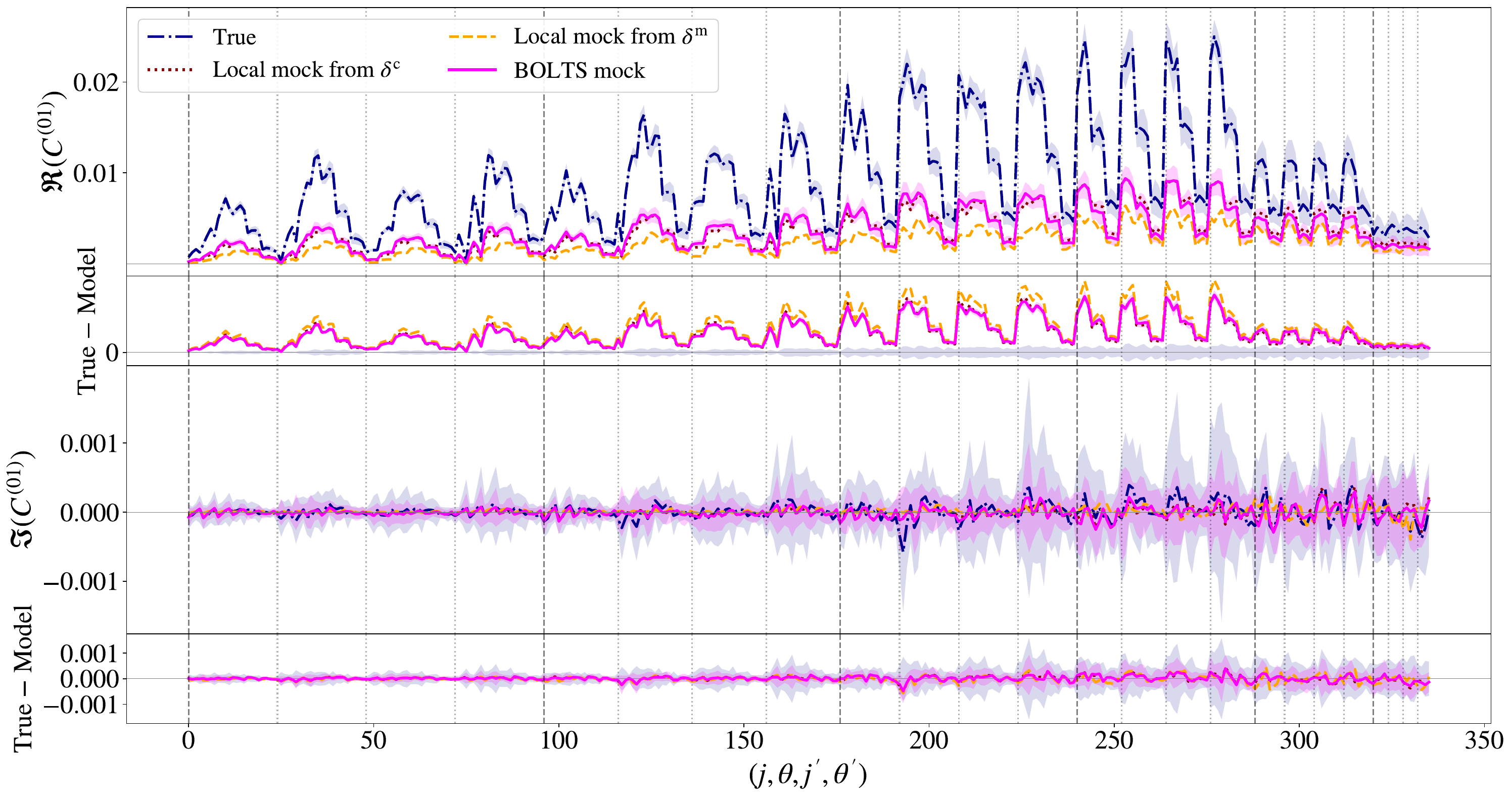}
    \caption{$C^{(01)}$ wavelet phase harmonic of the same bias field, $\delta^{\mathrm{b}}$, measured on the same twelve $10 \times 10\,\deg$ cutouts of the band-limited projected bias fields, $\delta^{\mathrm{b}}$, with $z\in[1.50, \, 1.55)$ for objects with a halo mass, $M_{\mathrm{h}} \in [10^{13}, \, \infty) \, \mathrm{M}_{\odot}$, as shown in Figure~\ref{fig:cartesian_maps}. The top panels show the measurement and the respective residuals for the real part of $C^{(01)}$, while the bottom two panels show its imaginary part. All lines represent the same models as in Figure~\ref{fig:wph_c01}. In the high redshift and high halo mass regime, the expectation of the bias field from the BOLTS model converges to a local and linear galaxy bias model.}
    \label{fig:wph_c01_highz}
\end{figure*}

Figures~\ref{fig:wph_c01} and~\ref{fig:wph_c01_highz} show the cross-scale $C_{01}$ coefficients, which capture the non-linear mode coupling between wavelet scales. For the low-mass, low-redshift regime (Figure~\ref{fig:wph_c01}), where both the stochastic and deterministic biases are important, BOLTS recovers the measured $C_{01}$ coefficients from \texttt{FLAMINGO} within the window scatter in the real part ($1.2\sigma$, with a relative residual of 0.26) and in the imaginary part, whereas the local mock from $\delta^{\mathrm{c}}$ exceeds the real part by a factor of 2 to 3 ($9\sigma$) and the local mock from $\delta^{\mathrm{m}}$ by 37\% ($1.7\sigma$). The gap is largest at the smallest scales, where the non-linear coupling is strongest and the local bias approximation is poorest.

In the high-redshift, high-mass regime (Figure~\ref{fig:wph_c01_highz}), the differences between the models diminish substantially. This convergence reflects the reduced non-locality and stochasticity at high masses and high redshifts: the bias parameters $\beta_{\ell}$ and $\sigma_{\ell}$ become approximately constant in $\ell$ (see Figure~\ref{fig:beta_sigma_ell_z}), so that the BOLTS model effectively reduces to a local, deterministic linear bias. In this regime BOLTS still reproduces $S_{11}$ (within $0.1\sigma$), but the real parts of $S_{01}$ and $C_{01}$ of all three models fall below the truth, BOLTS and the local mock from $\delta^{\mathrm{c}}$ by the same amount ($4$-$5\sigma$ of the window scatter, with relative residuals of 0.58 and 0.60) and the local mock from $\delta^{\mathrm{m}}$ by more (up to 0.75). In the calibration, at $\ell \sim 10^3$, the Poisson term $4\pi/N_{\mathrm{g}}$ is half of $C^{\mathrm{gg}}_{\ell}$ and $\sigma_{\ell}^2 C^{\mathrm{cc}}_{\ell}$ is the other half, i.e. the stochastic power of this bin is the discreteness of the galaxies themselves, which are the very objects that dominate the mass-weighted $\delta^{\mathrm{c}}$ deposits (Equation~\ref{eqn:beta_l_limit}). That part of the residual is deterministic point-source structure coherent with $\delta^{\mathrm{c}}$, i.e. not an independent random field. The Gaussian $\gamma_{\ell m}$ replaces it by one and dilutes the amplitude-phase coherence that $S_{01}$ and $C_{01}$ measure. A deterministic treatment of the self-pair term in the forward model would remove this limitation and is left to future work. This highlights that the BOLTS model is most accurate in regimes where the stochasticity is genuine, i.e. where it exceeds the discreteness of the tracer sample. 

The $C_{\mathrm{phase}}$ moments, which test phase alignment across scales, behave consistently (BOLTS agrees within $0.8\sigma$, while there is $5.4\sigma$ agreement for the local mock from $\delta^{\mathrm{c}}$ in the low-mass bin, and $2.6$-$2.8\sigma$ agreement for both in the high-mass bin). For the lower mass bins at low redshift the BOLTS mocks track the measured coefficients most closely of all models, while the local mock from $\delta^{\mathrm{m}}$ shows the largest deviations, and the local mock from $\delta^{\mathrm{c}}$ over-predicts the strongest peaks by roughly the $b^3$ rescaling. In the high-mass, high-redshift regime this trend is partially reverted. The strongest phase-alignment peaks of the truth are suppressed in the BOLTS mocks. The Gaussian stochastic term, which dominates this bin's variance, dilutes cross-scale phase coherence, while the deterministic $\delta^{\mathrm{c}}$ mock, which preserves the input field's phases exactly, remains closer to the truth there. This mirrors the behavior of $\Re \,C_{01}$ in Figures~\ref{fig:wph_c01} and~\ref{fig:wph_c01_highz}.

\subsection{Effect of mass cuts and resolution of the calibration sample}\label{calibration:mass_cuts_resol}

The three halo mass bins probe distinct regimes (see Figure~\ref{fig:beta_sigma_ell_z}, Sections~\ref{calibration:bispectrum} and~\ref{calibration:wph}). For $M_{\mathrm{h}} > 10^{13} \, \mathrm{M}_{\odot}$ the deterministic bias is large and approximately scale-independent at low to intermediate redshifts, so BOLTS closely approaches a simple linear deterministic bias, while at $z \gtrsim 2$ the stochastic term dominates over a broad range of scales. For the intermediate bin ($M_{\mathrm{h}} \in [10^{12}, 10^{13}) \, \mathrm{M}_{\odot}$), $\beta_{\ell}$ develops a clear scale dependence and the stochastic term grows. The residuals with respect to \texttt{FLAMINGO} grow accordingly, although BOLTS is still more accurate than local bias models in the bispectrum.

The lowest mass bin ($M_{\mathrm{h}} \in [10^{11.5}, 10^{12})\, \mathrm{M}_{\odot}$) represents the most challenging regime for the model, because the stellar-mass selection is incomplete there. Only 12–44\% of these halos host a galaxy above the stellar-mass threshold, so which halos enter the sample is itself a stochastic outcome of galaxy formation. This selection stochasticity adds to the intrinsic bias scatter, and the effective (median) halo mass of the bin sits near its upper edge. Nevertheless, BOLTS continues to outperform local bias models across most scales and redshifts, so the model can be applied down to $M_{\mathrm{h}} = 10^{11.5}\,\mathrm{M}_{\odot}$ with these caveats stated.

Throughout this work we calibrate at $N_{\mathrm{side}} = 512$ (a pixel scale of $\approx 7\,\mathrm{arcmin}$), which sets the maximum multipole $\ell \le 2 N_{\mathrm{side}}$ of the calibration. We have verified that calibrating at $N_{\mathrm{side}} = 1024$ does not significantly alter the measured bias parameters. Resolution and pixelisation effects, including their impact on the measured stochasticity, are quantified in Appendix~\ref{appendix:order_of_ops}.

\section{BOLTS Bias as a forward model}\label{forward_model}

Having validated the BOLTS model against hydrodynamical simulations, we now describe a semi-empirical extension that enables its use as an efficient forward model. The per-$\ell$ parameterisation of $\beta_{\ell}$ and $\sigma_{\ell}$ has $\mathcal{O}(\ell_{\mathrm{max}})$ free parameters per redshift shell. For forward modelling applications, such as simulation-based inference, it is advantageous to compress this information into smooth, low-dimensional fitting functions that can be interpolated across redshift.

\begin{figure}
    \centering
    \includegraphics[width=0.9\linewidth]{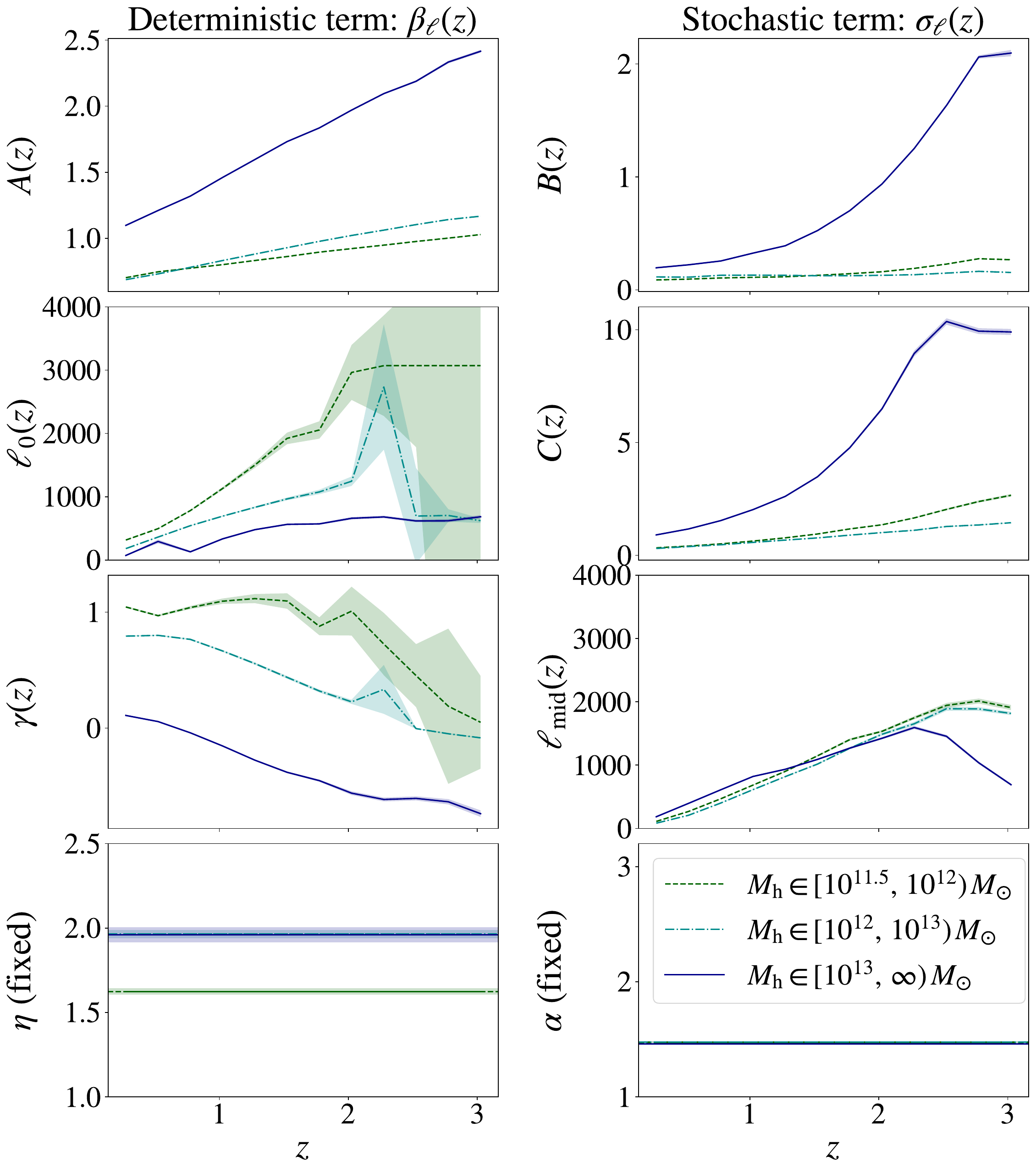}
    \caption{Best-fit parameters of the fitting functions for the deterministic bias $\beta_{\ell}$ (Equation~\ref{eqn:beta_l_fit}) and the stochastic bias $\sigma_{\ell}$ (Equation~\ref{eqn:sigma_l_fit}) as a function of the mean redshift of each shell for three halo mass ranges (with styles as in Figure~\ref{fig:kernel_width}). All fits are based on the mapping from $\delta^{\mathrm{c}}$ to $\delta^{\mathrm{g}}$ in the `L2p8\_m9' simulation. The left panel shows the parameters of $\beta_{\ell}$ ($A$, $\ell_0$, $\gamma$, and $\eta$) and the right panel shows the parameters of $\sigma_{\ell}$ ($B$, $C$, $\ell_{\mathrm{mid}}$, and $\alpha$). Both $\eta$ and $\alpha$ are kept constant as a function of redshift. The shaded bands are 1$\sigma$ parameter uncertainties propagated from the fit covariance. The fits are not smoothed nor are any unstable fits removed, so occasional off-scale excursions reflect genuine fit degeneracies.}
    \label{fig:params_per_z}
\end{figure}

We find that the deterministic bias $\beta_{\ell}$ is well described by a generalised Lorentzian profile,
\begin{equation}
    \beta_{\ell} = A\, \Bigg[1 + \bigg(\frac{\ell}{\ell_0}\bigg)^{\eta}\Bigg]^{-\frac{\gamma}{\eta}},
    \label{eqn:beta_l_fit}
\end{equation}
\noindent where $A$ sets the large-scale amplitude, $\ell_0$ is the scale-break position, $\eta$ controls the steepness of the transition, and $\gamma$ determines the asymptotic power-law slope. This functional form is motivated by the measured behaviour of $\beta_{\ell}$ across redshift and halo mass, which is approximately constant at low~$\ell$ and declines as a power law at high~$\ell$.

The stochastic amplitude $\sigma_{\ell}$ follows a complementary pattern, rising monotonically from a floor at large scales towards smaller scales where stochastic galaxy formation processes dominate. We parametrise this behaviour with a sigmoidal function,
\begin{equation}
    \sigma_{\ell} = B +
    C \Bigg[1 + \bigg( \frac{\ell_{\mathrm{mid}}}{\ell}\bigg)^{\alpha} \Bigg]^{-1},
    \label{eqn:sigma_l_fit}
\end{equation}
\noindent where $B$ sets the baseline stochasticity, $C$ governs the amplitude of the scale-dependent rise, $\ell_{\mathrm{mid}}$ marks the characteristic transition scale, and $\alpha$ determines the steepness of the sigmoid.

The fitting-function parameters exhibit clear trends with both redshift and halo mass (see Figure~\ref{fig:params_per_z}). We keep both $\eta$ and $\alpha$ fixed for all redshift, while still fitting them to the mean best fit over all epochs. When allowing $\eta$ and $\alpha$ to vary with redshift, we find that both are broadly consistent with a constant in redshift with $\eta = 1.62/1.97/1.96$ and $\alpha = 1.47/1.48/1.46$ for the three mass bins. In fact, we note that the steepness of the sigmoid of $\sigma_{\ell}$ appears to be unaffected by the choice of redshift or halo mass. The large-scale amplitude $A$ of the deterministic bias increases with halo mass and redshift, consistent with the well-established trend that more massive and higher-redshift tracers are more strongly biased \citep{mo1996analytic, tinker2010toward}. The break scale $\ell_0$ shifts to higher $\ell$ at higher redshifts. This drift is mostly due to projection geometry: the comoving break $k_0 = \ell_0/\chi(z)$ stays within $\sim 0.17$-$0.28\, \mathrm{Mpc}^{-1}$ for the two lower mass bins over $0.25 \le z \le 2$, while comoving distance, $\chi$, grows five-fold. The stochastic transition evolves as well: $\ell_{\mathrm{mid}}/\chi$ rises from $\sim 0.10$ to $0.28\,\mathrm{Mpc}^{-1}$ between $z = 0.3$ and $1.5$, since at earlier epochs stochasticity reaches to smaller comoving scales. The stochastic parameters exhibit a similar trend, with both $B$ and $C$ decreasing with increasing halo mass, which confirms that stochastic bias is predominantly driven by the galaxy-halo connection in lower-mass halos where baryonic processes introduce most of the scatter.

The remaining six parameters vary smoothly with redshift, enabling the efficient generation of galaxy density fields for arbitrary redshifts within $0 < z \leq 3$ by interpolation alone, without recalibrating the model on the full power spectra. Additionally, these analytical functions capture the cross-$\ell$ couplings that are explicitly ignored in the assumptions of BOLTS in Equation~\ref{eq:general_bias_model}. This property makes BOLTS particularly well suited for forward-modelling applications in simulation-based inference pipelines (e.g. \citealt{fluri2022full, lin2023simulation, lemos2023simbig, gatti2024des, vonwietersheimkramsta2024kidssbi, jeffrey2025dark, thomsen2026des, saoulis2026field}), where many realisations of galaxy fields must be drawn at varying cosmologies or galaxy selection functions. 

\begin{figure}
    \centering
    \includegraphics[trim={0.8cm 0 0.8cm 0},clip,width=0.45\linewidth]{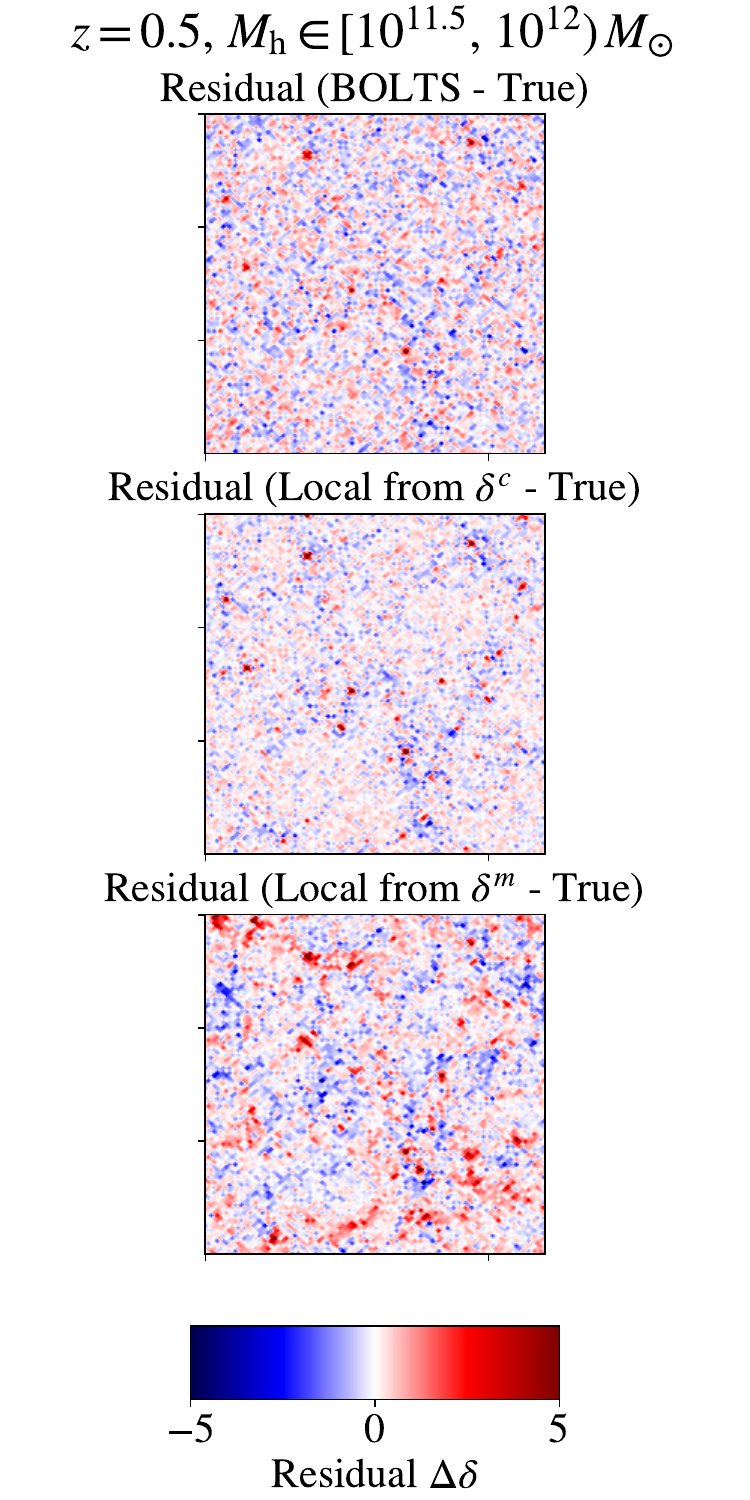}
    \includegraphics[trim={0.8cm 0 0.8cm 0},clip,width=0.45\linewidth]{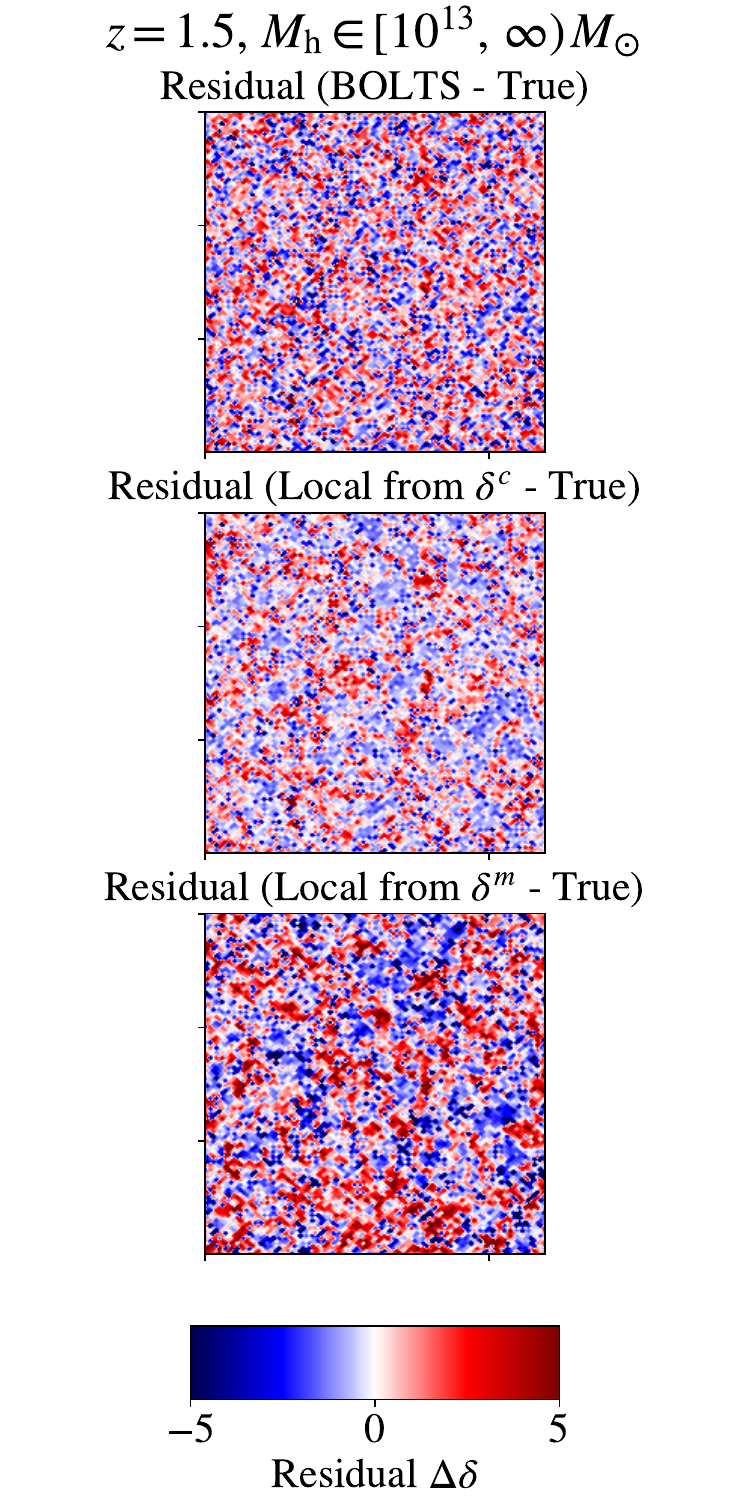}
    \caption{Cartesian projection of the 10 $\deg$ by 10 $\deg$ cutouts from the HealPix maps of the residuals from subtracting the biased density field, $\delta^{\mathrm{b}}$, as modelled by a galaxy bias model, from the truth, as simulated in \texttt{FLAMINGO}. Both columns show the same maps as in Figure~\ref{fig:cartesian_maps}. The top and bottom panels show the residuals with respect to \texttt{FLAMINGO} when comparing the biased density field, $\delta^{\mathrm{b}}$, against the prediction from the BOLTS model, against the `Local from $\delta^{\mathrm{c}}$' model (local deterministic linear bias applied to the collapsed matter density contrast field), and against the `Local from $\delta^{\mathrm{m}}$' model (local deterministic linear bias applied to the matter density contrast field).}
    \label{fig:cartesian_redsidual_maps}
\end{figure}

At the map level, the non-local stochastic model yields a marked improvement over local linear bias. The residuals between the BOLTS-generated field and the true \texttt{FLAMINGO} galaxy field are substantially smaller than those from a local model for both the low-mass, low-redshift regime and the high-mass, high-redshift regime (Figure~\ref{fig:cartesian_redsidual_maps}). The local model leaves spatially coherent residuals tracing the filamentary structure of the cosmic web, indicating that its failure is systematic rather than purely stochastic. In contrast, the BOLTS residuals are consistent with the expected scatter from the stochastic contribution, lacking large-scale correlations with the underlying density field.

\begin{figure*}
    \centering
    \includegraphics[trim={0cm 0cm 0cm 1cm},clip,width=0.9\linewidth]{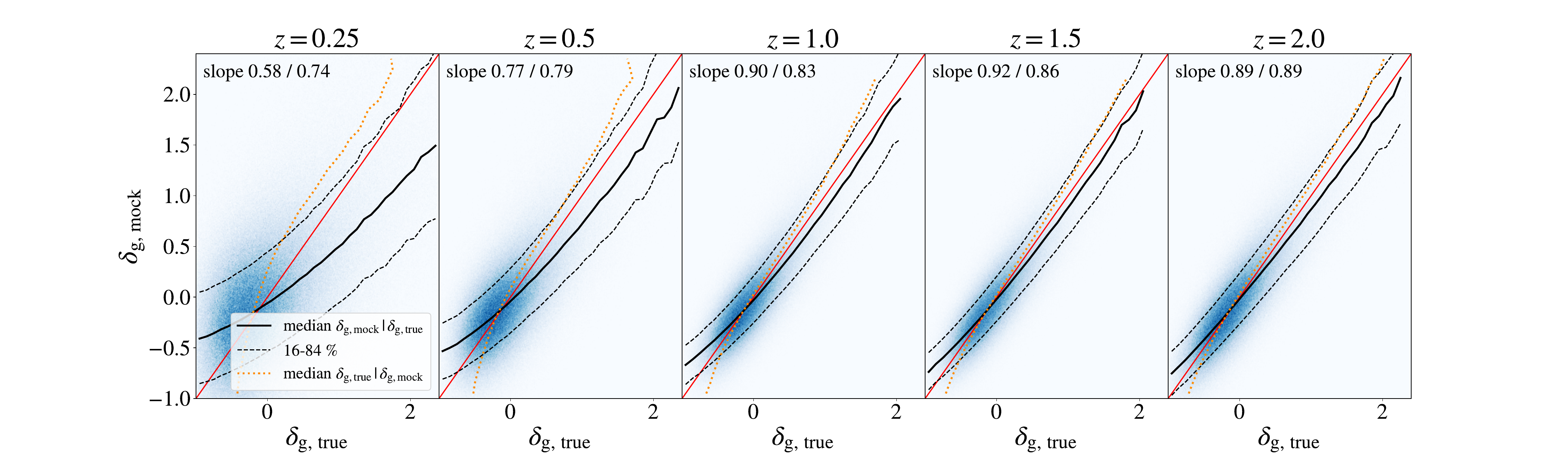}
    \caption{Pixel-by-pixel comparison of the true galaxy overdensity, $\delta^{\mathrm{g}}$, measured from the \texttt{FLAMINGO} `L2p8\_m9' simulation against the BOLTS mock prediction for galaxies with halo mass $M_{\mathrm{h}} \in [10^{11.5}, 10^{12}) \, \mathrm{M}_{\odot}$ at $N_{\mathrm{side}} = 256$. The lower $N_{\mathrm{side}}$ is chosen to minimise shot noise in the pixel-by-pixel comparison given the galaxy density in the simulation. Each panel corresponds to a different redshift shell spanning $z = 0$ to $z = 3$. The colour scale indicates pixel density, the black line shows the median of $P(\delta_{\mathrm{g,mock}} | \delta_{\mathrm{g,true} })$ (with the dashed lines showing the $1\sigma$ intervals), the orange dotted line showing the median of $P(\delta_{\mathrm{g,true}} | \delta_{\mathrm{g,mock} })$, and the red solid line marking the one-to-one relation. The slope values on the top left corner of each panel give the slope of the medians of $P(\delta_{\mathrm{g,mock}} | \delta_{\mathrm{g,true} })$ and $P(\delta_{\mathrm{g,true}} | \delta_{\mathrm{g,mock} })$, respectively.}
    \label{fig:mapping}
\end{figure*}

Figure~\ref{fig:mapping} compares BOLTS to \texttt{FLAMINGO} at the pixel level. At each redshift, it shows the two-dimensional histogram of the true against the mock $\delta^{\mathrm{g}}$, with the median mock value at fixed true value, and its transpose, overlaid. The mock is an independent draw from the calibrated model, not a reconstruction of the same sky. The two maps share only the deterministic component $\beta_{\ell}\delta^{\mathrm{c}}_{\ell m}$, and each carries its own stochastic part. We note that each median is shallower than the one-to-one line by the fraction of the variance the two maps share, $r = \sum_{\ell}(2\ell+1)\,\beta_{\ell}^{2}C^{\mathrm{cc}}_{\ell}\big/ \sum_{\ell}(2\ell+1)(\beta_{\ell}^{2}+\sigma_{\ell}^{2}) C^{\mathrm{cc}}_{\ell}$. The measured slopes rise from $0.58$ at $z=0.25$ to $\sim 0.86$ at $z \ge 1$, in agreement with this prediction, and a median on the one-to-one line would instead imply $\sigma_{\ell}=0$. This attenuation is the conditional-mean/scatter decomposition of stochastic biasing \citep{dekel1999stochastic}. The residual asymmetries are due to the non-Gaussianity limits discussed in Appendix~\ref{appendix:two-point}. The true field obeys $\delta \ge -1$ while the Gaussian stochastic term lets the mock dip below it ($3.5$\% of pixels at $z=0.25$, $\lesssim 0.5$ per cent beyond).

Given the fits provided here for $\beta_{\ell}$ and $\sigma_{\ell}$ for the mapping of $\delta^{\mathrm{c}}$ to $\delta^{\mathrm{g}}$, the forward model requires only the collapsed matter density field. Where halo catalogues exist, $\delta^{\mathrm{c}}$ is computed directly (Section~\ref{flamingo}). For applications where only a matter field $\delta^{\mathrm{m}}$ is available, for example, lognormal simulators such as \textsc{GLASS} \citep{tessore2023glass}, we have tested a set of approximate constructions on projected shells, spanning binary masks (thresholding based on virial radii calibrated to the collapsed mass fraction or two-dimensional Hessian `knot' classifications similar to \citealt{hahn2007properties}) as well as physically calibrated maps built from the conditional excursion-set collapsed fraction \citep{lacey1993merger, mo1996analytic}. Evaluated against the true $\delta^{\mathrm{c}}$ of \texttt{FLAMINGO}, none of the binary mask methods is accurate enough. The two-dimensional projection makes the pixel-level collapsed fraction a smooth function of the projected density. This fails at capturing the two-point statistics of the field beyond the largest scales, while also not capturing any of the higher-order moments of the field. Approximating $\delta^{\mathrm{c}}$ with a conditional excursion-set weighting reproduces the two-point statistics of the field accurately. Additionally, extending the excursion set by Poisson-sampling halo counts per pixel from the conditional mass function reproduces the angular power spectrum, discreteness level, and the heavy tail in the one-point function of the density field, simultaneously. Nevertheless, even the most accurate approximations have a maximum cross-correlation $\sim 0.5$–$0.6$ with the true  $\delta^{\mathrm{c}}$ field. At a fixed projected matter density, gravitational collapse retains non-local scatter that cannot be captured in a deterministic local map. In other words, the local-in-matter-density biasing \citep{bartlett2024bye} also fails in projection and limits the accuracy that can be achieved by these approximate methods. The BOLTS model applied to such approximate collapsed matter density fields therefore reproduces the statistics of the galaxy field in the ensemble rather than its per-sky realisation. A detailed analysis of approximations of the $\delta^{\mathrm{c}}$ field or the dark matter halo field from the matter field, $\delta^{\mathrm{m}}$, is beyond the scope of this work and will be explored in future work.

Additionally, since the BOLTS model is defined for any bias mapping $\delta \rightarrow \delta^{\mathrm{b}}$, the fitting functions for $\beta_{\ell}$ and $\sigma_{\ell}$ can be applied to other input fields beyond $\delta^{\mathrm{c}}$, if the mapping is approximately linear while still exhibiting stochasticity and non-locality. The mapping from $\delta^{\mathrm{m}}$ to $\delta^{\mathrm{c}}$ could in principle be modelled with a separate BOLTS parameterisation, for example, and the resulting $\beta_{\ell}$ and $\sigma_{\ell}$ could be combined with those for the $\delta^{\mathrm{c}}$ to $\delta^{\mathrm{g}}$ mapping to yield an effective bias model from $\delta^{\mathrm{m}}$ to $\delta^{\mathrm{g}}$. However, we know that the mapping from $\delta^{\mathrm{m}}$ to $\delta^{\mathrm{c}}$ is dominated by non-linear gravitational collapse, BOLTS cannot capture such non-linearities while also not necessarily incorporating additional contributions, such as assembly bias, unless it is explicitly conditioned on certain galaxy properties.

\section{Conclusions}\label{conclusion}

We have presented the Bias of Overdensities in Large-scale Tracers on the Sphere (BOLTS), an analytical model for non-local stochastic field-level galaxy bias in harmonic space on the sphere. Homogeneity and realness of the matter and tracer fields require $\langle b_{\ell m} \rangle = \beta_{\ell} \in \mathbb{R}$, and, as we show in Appendix~\ref{appendix:stochasticity_in_n}, scale-dependent stochasticity cannot arise in the tracer sampling step alone. It must enter at the field-level mapping $\delta \rightarrow \delta^{\mathrm{b}}$. The simplest physically consistent linear bias model is therefore both non-local and stochastic, with a zero-mean complex Gaussian term $\gamma_{\ell m}$ whose real and imaginary parts are uncorrelated ($\rho = 0$, a consequence of isotropy) and share the total variance $\sigma_{\ell}^2$ equally (exact for the stochastic residual field, and measured to hold at the per-cent level for the multiplicative coefficient, see Equation~\ref{eqn:final_model}).

Despite the added complexity of non-locality and stochasticity, the model remains analytically tractable, since $\beta_{\ell}$ and $\sigma_{\ell}^2$ allow for closed-form least-squares estimators expressible as ratios of the cross- and auto-power spectra of the tracer and matter fields (Equations~\ref{eqn:beta_ml} and~\ref{eqn:sigma_ml}), making calibration from simulations or observations straightforward and computationally inexpensive.

We have validated BOLTS against the \texttt{FLAMINGO} `L2p8\_m9' hydrodynamical simulation across three halo mass ranges and out to $z = 3$. When calibrated on projected power spectra alone, as shown in Figure~\ref{fig:bispectrum}, the model recovers the equilateral angular bispectrum down to comoving scales of $\sim 3$-$4.5\,\mathrm{Mpc}$ at $z \le 0.5$ for $M_{\mathrm{h}} > 10^{13}\,\mathrm{M}_{\odot}$ (degrading to $50$-$90\,\mathrm{Mpc}$ by $z = 2$-$3$ as this sample becomes sparse), and down to $\sim 11$-$27\,\mathrm{Mpc}$ at $z \le 1.5$ and $\sim 9$-$19\,\mathrm{Mpc}$ at $2 \le z \le 3$ for $10^{11.5}\,\mathrm{M}_{\odot} < M_{\mathrm{h}} < 10^{13}\,\mathrm{M}_{\odot}$. The analytical bispectrum prediction (Equation~\ref{eqn:analytical_blll}) agrees with the mock measurements at all halo masses and redshifts.

Beyond two- and three-point statistics, the wavelet phase harmonic (WPH) moments measured from BOLTS mocks shown in Figures~\ref{fig:wph_s01s11},~\ref{fig:wph_c01}, and~\ref{fig:wph_c01_highz} match \texttt{FLAMINGO} within the sampling scatter of the measurement across all WPH coefficient types ($S_{11}$, $S_{01}$, $C_{01}$, $C_{\mathrm{phase}}$) in the low-mass, low-redshift regime where non-locality and stochasticity are most pronounced, while local bias models deviate by 5-9$\sigma$ in the high-mass, high-redshift regime, where the stochastic power is the discreteness of the tracers, BOLTS and a deterministic local bias of $\delta^{\mathrm{c}}$ are equally accurate. At high redshift and high halo mass, where the bias becomes approximately scale-independent, BOLTS converges smoothly to a local deterministic model on large scales. On small scales in this regime, the stochastic term can erase phase correlations and the discreteness of the collapsed field raises the measured bias, reducing the accuracy of some higher-order statistics.

As exemplified by Figure~\ref{fig:cartesian_redsidual_maps}, we find that this agreement extends to the real-space galaxy fields. Across halo masses and redshifts, BOLTS produces galaxy fields which are statistically more consistent \texttt{FLAMINGO} than any linear local bias model.

For practical forward-modelling applications, we have introduced analytical fitting functions that compress the per-$\ell$ parameterisation into eight parameters per mass bin (two of which are fixed as a function of redshift; see Figure~\ref{fig:params_per_z}). The scale dependence of $\beta_{\ell}$ and $\sigma_{\ell}$ is well captured by the generalised Lorentzian (Equation~\ref{eqn:beta_l_fit}) and sigmoidal (Equation~\ref{eqn:sigma_l_fit}) forms, whose parameters vary smoothly with redshift and halo mass. This makes BOLTS a practical simulator for simulation-based inference pipelines, while the same parameterisation can serve as an analytical prescription for galaxy bias in standard likelihood-based analyses, with the values quoted here as informative priors on the nuisance parameters.

The BOLTS framework is designed to improve the robustness of joint galaxy clustering and galaxy-galaxy lensing consistency tests by separating the non-local, stochastic component of galaxy bias from the deterministic, large-scale contribution. It is naturally suited to projected analyses on the sphere, as required by wide-field surveys such as \textit{Euclid} \citep{laureijs2011euclid}, \textit{Rubin} LSST \citep{lsst2009lsst}, and \emph{Roman} \citep{spergel2015wide}. The model can be conditioned on galaxy properties such as stellar mass, star-formation rate, or colour, enabling the study of assembly bias within the same formalism. This conditioning, along with the exploration of the cosmology dependence of the BOLTS parameters, will be the subject of future work.

Several limitations of the present analysis should be noted. The calibration and validation rest on a single \texttt{FLAMINGO} run (L2p8\_m9), i.e. a single cosmology (D3A, based on \citealt{planck2020results}) and a single sub-grid feedback calibration, so the sensitivity of $\beta_{\ell}$ and $\sigma_{\ell}$ to cosmological parameters or to alternative feedback prescriptions has not been quantified. Since feedback alters the small-scale matter power spectrum and the halo mass function, $\sigma_{\ell}$ is plausibly the more sensitive of the two. Quantifying these dependencies with the broader \texttt{FLAMINGO} suite is essential before BOLTS enters cosmological inference pipelines, and is left to future work. In addition, the mock uncertainties quoted here vary only the stochastic bias draws on a single realisation of the matter field, and therefore contain no cosmic variance.

By construction, BOLTS is a linear mode-by-mode mapping in harmonic space, so it cannot capture contributions that couple different $(\ell, m)$, as induced by non-linear gravitational evolution and galaxy formation. Its success with the bispectrum and WPH statistics indicates that much of the apparently non-linear structure in the galaxy field is phenomenologically well approximated by a stochastic linear mapping once non-locality has been accounted for. The residual WPH discrepancies at the highest halo masses and redshifts (Section~\ref{calibration:wph}) indicate where the discreteness of the tracers, rather than genuine stochasticity, become non-negligible, and where computationally tractable higher-order extensions would be a natural development.

The model treats each redshift shell independently: the stochastic draws of distinct shells are uncorrelated by construction, whereas projection and the line-of-sight extent of structure correlate adjacent thin shells. For tomographic analyses such as $3{\times}2\mathrm{pt}$, the framework can be extended by promoting the per-shell stochastic variance to a cross-shell covariance. We defer this to future work.

We do not distinguish between central and satellite galaxies, so a given halo-mass bin mixes the two populations. In initial tests, satellites are more strongly clustered at small scales and more stochastic than centrals. Since conditioning on centrality works exactly like conditioning on mass, BOLTS can be calibrated separately for the two populations, providing an interface to decompositions similar to halo occupation distribution models in future extensions.

BOLTS is complementary to three-dimensional bias frameworks. Lagrangian Perturbation Theory (LPT) and Hybrid Effective Field Theory (HEFT) approaches \citep{modi2020simulations, kokron2021cosmology, pellejero2022modelling} advect Lagrangian bias operators forward, generating non-local Eulerian contributions and reaching $k \sim 0.5$-$0.6\,h\,\mathrm{Mpc}^{-1}$ in three dimensions, while cosmic-web-classified effective bias models \citep{coloma2024hierarchical} offer another approach to non-locality. However, applying either to projected surveys requires constructing and integrating full lightcones at every likelihood evaluation. BOLTS trades their physical operator basis for a phenomenological description calibrated at the level of angular power spectra, operating directly on the projected sphere. LPT/HEFT suit spectroscopic three-dimensional analyses, while BOLTS provides the rapid on-sphere mock generation required by photometric surveys and simulation-based inference.

\section*{Acknowledgements}
The authors thank Richard Massey, Matthew Ho, Benjamin Wandelt, Matthieu Schaller, Jason McEwen, Jess Doppel, and Willem Elbers for helpful discussions. MvWK acknowledges the support by UK STFC (grant no. ST/X001075/1), the UK Space Agency (grant no. ST/X001997/1), and InnovateUK (grant no. TS/Y014693/1). BJ, QH, and NJ acknowledge support by the ERC-selected UKRI Frontier Research Grant EP/Y03015X/1. This work used the DiRAC@Durham facility managed by the Institute for Computational Cosmology on behalf of the STFC DiRAC HPC Facility (www.dirac.ac.uk). The equipment was funded by BEIS capital funding via STFC capital grants ST/K00042X/1, ST/P002293/1, ST/R002371/1 and ST/S002502/1, Durham University and STFC operations grant ST/R000832/1. DiRAC is part of the National e-Infrastructure. The work presented in this paper made use of the following software packages: \texttt{Numpy} \citep{harris2020array}, \texttt{SciPy} \citep{virtanen2020scipy}, \texttt{AstroPy} \citep{astropy2018astropy}, \texttt{HealPy} \citep{zonca2019healpy}, \texttt{PyTorch} \citep{paszke2017automatic}, \texttt{Jupyter Lab} \citep{kluyver2016jupyter}, \texttt{numba} \citep{lam2015numba}, and \texttt{Matplotlib} \citep{hunter2007matplotlib}.

During the preparation of this work, the authors utilised Claude Code (Anthropic, \url{https://anthropic.com}) to assist with software 
development, specifically for optimising and debugging the python code based on original prototype scripts. The tool was used exclusively for coding assistance, while no text or scientific interpretation within this manuscript was generated by Artificial Intelligence. The authors accept full accountability for the accuracy and reproducibility of all software and scientific outputs.

\section*{Data Availability}

The \texttt{FLAMINGO} simulation data used in this work are publicly available through the FLAMINGO data release \citep{helly2026flamingo}. The BOLTS calibration code and the scripts to reproduce the figures presented in this paper will be made available at \url{https://github.com/mwiet} upon publication. Derived data products, including the calibrated bias parameters and synthetic galaxy field realisations, will be shared upon reasonable request to the corresponding author.



\bibliographystyle{mnras}
\bibliography{example} 

\begin{thebibliography}{}
\makeatletter
\relax
\def\mn@urlcharsother{\let\do\@makeother \do\$\do\&\do\#\do\^\do\_\do\%\do\~}
\def\mn@doi{\begingroup\mn@urlcharsother \@ifnextchar [ {\mn@doi@} {\mn@doi@[]}}
\def\mn@doi@[#1]#2{\def\@tempa{#1}\ifx\@tempa\@empty \href {http://dx.doi.org/#2} {doi:#2}\else \href {http://dx.doi.org/#2} {#1}\fi \endgroup}
\def\mn@eprint#1#2{\mn@eprint@#1:#2::\@nil}
\def\mn@eprint@arXiv#1{\href {http://arxiv.org/abs/#1} {{\tt arXiv:#1}}}
\def\mn@eprint@dblp#1{\href {http://dblp.uni-trier.de/rec/bibtex/#1.xml} {dblp:#1}}
\def\mn@eprint@#1:#2:#3:#4\@nil{\def\@tempa {#1}\def\@tempb {#2}\def\@tempc {#3}\ifx \@tempc \@empty \let \@tempc \@tempb \let \@tempb \@tempa \fi \ifx \@tempb \@empty \def\@tempb {arXiv}\fi \@ifundefined {mn@eprint@\@tempb}{\@tempb:\@tempc}{\expandafter \expandafter \csname mn@eprint@\@tempb\endcsname \expandafter{\@tempc}}}

\bibitem[\protect\citeauthoryear{{Abdul Karim} et~al.,}{{Abdul Karim} et~al.}{2025}]{desi2025results}
{Abdul Karim} M.,  et~al., 2025, \mn@doi [\prd] {10.1103/tr6y-kpc6}, \href {https://ui.adsabs.harvard.edu/abs/2025PhRvD.112h3515A} {112, 083515}

\bibitem[\protect\citeauthoryear{{Allys}, {Marchand}, {Cardoso}, {Villaescusa-Navarro}, {Ho}  \& {Mallat}}{{Allys} et~al.}{2020}]{allys2020new}
{Allys} E.,  {Marchand} T.,  {Cardoso} J.-F.,  {Villaescusa-Navarro} F.,  {Ho} S.,   {Mallat} S.,  2020, \mn@doi [\prd] {10.1103/PhysRevD.102.103506}, \href {https://ui.adsabs.harvard.edu/abs/2020PhRvD.102j3506A} {102, 103506}

\bibitem[\protect\citeauthoryear{{Artale}, {Zehavi}, {Contreras}  \& {Norberg}}{{Artale} et~al.}{2018}]{artale2018impact}
{Artale} M.~C.,  {Zehavi} I.,  {Contreras} S.,   {Norberg} P.,  2018, \mn@doi [\mnras] {10.1093/mnras/sty2110}, \href {https://ui.adsabs.harvard.edu/abs/2018MNRAS.480.3978A} {480, 3978}

\bibitem[\protect\citeauthoryear{{Astropy Collaboration} et~al.,}{{Astropy Collaboration} et~al.}{2018}]{astropy2018astropy}
{Astropy Collaboration} et~al., 2018, \mn@doi [\aj] {10.3847/1538-3881/aabc4f}, \href {https://ui.adsabs.harvard.edu/abs/2018AJ....156..123A} {156, 123}

\bibitem[\protect\citeauthoryear{{Ata}, {Kitaura}  \& {M{\"u}ller}}{{Ata} et~al.}{2015}]{Ata2015Bayesian}
{Ata} M.,  {Kitaura} F.-S.,   {M{\"u}ller} V.,  2015, \mn@doi [\mnras] {10.1093/mnras/stu2347}, \href {https://ui.adsabs.harvard.edu/abs/2015MNRAS.446.4250A} {446, 4250}

\bibitem[\protect\citeauthoryear{{Baldauf}, {Seljak}, {Smith}, {Hamaus}  \& {Desjacques}}{{Baldauf} et~al.}{2013}]{baldauf2013halo}
{Baldauf} T.,  {Seljak} U.,  {Smith} R.~E.,  {Hamaus} N.,   {Desjacques} V.,  2013, \mn@doi [\prd] {10.1103/PhysRevD.88.083507}, \href {https://ui.adsabs.harvard.edu/abs/2013PhRvD..88h3507B} {88, 083507}

\bibitem[\protect\citeauthoryear{{Baldi} \& {Marinucci}}{{Baldi} \& {Marinucci}}{2006}]{baldi2007characterizations}
{Baldi} P.,  {Marinucci} D.,  2006, \mn@doi [arXiv Mathematics e-prints] {10.48550/arXiv.math/0606709}, \href {https://ui.adsabs.harvard.edu/abs/2006math......6709B} {p. math/0606709}

\bibitem[\protect\citeauthoryear{{Bartlett}, {Ho}  \& {Wandelt}}{{Bartlett} et~al.}{2024}]{bartlett2024bye}
{Bartlett} D.~J.,  {Ho} M.,   {Wandelt} B.~D.,  2024, \mn@doi [\apjl] {10.3847/2041-8213/ad97b9}, \href {https://ui.adsabs.harvard.edu/abs/2024ApJ...977L..44B} {977, L44}

\bibitem[\protect\citeauthoryear{{Berlind} \& {Weinberg}}{{Berlind} \& {Weinberg}}{2002}]{Berlind2002halo}
{Berlind} A.~A.,  {Weinberg} D.~H.,  2002, \mn@doi [\apj] {10.1086/341469}, \href {https://ui.adsabs.harvard.edu/abs/2002ApJ...575..587B} {575, 587}

\bibitem[\protect\citeauthoryear{{Bernardeau}, {Colombi}, {Gazta{\~n}aga}  \& {Scoccimarro}}{{Bernardeau} et~al.}{2002}]{Bernardeau2002large}
{Bernardeau} F.,  {Colombi} S.,  {Gazta{\~n}aga} E.,   {Scoccimarro} R.,  2002, \mn@doi [\physrep] {10.1016/S0370-1573(02)00135-7}, \href {https://ui.adsabs.harvard.edu/abs/2002PhR...367....1B} {367, 1}

\bibitem[\protect\citeauthoryear{{Booth} \& {Schaye}}{{Booth} \& {Schaye}}{2009}]{booth2009cosmological}
{Booth} C.~M.,  {Schaye} J.,  2009, \mn@doi [\mnras] {10.1111/j.1365-2966.2009.15043.x}, \href {https://ui.adsabs.harvard.edu/abs/2009MNRAS.398...53B} {398, 53}

\bibitem[\protect\citeauthoryear{{Borrow}, {Schaller}, {Bower}  \& {Schaye}}{{Borrow} et~al.}{2022}]{borrow2022sphenix}
{Borrow} J.,  {Schaller} M.,  {Bower} R.~G.,   {Schaye} J.,  2022, \mn@doi [\mnras] {10.1093/mnras/stab3166}, \href {https://ui.adsabs.harvard.edu/abs/2022MNRAS.511.2367B} {511, 2367}

\bibitem[\protect\citeauthoryear{{Bose}, {Eisenstein}, {Hernquist}, {Pillepich}, {Nelson}, {Marinacci}, {Springel}  \& {Vogelsberger}}{{Bose} et~al.}{2019}]{bose2019revealing}
{Bose} S.,  {Eisenstein} D.~J.,  {Hernquist} L.,  {Pillepich} A.,  {Nelson} D.,  {Marinacci} F.,  {Springel} V.,   {Vogelsberger} M.,  2019, \mn@doi [\mnras] {10.1093/mnras/stz2546}, \href {https://ui.adsabs.harvard.edu/abs/2019MNRAS.490.5693B} {490, 5693}

\bibitem[\protect\citeauthoryear{{Carrasco}, {Hertzberg}  \& {Senatore}}{{Carrasco} et~al.}{2012}]{Carrasco2012effective}
{Carrasco} J. J.~M.,  {Hertzberg} M.~P.,   {Senatore} L.,  2012, \mn@doi [Journal of High Energy Physics] {10.1007/JHEP09(2012)082}, \href {https://ui.adsabs.harvard.edu/abs/2012JHEP...09..082C} {2012, 82}

\bibitem[\protect\citeauthoryear{{Casas-Miranda}, {Mo}, {Sheth}  \& {Boerner}}{{Casas-Miranda} et~al.}{2002}]{casas-miranda2002on}
{Casas-Miranda} R.,  {Mo} H.~J.,  {Sheth} R.~K.,   {Boerner} G.,  2002, \mn@doi [\mnras] {10.1046/j.1365-8711.2002.05378.x}, \href {https://ui.adsabs.harvard.edu/abs/2002MNRAS.333..730C} {333, 730}

\bibitem[\protect\citeauthoryear{{Chaikin}, {Schaye}, {Schaller}, {Ben{\'\i}tez-Llambay}, {Nobels}  \& {Ploeckinger}}{{Chaikin} et~al.}{2023}]{chaikin2023thermal}
{Chaikin} E.,  {Schaye} J.,  {Schaller} M.,  {Ben{\'\i}tez-Llambay} A.,  {Nobels} F. S.~J.,   {Ploeckinger} S.,  2023, \mn@doi [\mnras] {10.1093/mnras/stad1626}, \href {https://ui.adsabs.harvard.edu/abs/2023MNRAS.523.3709C} {523, 3709}

\bibitem[\protect\citeauthoryear{{Coloma-Nadal}, {Kitaura}, {Garc{\'\i}a-Farieta}, {Sinigaglia}, {Favole}  \& {Forero S{\'a}nchez}}{{Coloma-Nadal} et~al.}{2024}]{coloma2024hierarchical}
{Coloma-Nadal} J.~M.,  {Kitaura} F.-S.,  {Garc{\'\i}a-Farieta} J.~E.,  {Sinigaglia} F.,  {Favole} G.,   {Forero S{\'a}nchez} D.,  2024, \mn@doi [\jcap] {10.1088/1475-7516/2024/07/083}, \href {https://ui.adsabs.harvard.edu/abs/2024JCAP...07..083C} {2024, 083}

\bibitem[\protect\citeauthoryear{{Conroy}, {Wechsler}  \& {Kravtsov}}{{Conroy} et~al.}{2006}]{conroy2006modeling}
{Conroy} C.,  {Wechsler} R.~H.,   {Kravtsov} A.~V.,  2006, \mn@doi [\apj] {10.1086/503602}, \href {https://ui.adsabs.harvard.edu/abs/2006ApJ...647..201C} {647, 201}

\bibitem[\protect\citeauthoryear{{Cooray} \& {Sheth}}{{Cooray} \& {Sheth}}{2002}]{cooray2002halo}
{Cooray} A.,  {Sheth} R.,  2002, \mn@doi [\physrep] {10.1016/S0370-1573(02)00276-4}, \href {https://ui.adsabs.harvard.edu/abs/2002PhR...372....1C} {372, 1}

\bibitem[\protect\citeauthoryear{{Croton}, {Gao}  \& {White}}{{Croton} et~al.}{2007}]{croton2007halo}
{Croton} D.~J.,  {Gao} L.,   {White} S. D.~M.,  2007, \mn@doi [\mnras] {10.1111/j.1365-2966.2006.11230.x}, \href {https://ui.adsabs.harvard.edu/abs/2007MNRAS.374.1303C} {374, 1303}

\bibitem[\protect\citeauthoryear{{Dalal}, {White}, {Bond}  \& {Shirokov}}{{Dalal} et~al.}{2008}]{dalal2008halo}
{Dalal} N.,  {White} M.,  {Bond} J.~R.,   {Shirokov} A.,  2008, \mn@doi [\apj] {10.1086/591512}, \href {https://ui.adsabs.harvard.edu/abs/2008ApJ...687...12D} {687, 12}

\bibitem[\protect\citeauthoryear{{Dalal} et~al.,}{{Dalal} et~al.}{2023}]{dalal2023hsc}
{Dalal} R.,  et~al., 2023, \mn@doi [\prd] {10.1103/PhysRevD.108.123519}, \href {https://ui.adsabs.harvard.edu/abs/2023PhRvD.108l3519D} {108, 123519}

\bibitem[\protect\citeauthoryear{{Dekel} \& {Lahav}}{{Dekel} \& {Lahav}}{1999}]{dekel1999stochastic}
{Dekel} A.,  {Lahav} O.,  1999, \mn@doi [\apj] {10.1086/307428}, \href {https://ui.adsabs.harvard.edu/abs/1999ApJ...520...24D} {520, 24}

\bibitem[\protect\citeauthoryear{{Desjacques}, {Jeong}  \& {Schmidt}}{{Desjacques} et~al.}{2018}]{Desjacques2018large}
{Desjacques} V.,  {Jeong} D.,   {Schmidt} F.,  2018, \mn@doi [\physrep] {10.1016/j.physrep.2017.12.002}, \href {https://ui.adsabs.harvard.edu/abs/2018PhR...733....1D} {733, 1}

\bibitem[\protect\citeauthoryear{{Elbers}, {Frenk}, {Jenkins}, {Li}  \& {Pascoli}}{{Elbers} et~al.}{2021}]{elbers2021optimal}
{Elbers} W.,  {Frenk} C.~S.,  {Jenkins} A.,  {Li} B.,   {Pascoli} S.,  2021, \mn@doi [\mnras] {10.1093/mnras/stab2260}, \href {https://ui.adsabs.harvard.edu/abs/2021MNRAS.507.2614E} {507, 2614}

\bibitem[\protect\citeauthoryear{{Erdo{\v{g}}du} et~al.,}{{Erdo{\v{g}}du} et~al.}{2004}]{erdogdu2004the}
{Erdo{\v{g}}du} P.,  et~al., 2004, \mn@doi [\mnras] {10.1111/j.1365-2966.2004.07984.x}, \href {https://ui.adsabs.harvard.edu/abs/2004MNRAS.352..939E} {352, 939}

\bibitem[\protect\citeauthoryear{{Euclid Collaboration} et~al.,}{{Euclid Collaboration} et~al.}{2025}]{euclid2025mission}
{Euclid Collaboration} et~al., 2025, \mn@doi [\aap] {10.1051/0004-6361/202450810}, \href {https://ui.adsabs.harvard.edu/abs/2025A&A...697A...1E} {697, A1}

\bibitem[\protect\citeauthoryear{{Fluri}, {Kacprzak}, {Lucchi}, {Schneider}, {Refregier}  \& {Hofmann}}{{Fluri} et~al.}{2022}]{fluri2022full}
{Fluri} J.,  {Kacprzak} T.,  {Lucchi} A.,  {Schneider} A.,  {Refregier} A.,   {Hofmann} T.,  2022, \mn@doi [\prd] {10.1103/PhysRevD.105.083518}, \href {https://ui.adsabs.harvard.edu/abs/2022PhRvD.105h3518F} {105, 083518}

\bibitem[\protect\citeauthoryear{{Forouhar Moreno}, {Helly}, {McGibbon}, {Schaye}, {Schaller}, {Han}, {Kugel}  \& {Bah{\'e}}}{{Forouhar Moreno} et~al.}{2025}]{forouhar2025assessing}
{Forouhar Moreno} V.~J.,  {Helly} J.,  {McGibbon} R.,  {Schaye} J.,  {Schaller} M.,  {Han} J.,  {Kugel} R.,   {Bah{\'e}} Y.~M.,  2025, \mn@doi [\mnras] {10.1093/mnras/staf1478}, \href {https://ui.adsabs.harvard.edu/abs/2025MNRAS.543.1339F} {543, 1339}

\bibitem[\protect\citeauthoryear{{Friedrich}, {Halder}, {Boyle}, {Uhlemann}, {Britt}, {Codis}, {Gruen}  \& {Hahn}}{{Friedrich} et~al.}{2022}]{friedrich2022pdf}
{Friedrich} O.,  {Halder} A.,  {Boyle} A.,  {Uhlemann} C.,  {Britt} D.,  {Codis} S.,  {Gruen} D.,   {Hahn} C.,  2022, \mn@doi [\mnras] {10.1093/mnras/stab3703}, \href {https://ui.adsabs.harvard.edu/abs/2022MNRAS.510.5069F} {510, 5069}

\bibitem[\protect\citeauthoryear{{Frusciante} \& {Sheth}}{{Frusciante} \& {Sheth}}{2012}]{Frusciante2012lagrangian}
{Frusciante} N.,  {Sheth} R.~K.,  2012, \mn@doi [\jcap] {10.1088/1475-7516/2012/11/016}, \href {https://ui.adsabs.harvard.edu/abs/2012JCAP...11..016F} {2012, 016}

\bibitem[\protect\citeauthoryear{{Fry} \& {Gaztanaga}}{{Fry} \& {Gaztanaga}}{1993}]{fry1993biasing}
{Fry} J.~N.,  {Gaztanaga} E.,  1993, \mn@doi [\apj] {10.1086/173015}, \href {https://ui.adsabs.harvard.edu/abs/1993ApJ...413..447F} {413, 447}

\bibitem[\protect\citeauthoryear{{Gao} \& {White}}{{Gao} \& {White}}{2007}]{gao2007assembly}
{Gao} L.,  {White} S. D.~M.,  2007, \mn@doi [\mnras] {10.1111/j.1745-3933.2007.00292.x}, \href {https://ui.adsabs.harvard.edu/abs/2007MNRAS.377L...5G} {377, L5}

\bibitem[\protect\citeauthoryear{{Gatti} et~al.,}{{Gatti} et~al.}{2024}]{gatti2024des}
{Gatti} M.,  et~al., 2024, \mn@doi [\prd] {10.1103/PhysRevD.109.063534}, \href {https://ui.adsabs.harvard.edu/abs/2024PhRvD.109f3534G} {109, 063534}

\bibitem[\protect\citeauthoryear{{Ginzburg}, {Desjacques}  \& {Chan}}{{Ginzburg} et~al.}{2017}]{Ginzburg2017shot}
{Ginzburg} D.,  {Desjacques} V.,   {Chan} K.~C.,  2017, \mn@doi [\prd] {10.1103/PhysRevD.96.083528}, \href {https://ui.adsabs.harvard.edu/abs/2017PhRvD..96h3528G} {96, 083528}

\bibitem[\protect\citeauthoryear{{G{\'o}rski}, {Hivon}, {Banday}, {Wandelt}, {Hansen}, {Reinecke}  \& {Bartelmann}}{{G{\'o}rski} et~al.}{2005}]{gorski2005healpix}
{G{\'o}rski} K.~M.,  {Hivon} E.,  {Banday} A.~J.,  {Wandelt} B.~D.,  {Hansen} F.~K.,  {Reinecke} M.,   {Bartelmann} M.,  2005, \mn@doi [\apj] {10.1086/427976}, \href {https://ui.adsabs.harvard.edu/abs/2005ApJ...622..759G} {622, 759}

\bibitem[\protect\citeauthoryear{{Hadzhiyska}, {Bose}, {Eisenstein}, {Hernquist}  \& {Spergel}}{{Hadzhiyska} et~al.}{2020}]{Hadzhiyska2020limitations}
{Hadzhiyska} B.,  {Bose} S.,  {Eisenstein} D.,  {Hernquist} L.,   {Spergel} D.~N.,  2020, \mn@doi [\mnras] {10.1093/mnras/staa623}, \href {https://ui.adsabs.harvard.edu/abs/2020MNRAS.493.5506H} {493, 5506}

\bibitem[\protect\citeauthoryear{{Hadzhiyska} et~al.,}{{Hadzhiyska} et~al.}{2023}]{Hadzhiyska2023millennium}
{Hadzhiyska} B.,  et~al., 2023, \mn@doi [\mnras] {10.1093/mnras/stad731}, \href {https://ui.adsabs.harvard.edu/abs/2023MNRAS.524.2507H} {524, 2507}

\bibitem[\protect\citeauthoryear{{Hahn}, {Porciani}, {Carollo}  \& {Dekel}}{{Hahn} et~al.}{2007}]{hahn2007properties}
{Hahn} O.,  {Porciani} C.,  {Carollo} C.~M.,   {Dekel} A.,  2007, \mn@doi [\mnras] {10.1111/j.1365-2966.2006.11318.x}, \href {https://ui.adsabs.harvard.edu/abs/2007MNRAS.375..489H} {375, 489}

\bibitem[\protect\citeauthoryear{{Hamaus}, {Seljak}, {Desjacques}, {Smith}  \& {Baldauf}}{{Hamaus} et~al.}{2010}]{hamaus2010minimizing}
{Hamaus} N.,  {Seljak} U.,  {Desjacques} V.,  {Smith} R.~E.,   {Baldauf} T.,  2010, \mn@doi [\prd] {10.1103/PhysRevD.82.043515}, \href {https://ui.adsabs.harvard.edu/abs/2010PhRvD..82d3515H} {82, 043515}

\bibitem[\protect\citeauthoryear{{Han}, {Cole}, {Frenk}, {Benitez-Llambay}  \& {Helly}}{{Han} et~al.}{2018}]{han2018hbt}
{Han} J.,  {Cole} S.,  {Frenk} C.~S.,  {Benitez-Llambay} A.,   {Helly} J.,  2018, \mn@doi [\mnras] {10.1093/mnras/stx2792}, \href {https://ui.adsabs.harvard.edu/abs/2018MNRAS.474..604H} {474, 604}

\bibitem[\protect\citeauthoryear{{Harris} et~al.,}{{Harris} et~al.}{2020}]{harris2020array}
{Harris} C.~R.,  et~al., 2020, \mn@doi [\nat] {10.1038/s41586-020-2649-2}, \href {https://ui.adsabs.harvard.edu/abs/2020Natur.585..357H} {585, 357}

\bibitem[\protect\citeauthoryear{{Hearin}, {Zentner}, {van den Bosch}, {Campbell}  \& {Tollerud}}{{Hearin} et~al.}{2016}]{hearin2016introducing}
{Hearin} A.~P.,  {Zentner} A.~R.,  {van den Bosch} F.~C.,  {Campbell} D.,   {Tollerud} E.,  2016, \mn@doi [\mnras] {10.1093/mnras/stw840}, \href {https://ui.adsabs.harvard.edu/abs/2016MNRAS.460.2552H} {460, 2552}

\bibitem[\protect\citeauthoryear{{Helly} et~al.,}{{Helly} et~al.}{2026}]{helly2026flamingo}
{Helly} J.~C.,  et~al., 2026, \mn@doi [Astronomy and Computing] {10.1016/j.ascom.2026.101159}, \href {https://ui.adsabs.harvard.edu/abs/2026A&C....5701159H} {57, 101159}

\bibitem[\protect\citeauthoryear{{Hunter}}{{Hunter}}{2007}]{hunter2007matplotlib}
{Hunter} J.~D.,  2007, \mn@doi [Computing in Science and Engineering] {10.1109/MCSE.2007.55}, \href {https://ui.adsabs.harvard.edu/abs/2007CSE.....9...90H} {9, 90}

\bibitem[\protect\citeauthoryear{{Jasche}, {Kitaura}, {Wandelt}  \& {En{\ss}lin}}{{Jasche} et~al.}{2010}]{jasche2010bayesian}
{Jasche} J.,  {Kitaura} F.~S.,  {Wandelt} B.~D.,   {En{\ss}lin} T.~A.,  2010, \mn@doi [\mnras] {10.1111/j.1365-2966.2010.16610.x}, \href {https://ui.adsabs.harvard.edu/abs/2010MNRAS.406...60J} {406, 60}

\bibitem[\protect\citeauthoryear{{Jasche}, {Leclercq}  \& {Wandelt}}{{Jasche} et~al.}{2015}]{jasche2015past}
{Jasche} J.,  {Leclercq} F.,   {Wandelt} B.~D.,  2015, \mn@doi [\jcap] {10.1088/1475-7516/2015/01/036}, \href {https://ui.adsabs.harvard.edu/abs/2015JCAP...01..036J} {2015, 036}

\bibitem[\protect\citeauthoryear{{Jeffrey} et~al.,}{{Jeffrey} et~al.}{2025}]{jeffrey2025dark}
{Jeffrey} N.,  et~al., 2025, \mn@doi [\mnras] {10.1093/mnras/stae2629}, \href {https://ui.adsabs.harvard.edu/abs/2025MNRAS.536.1303J} {536, 1303}

\bibitem[\protect\citeauthoryear{{Kaiser}}{{Kaiser}}{1984}]{kaiser1984on}
{Kaiser} N.,  1984, \mn@doi [\apjl] {10.1086/184341}, \href {https://ui.adsabs.harvard.edu/abs/1984ApJ...284L...9K} {284, L9}

\bibitem[\protect\citeauthoryear{{Kitaura}, {Jasche}, {Li}, {En{\ss}lin}, {Metcalf}, {Wandelt}, {Lemson}  \& {White}}{{Kitaura} et~al.}{2009}]{kitaura2009cosmic}
{Kitaura} F.~S.,  {Jasche} J.,  {Li} C.,  {En{\ss}lin} T.~A.,  {Metcalf} R.~B.,  {Wandelt} B.~D.,  {Lemson} G.,   {White} S. D.~M.,  2009, \mn@doi [\mnras] {10.1111/j.1365-2966.2009.15470.x}, \href {https://ui.adsabs.harvard.edu/abs/2009MNRAS.400..183K} {400, 183}

\bibitem[\protect\citeauthoryear{{Kluyver} et~al.,}{{Kluyver} et~al.}{2016}]{kluyver2016jupyter}
{Kluyver} T.,  et~al., 2016, in , IOS Press.
IOS Press, pp 87--90, \mn@doi{10.3233/978-1-61499-649-1-87}

\bibitem[\protect\citeauthoryear{{Kokron}, {DeRose}, {Chen}, {White}  \& {Wechsler}}{{Kokron} et~al.}{2021}]{kokron2021cosmology}
{Kokron} N.,  {DeRose} J.,  {Chen} S.-F.,  {White} M.,   {Wechsler} R.~H.,  2021, \mn@doi [\mnras] {10.1093/mnras/stab1358}, \href {https://ui.adsabs.harvard.edu/abs/2021MNRAS.505.1422K} {505, 1422}

\bibitem[\protect\citeauthoryear{{Kokron}, {DeRose}, {Chen}, {White}  \& {Wechsler}}{{Kokron} et~al.}{2022}]{Kokron2022priors}
{Kokron} N.,  {DeRose} J.,  {Chen} S.-F.,  {White} M.,   {Wechsler} R.~H.,  2022, \mn@doi [\mnras] {10.1093/mnras/stac1420}, \href {https://ui.adsabs.harvard.edu/abs/2022MNRAS.514.2198K} {514, 2198}

\bibitem[\protect\citeauthoryear{{Kugel} et~al.,}{{Kugel} et~al.}{2023}]{kugel2023flamingo}
{Kugel} R.,  et~al., 2023, \mn@doi [\mnras] {10.1093/mnras/stad2540}, \href {https://ui.adsabs.harvard.edu/abs/2023MNRAS.526.6103K} {526, 6103}

\bibitem[\protect\citeauthoryear{{LSST Science Collaboration} et~al.,}{{LSST Science Collaboration} et~al.}{2009}]{lsst2009lsst}
{LSST Science Collaboration} et~al., 2009, \mn@doi [arXiv e-prints] {10.48550/arXiv.0912.0201}, \href {https://ui.adsabs.harvard.edu/abs/2009arXiv0912.0201L} {p. arXiv:0912.0201}

\bibitem[\protect\citeauthoryear{{Lacey} \& {Cole}}{{Lacey} \& {Cole}}{1993}]{lacey1993merger}
{Lacey} C.,  {Cole} S.,  1993, \mn@doi [\mnras] {10.1093/mnras/262.3.627}, \href {https://ui.adsabs.harvard.edu/abs/1993MNRAS.262..627L} {262, 627}

\bibitem[\protect\citeauthoryear{{Lam}, {Pitrou}  \& {Seibert}}{{Lam} et~al.}{2015}]{lam2015numba}
{Lam} S.~K.,  {Pitrou} A.,   {Seibert} S.,  2015, in Proc. Second Workshop on the LLVM Compiler Infrastructure in HPC. pp~1--6, \mn@doi{10.1145/2833157.2833162}

\bibitem[\protect\citeauthoryear{{Laureijs} et~al.,}{{Laureijs} et~al.}{2011}]{laureijs2011euclid}
{Laureijs} R.,  et~al., 2011, \mn@doi [arXiv e-prints] {10.48550/arXiv.1110.3193}, \href {https://ui.adsabs.harvard.edu/abs/2011arXiv1110.3193L} {p. arXiv:1110.3193}

\bibitem[\protect\citeauthoryear{{Lazeyras} \& {Schmidt}}{{Lazeyras} \& {Schmidt}}{2018}]{lazeyras2018beyond}
{Lazeyras} T.,  {Schmidt} F.,  2018, \mn@doi [\jcap] {10.1088/1475-7516/2018/09/008}, \href {https://ui.adsabs.harvard.edu/abs/2018JCAP...09..008L} {2018, 008}

\bibitem[\protect\citeauthoryear{{Lazeyras} \& {Schmidt}}{{Lazeyras} \& {Schmidt}}{2019}]{lazeyras2019robust}
{Lazeyras} T.,  {Schmidt} F.,  2019, \mn@doi [\jcap] {10.1088/1475-7516/2019/11/041}, \href {https://ui.adsabs.harvard.edu/abs/2019JCAP...11..041L} {2019, 041}

\bibitem[\protect\citeauthoryear{{Lemos} et~al.,}{{Lemos} et~al.}{2023}]{lemos2023simbig}
{Lemos} P.,  et~al., 2023, in Machine Learning for Astrophysics. p.~18 (\mn@eprint {arXiv} {2310.15256}), \mn@doi{10.48550/arXiv.2310.15256}

\bibitem[\protect\citeauthoryear{{Lin}, {von wietersheim-Kramsta}, {Joachimi}  \& {Feeney}}{{Lin} et~al.}{2023}]{lin2023simulation}
{Lin} K.,  {von wietersheim-Kramsta} M.,  {Joachimi} B.,   {Feeney} S.,  2023, \mn@doi [\mnras] {10.1093/mnras/stad2262}, \href {https://ui.adsabs.harvard.edu/abs/2023MNRAS.524.6167L} {524, 6167}

\bibitem[\protect\citeauthoryear{{Mallat}, {Zhang}  \& {Rochette}}{{Mallat} et~al.}{2018}]{mallat2019phase}
{Mallat} S.,  {Zhang} S.,   {Rochette} G.,  2018, \mn@doi [arXiv e-prints] {10.48550/arXiv.1810.12136}, \href {https://ui.adsabs.harvard.edu/abs/2018arXiv181012136M} {p. arXiv:1810.12136}

\bibitem[\protect\citeauthoryear{{Matsubara}}{{Matsubara}}{1999}]{matsubara1999stochasticity}
{Matsubara} T.,  1999, \mn@doi [\apj] {10.1086/307931}, \href {https://ui.adsabs.harvard.edu/abs/1999ApJ...525..543M} {525, 543}

\bibitem[\protect\citeauthoryear{{Matsubara}}{{Matsubara}}{2011}]{matsubara2011nonlinear}
{Matsubara} T.,  2011, \mn@doi [\prd] {10.1103/PhysRevD.83.083518}, \href {https://ui.adsabs.harvard.edu/abs/2011PhRvD..83h3518M} {83, 083518}

\bibitem[\protect\citeauthoryear{{McDonald} \& {Roy}}{{McDonald} \& {Roy}}{2009}]{McDonald2009clustering}
{McDonald} P.,  {Roy} A.,  2009, \mn@doi [\jcap] {10.1088/1475-7516/2009/08/020}, \href {https://ui.adsabs.harvard.edu/abs/2009JCAP...08..020M} {2009, 020}

\bibitem[\protect\citeauthoryear{{Mirbabayi}, {Schmidt}  \& {Zaldarriaga}}{{Mirbabayi} et~al.}{2015}]{mirbabayi2015biased}
{Mirbabayi} M.,  {Schmidt} F.,   {Zaldarriaga} M.,  2015, \mn@doi [\jcap] {10.1088/1475-7516/2015/07/030}, \href {https://ui.adsabs.harvard.edu/abs/2015JCAP...07..030M} {2015, 030}

\bibitem[\protect\citeauthoryear{{Mo} \& {White}}{{Mo} \& {White}}{1996}]{mo1996analytic}
{Mo} H.~J.,  {White} S.~D.~M.,  1996, \mn@doi [\mnras] {10.1093/mnras/282.2.347}, \href {https://ui.adsabs.harvard.edu/abs/1996MNRAS.282..347M} {282, 347}

\bibitem[\protect\citeauthoryear{{Modi}, {Feng}  \& {Seljak}}{{Modi} et~al.}{2018}]{modi2018cosmological}
{Modi} C.,  {Feng} Y.,   {Seljak} U.,  2018, \mn@doi [\jcap] {10.1088/1475-7516/2018/10/028}, \href {https://ui.adsabs.harvard.edu/abs/2018JCAP...10..028M} {2018, 028}

\bibitem[\protect\citeauthoryear{{Modi}, {Chen}  \& {White}}{{Modi} et~al.}{2020}]{modi2020simulations}
{Modi} C.,  {Chen} S.-F.,   {White} M.,  2020, \mn@doi [\mnras] {10.1093/mnras/staa251}, \href {https://ui.adsabs.harvard.edu/abs/2020MNRAS.492.5754M} {492, 5754}

\bibitem[\protect\citeauthoryear{{Neyrinck}}{{Neyrinck}}{2014}]{Neyrinck2014Transformationally}
{Neyrinck} M.~C.,  2014, in {Heavens} A.,  {Starck} J.-L.,   {Krone-Martins} A.,  eds,  IAU Symposium Vol. 306, Statistical Challenges in 21st Century Cosmology. pp 251--254 (\mn@eprint {arXiv} {1407.4815}), \mn@doi{10.1017/S1743921314013702}

\bibitem[\protect\citeauthoryear{{Pandey}, {S{\'a}nchez}, {Jain}  \& {LSST Dark Energy Science Collaboration}}{{Pandey} et~al.}{2025}]{Pandey2025cosmology}
{Pandey} S.,  {S{\'a}nchez} C.,  {Jain} B.,   {LSST Dark Energy Science Collaboration} 2025, \mn@doi [\prd] {10.1103/PhysRevD.111.023527}, \href {https://ui.adsabs.harvard.edu/abs/2025PhRvD.111b3527P} {111, 023527}

\bibitem[\protect\citeauthoryear{Paszke et~al.,}{Paszke et~al.}{2017}]{paszke2017automatic}
Paszke A.,  et~al., 2017

\bibitem[\protect\citeauthoryear{{Paviot}, {Rocher}, {Codis}, {de Mattia}, {Jullo}  \& {de la Torre}}{{Paviot} et~al.}{2024}]{paviot2024impact}
{Paviot} R.,  {Rocher} A.,  {Codis} S.,  {de Mattia} A.,  {Jullo} E.,   {de la Torre} S.,  2024, \mn@doi [\aap] {10.1051/0004-6361/202449574}, \href {https://ui.adsabs.harvard.edu/abs/2024A&A...690A.221P} {690, A221}

\bibitem[\protect\citeauthoryear{{Peacock} \& {Smith}}{{Peacock} \& {Smith}}{2000}]{peacock2000halo}
{Peacock} J.~A.,  {Smith} R.~E.,  2000, \mn@doi [\mnras] {10.1046/j.1365-8711.2000.03779.x}, \href {https://ui.adsabs.harvard.edu/abs/2000MNRAS.318.1144P} {318, 1144}

\bibitem[\protect\citeauthoryear{{Pellejero Iba{\~n}ez}, {St{\"u}cker}, {Angulo}, {Zennaro}, {Contreras}  \& {Aric{\`o}}}{{Pellejero Iba{\~n}ez} et~al.}{2022}]{pellejero2022modelling}
{Pellejero Iba{\~n}ez} M.,  {St{\"u}cker} J.,  {Angulo} R.~E.,  {Zennaro} M.,  {Contreras} S.,   {Aric{\`o}} G.,  2022, \mn@doi [\mnras] {10.1093/mnras/stac1602}, \href {https://ui.adsabs.harvard.edu/abs/2022MNRAS.514.3993P} {514, 3993}

\bibitem[\protect\citeauthoryear{{Planck Collaboration} et~al.,}{{Planck Collaboration} et~al.}{2020}]{planck2020results}
{Planck Collaboration} et~al., 2020, \mn@doi [\aap] {10.1051/0004-6361/201833880}, \href {https://ui.adsabs.harvard.edu/abs/2020A&A...641A...1P} {641, A1}

\bibitem[\protect\citeauthoryear{{Ploeckinger} \& {Schaye}}{{Ploeckinger} \& {Schaye}}{2020}]{ploeckinger2020radiative}
{Ploeckinger} S.,  {Schaye} J.,  2020, \mn@doi [\mnras] {10.1093/mnras/staa2172}, \href {https://ui.adsabs.harvard.edu/abs/2020MNRAS.497.4857P} {497, 4857}

\bibitem[\protect\citeauthoryear{Raikov}{Raikov}{1937}]{raikov1937decomposition}
Raikov D.,  1937, in Dokl. Akad. Nauk SSSR. pp 9--12

\bibitem[\protect\citeauthoryear{{Ramanah}, {Lavaux}, {Jasche}  \& {Wandelt}}{{Ramanah} et~al.}{2019}]{ramanah2019cosmological}
{Ramanah} D.~K.,  {Lavaux} G.,  {Jasche} J.,   {Wandelt} B.~D.,  2019, \mn@doi [\aap] {10.1051/0004-6361/201834117}, \href {https://ui.adsabs.harvard.edu/abs/2019A&A...621A..69R} {621, A69}

\bibitem[\protect\citeauthoryear{{Reddick}, {Wechsler}, {Tinker}  \& {Behroozi}}{{Reddick} et~al.}{2013}]{Reddick2013connection}
{Reddick} R.~M.,  {Wechsler} R.~H.,  {Tinker} J.~L.,   {Behroozi} P.~S.,  2013, \mn@doi [\apj] {10.1088/0004-637X/771/1/30}, \href {https://ui.adsabs.harvard.edu/abs/2013ApJ...771...30R} {771, 30}

\bibitem[\protect\citeauthoryear{{Regaldo-Saint Blancard}, {Allys}, {Boulanger}, {Levrier}  \& {Jeffrey}}{{Regaldo-Saint Blancard} et~al.}{2021}]{regaldo2021new}
{Regaldo-Saint Blancard} B.,  {Allys} E.,  {Boulanger} F.,  {Levrier} F.,   {Jeffrey} N.,  2021, \mn@doi [\aap] {10.1051/0004-6361/202140503}, \href {https://ui.adsabs.harvard.edu/abs/2021A&A...649L..18R} {649, L18}

\bibitem[\protect\citeauthoryear{{R{\'e}galdo-Saint Blancard}, {Allys}, {Auclair}, {Boulanger}, {Eickenberg}, {Levrier}, {Vacher}  \& {Zhang}}{{R{\'e}galdo-Saint Blancard} et~al.}{2023}]{regaldo2023generative}
{R{\'e}galdo-Saint Blancard} B.,  {Allys} E.,  {Auclair} C.,  {Boulanger} F.,  {Eickenberg} M.,  {Levrier} F.,  {Vacher} L.,   {Zhang} S.,  2023, \mn@doi [\apj] {10.3847/1538-4357/aca538}, \href {https://ui.adsabs.harvard.edu/abs/2023ApJ...943....9R} {943, 9}

\bibitem[\protect\citeauthoryear{{Rubira} \& {Schmidt}}{{Rubira} \& {Schmidt}}{2025}]{rubira2025non}
{Rubira} H.,  {Schmidt} F.,  2025, \mn@doi [arXiv e-prints] {10.48550/arXiv.2511.05484}, \href {https://ui.adsabs.harvard.edu/abs/2025arXiv251105484R} {p. arXiv:2511.05484}

\bibitem[\protect\citeauthoryear{{Sanchez-Cid} et~al.,}{{Sanchez-Cid} et~al.}{2026}]{des2026dark}
{Sanchez-Cid} D.,  et~al., 2026, \mn@doi [arXiv e-prints] {10.48550/arXiv.2601.14859}, \href {https://ui.adsabs.harvard.edu/abs/2026arXiv260114859S} {p. arXiv:2601.14859}

\bibitem[\protect\citeauthoryear{{Saoulis}, {Lin}, {Jeffrey}, {von Wietersheim-Kramsta}, {Piras}, {Spurio Mancini}, {Ferreira}  \& {Joachimi}}{{Saoulis} et~al.}{2026}]{saoulis2026field}
{Saoulis} A.~A.,  {Lin} K.,  {Jeffrey} N.,  {von Wietersheim-Kramsta} M.,  {Piras} D.,  {Spurio Mancini} A.,  {Ferreira} A. M.~G.,   {Joachimi} B.,  2026, \mn@doi [arXiv e-prints] {10.48550/arXiv.2606.23346}, \href {https://ui.adsabs.harvard.edu/abs/2026arXiv260623346S} {p. arXiv:2606.23346}

\bibitem[\protect\citeauthoryear{{Schaller} et~al.,}{{Schaller} et~al.}{2024}]{schaller2024swift}
{Schaller} M.,  et~al., 2024, \mn@doi [\mnras] {10.1093/mnras/stae922}, \href {https://ui.adsabs.harvard.edu/abs/2024MNRAS.530.2378S} {530, 2378}

\bibitem[\protect\citeauthoryear{{Schaye} \& {Dalla Vecchia}}{{Schaye} \& {Dalla Vecchia}}{2008}]{schaye2008relation}
{Schaye} J.,  {Dalla Vecchia} C.,  2008, \mn@doi [\mnras] {10.1111/j.1365-2966.2007.12639.x}, \href {https://ui.adsabs.harvard.edu/abs/2008MNRAS.383.1210S} {383, 1210}

\bibitem[\protect\citeauthoryear{{Schaye} et~al.,}{{Schaye} et~al.}{2023}]{schaye2023flamingo}
{Schaye} J.,  et~al., 2023, \mn@doi [\mnras] {10.1093/mnras/stad2419}, \href {https://ui.adsabs.harvard.edu/abs/2023MNRAS.526.4978S} {526, 4978}

\bibitem[\protect\citeauthoryear{{Schmoldt} et~al.,}{{Schmoldt} et~al.}{1999}]{schmoldt1999density}
{Schmoldt} I.~M.,  et~al., 1999, \mn@doi [\aj] {10.1086/301001}, \href {https://ui.adsabs.harvard.edu/abs/1999AJ....118.1146S} {118, 1146}

\bibitem[\protect\citeauthoryear{{Seljak}}{{Seljak}}{2000}]{seljak2000analytic}
{Seljak} U.,  2000, \mn@doi [\mnras] {10.1046/j.1365-8711.2000.03715.x}, \href {https://ui.adsabs.harvard.edu/abs/2000MNRAS.318..203S} {318, 203}

\bibitem[\protect\citeauthoryear{{Shiferaw}, {Kokron}  \& {Wechsler}}{{Shiferaw} et~al.}{2025}]{Shiferaw2025how}
{Shiferaw} M.,  {Kokron} N.,   {Wechsler} R.~H.,  2025, \mn@doi [\apj] {10.3847/1538-4357/ade66e}, \href {https://ui.adsabs.harvard.edu/abs/2025ApJ...989..218S} {989, 218}

\bibitem[\protect\citeauthoryear{{Spergel} et~al.,}{{Spergel} et~al.}{2015}]{spergel2015wide}
{Spergel} D.,  et~al., 2015, \mn@doi [arXiv e-prints] {10.48550/arXiv.1503.03757}, \href {https://ui.adsabs.harvard.edu/abs/2015arXiv150303757S} {p. arXiv:1503.03757}

\bibitem[\protect\citeauthoryear{{Szalay}}{{Szalay}}{1988}]{szalay1988constraints}
{Szalay} A.~S.,  1988, \mn@doi [\apj] {10.1086/166721}, \href {https://ui.adsabs.harvard.edu/abs/1988ApJ...333...21S} {333, 21}

\bibitem[\protect\citeauthoryear{{Taruya} \& {Soda}}{{Taruya} \& {Soda}}{1999}]{taruya1999stochastic}
{Taruya} A.,  {Soda} J.,  1999, \mn@doi [\apj] {10.1086/307612}, \href {https://ui.adsabs.harvard.edu/abs/1999ApJ...522...46T} {522, 46}

\bibitem[\protect\citeauthoryear{{Tessore}, {Loureiro}, {Joachimi}, {von Wietersheim-Kramsta}  \& {Jeffrey}}{{Tessore} et~al.}{2023}]{tessore2023glass}
{Tessore} N.,  {Loureiro} A.,  {Joachimi} B.,  {von Wietersheim-Kramsta} M.,   {Jeffrey} N.,  2023, \mn@doi [The Open Journal of Astrophysics] {10.21105/astro.2302.01942}, \href {https://ui.adsabs.harvard.edu/abs/2023OJAp....6E..11T} {6, 11}

\bibitem[\protect\citeauthoryear{{Thomsen} et~al.,}{{Thomsen} et~al.}{2026}]{thomsen2026des}
{Thomsen} A.,  et~al., 2026, \mn@doi [\prd] {10.1103/3sj1-1l9f}, \href {https://ui.adsabs.harvard.edu/abs/2026PhRvD.113h3501T} {113, 083501}

\bibitem[\protect\citeauthoryear{{Tinker}, {Weinberg}, {Zheng}  \& {Zehavi}}{{Tinker} et~al.}{2005}]{tinker2005on}
{Tinker} J.~L.,  {Weinberg} D.~H.,  {Zheng} Z.,   {Zehavi} I.,  2005, \mn@doi [\apj] {10.1086/432084}, \href {https://ui.adsabs.harvard.edu/abs/2005ApJ...631...41T} {631, 41}

\bibitem[\protect\citeauthoryear{{Tinker}, {Robertson}, {Kravtsov}, {Klypin}, {Warren}, {Yepes}  \& {Gottl{\"o}ber}}{{Tinker} et~al.}{2010}]{tinker2010toward}
{Tinker} J.~L.,  {Robertson} B.~E.,  {Kravtsov} A.~V.,  {Klypin} A.,  {Warren} M.~S.,  {Yepes} G.,   {Gottl{\"o}ber} S.,  2010, \mn@doi [\apj] {10.1088/0004-637X/724/2/878}, \href {https://ui.adsabs.harvard.edu/abs/2010ApJ...724..878T} {724, 878}

\bibitem[\protect\citeauthoryear{{Vale} \& {Ostriker}}{{Vale} \& {Ostriker}}{2006}]{vale2006non}
{Vale} A.,  {Ostriker} J.~P.,  2006, \mn@doi [\mnras] {10.1111/j.1365-2966.2006.10605.x}, \href {https://ui.adsabs.harvard.edu/abs/2006MNRAS.371.1173V} {371, 1173}

\bibitem[\protect\citeauthoryear{{Velliscig}, {van Daalen}, {Schaye}, {McCarthy}, {Cacciato}, {Le Brun}  \& {Dalla Vecchia}}{{Velliscig} et~al.}{2014}]{velliscig2014impact}
{Velliscig} M.,  {van Daalen} M.~P.,  {Schaye} J.,  {McCarthy} I.~G.,  {Cacciato} M.,  {Le Brun} A. M.~C.,   {Dalla Vecchia} C.,  2014, \mn@doi [\mnras] {10.1093/mnras/stu1044}, \href {https://ui.adsabs.harvard.edu/abs/2014MNRAS.442.2641V} {442, 2641}

\bibitem[\protect\citeauthoryear{{Virtanen} et~al.,}{{Virtanen} et~al.}{2020}]{virtanen2020scipy}
{Virtanen} P.,  et~al., 2020, \mn@doi [Nature Methods] {10.1038/s41592-019-0686-2}, \href {https://ui.adsabs.harvard.edu/abs/2020NatMe..17..261V} {17, 261}

\bibitem[\protect\citeauthoryear{{Wechsler} \& {Tinker}}{{Wechsler} \& {Tinker}}{2018}]{wechsler2018connection}
{Wechsler} R.~H.,  {Tinker} J.~L.,  2018, \mn@doi [\araa] {10.1146/annurev-astro-081817-051756}, \href {https://ui.adsabs.harvard.edu/abs/2018ARA&A..56..435W} {56, 435}

\bibitem[\protect\citeauthoryear{{Wiersma}, {Schaye}, {Theuns}, {Dalla Vecchia}  \& {Tornatore}}{{Wiersma} et~al.}{2009}]{wiersma2009chemical}
{Wiersma} R. P.~C.,  {Schaye} J.,  {Theuns} T.,  {Dalla Vecchia} C.,   {Tornatore} L.,  2009, \mn@doi [\mnras] {10.1111/j.1365-2966.2009.15331.x}, \href {https://ui.adsabs.harvard.edu/abs/2009MNRAS.399..574W} {399, 574}

\bibitem[\protect\citeauthoryear{{Wright} et~al.,}{{Wright} et~al.}{2025}]{wright2025kids}
{Wright} A.~H.,  et~al., 2025, \mn@doi [arXiv e-prints] {10.48550/arXiv.2503.19441}, \href {https://ui.adsabs.harvard.edu/abs/2025arXiv250319441W} {p. arXiv:2503.19441}

\bibitem[\protect\citeauthoryear{{Zehavi}, {Contreras}, {Padilla}, {Smith}, {Baugh}  \& {Norberg}}{{Zehavi} et~al.}{2018}]{zehavi2018impact}
{Zehavi} I.,  {Contreras} S.,  {Padilla} N.,  {Smith} N.~J.,  {Baugh} C.~M.,   {Norberg} P.,  2018, \mn@doi [\apj] {10.3847/1538-4357/aaa54a}, \href {https://ui.adsabs.harvard.edu/abs/2018ApJ...853...84Z} {853, 84}

\bibitem[\protect\citeauthoryear{{Zennaro}, {Angulo}, {Contreras}, {Pellejero-Ib{\'a}{\~n}ez}  \& {Maion}}{{Zennaro} et~al.}{2022}]{Zennaro2022priors}
{Zennaro} M.,  {Angulo} R.~E.,  {Contreras} S.,  {Pellejero-Ib{\'a}{\~n}ez} M.,   {Maion} F.,  2022, \mn@doi [\mnras] {10.1093/mnras/stac1673}, \href {https://ui.adsabs.harvard.edu/abs/2022MNRAS.514.5443Z} {514, 5443}

\bibitem[\protect\citeauthoryear{{Zentner}, {Hearin}  \& {van den Bosch}}{{Zentner} et~al.}{2014}]{Zentner2014galaxy}
{Zentner} A.~R.,  {Hearin} A.~P.,   {van den Bosch} F.~C.,  2014, \mn@doi [\mnras] {10.1093/mnras/stu1383}, \href {https://ui.adsabs.harvard.edu/abs/2014MNRAS.443.3044Z} {443, 3044}

\bibitem[\protect\citeauthoryear{{Zhang} \& {Mallat}}{{Zhang} \& {Mallat}}{2019}]{zhang2019maximum}
{Zhang} S.,  {Mallat} S.,  2019, \mn@doi [arXiv e-prints] {10.48550/arXiv.1911.10017}, \href {https://ui.adsabs.harvard.edu/abs/2019arXiv191110017Z} {p. arXiv:1911.10017}

\bibitem[\protect\citeauthoryear{{Zheng} et~al.,}{{Zheng} et~al.}{2005}]{Zheng2005theoretical}
{Zheng} Z.,  et~al., 2005, \mn@doi [\apj] {10.1086/466510}, \href {https://ui.adsabs.harvard.edu/abs/2005ApJ...633..791Z} {633, 791}

\bibitem[\protect\citeauthoryear{{Zheng}, {Coil}  \& {Zehavi}}{{Zheng} et~al.}{2007}]{zheng2007galaxy}
{Zheng} Z.,  {Coil} A.~L.,   {Zehavi} I.,  2007, \mn@doi [\apj] {10.1086/521074}, \href {https://ui.adsabs.harvard.edu/abs/2007ApJ...667..760Z} {667, 760}

\bibitem[\protect\citeauthoryear{{Zonca}, {Singer}, {Lenz}, {Reinecke}, {Rosset}, {Hivon}  \& {Gorski}}{{Zonca} et~al.}{2019}]{zonca2019healpy}
{Zonca} A.,  {Singer} L.,  {Lenz} D.,  {Reinecke} M.,  {Rosset} C.,  {Hivon} E.,   {Gorski} K.,  2019, \mn@doi [The Journal of Open Source Software] {10.21105/joss.01298}, \href {https://ui.adsabs.harvard.edu/abs/2019JOSS....4.1298Z} {4, 1298}

\bibitem[\protect\citeauthoryear{{van Daalen}, {Schaye}, {Booth}  \& {Dalla Vecchia}}{{van Daalen} et~al.}{2011}]{vandaalen2011effects}
{van Daalen} M.~P.,  {Schaye} J.,  {Booth} C.~M.,   {Dalla Vecchia} C.,  2011, \mn@doi [\mnras] {10.1111/j.1365-2966.2011.18981.x}, \href {https://ui.adsabs.harvard.edu/abs/2011MNRAS.415.3649V} {415, 3649}

\bibitem[\protect\citeauthoryear{{van Daalen}, {Schaye}, {McCarthy}, {Booth}  \& {Dalla Vecchia}}{{van Daalen} et~al.}{2014}]{vandaalen2014impact}
{van Daalen} M.~P.,  {Schaye} J.,  {McCarthy} I.~G.,  {Booth} C.~M.,   {Dalla Vecchia} C.,  2014, \mn@doi [\mnras] {10.1093/mnras/stu482}, \href {https://ui.adsabs.harvard.edu/abs/2014MNRAS.440.2997V} {440, 2997}

\bibitem[\protect\citeauthoryear{{von Wietersheim-Kramsta}, {Lin}, {Tessore}, {Joachimi}, {Loureiro}, {Reischke}  \& {Wright}}{{von Wietersheim-Kramsta} et~al.}{2025}]{vonwietersheimkramsta2024kidssbi}
{von Wietersheim-Kramsta} M.,  {Lin} K.,  {Tessore} N.,  {Joachimi} B.,  {Loureiro} A.,  {Reischke} R.,   {Wright} A.~H.,  2025, \mn@doi [\aap] {10.1051/0004-6361/202450487}, \href {https://ui.adsabs.harvard.edu/abs/2025A&A...694A.223V} {694, A223}

\makeatother
\end{thebibliography}



\appendix

\section{Stochasticity in number counts}\label{appendix:stochasticity_in_n}

In this appendix, we demonstrate that non-trivial scale-dependent stochasticity in the galaxy bias cannot originate from the tracer sampling step. The argument proceeds purely from the definition of the density contrast and the boundary condition on empty cells.

Consider a density field discretised into non-overlapping grid cells, where $\delta_i$ denotes the matter density contrast in cell~$i$. We wish to obtain the number of tracers~$N_i$ given the density field. By definition, the conditional expectation of~$N_i$ is
\begin{equation}
    \Ev[\big]{N_i}_{\delta_i^{\rm b}}
    = \bar{N} \, \Bigl(1 + \delta_i^{\rm b}\Bigr) \;,
\end{equation}
where $\bar{N}$ is the mean number of tracers per cell and $\delta_i^{\rm b}$ is a biased density contrast computed from $\delta_i$ via some mapping to be specified. The production of the tracer field therefore separates into two stages:
\begin{equation}
    \delta_i \to \delta_i^{\rm b} \to N_i \;.
\end{equation}
Stochasticity may in principle enter at either stage: the field-level bias mapping, where the BOLTS model operates, or the tracer sampling, where $N_i$ is drawn from a distribution conditioned on $\delta_i^{\rm b}$. We now show that only the first stage permits a non-trivial modification of the two-point statistics.

For two distinct cells $i \ne j$, the two-point function of the normalised counts is
\begin{equation}
    \Ev*{\frac{N_i - \bar{N}}{\bar{N}} \, \frac{N_j - \bar{N}}{\bar{N}}}_{\delta_i^{\rm b} \delta_j^{\rm b}}
    = \delta_i^{\rm b} \, \delta_j^{\rm b} + \frac{\Sigma_{ij}}{\bar{N}^2},
\end{equation}
where $\Sigma_{ij}$ is the covariance between $N_i$ and $N_j$ arising from the sampling step. For a stochastic linear bias $b$ to be encoded entirely in this stage, we require
\begin{equation}
    \Ev*{\frac{N_i - \bar{N}}{\bar{N}} \, \frac{N_j - \bar{N}}{\bar{N}}}
    = b \, \Ev[\Big]{\delta_i^{\rm b} \, \delta_j^{\rm b}}.
\end{equation}
Within the family of covariances at most bilinear in the density contrasts, the general form is
\begin{equation}
    \Sigma_{ij}
    = a \, \frac{\delta_i^{\rm b} + \delta_j^{\rm b}}{2}
    + (b - 1) \, \bar{N}^2 \, \delta_i^{\rm b} \, \delta_j^{\rm b}.
\end{equation}
A physical boundary condition constrains $a$. When $\delta_i = -1$, the cell is empty by definition and the count $N_i = 0$ is deterministic. The covariance with any other cell must therefore vanish, giving
\begin{equation}
    a
    = (b - 1) \, \bar{N}^2.
\end{equation}
However, applying the same boundary condition when only one of $\delta_i$ or $\delta_j$ equals $-1$ yields
\begin{equation}
    (b - 1) \, \bar{N}^2 \, (\delta_i^{\rm b} + 1)
    = 0,
\end{equation}
which holds for all $\delta_i^{\rm b}$ only if $b = 1$: the empty-cell constraint forbids any non-trivial bias in the sampling covariance.

Raikov's theorem \citep{raikov1937decomposition} offers a complementary intuition. If a Poisson variable $\xi$ admits a decomposition $\xi = \xi_1 + \xi_2$ into independent summands, each summand is itself a shifted Poisson variable. Any additive decomposition of the Poisson sampling of galaxy counts into independent components therefore only produces shifted Poisson variates and cannot generate the scale-dependent covariance required by a non-trivial bias. Scale-dependent stochasticity must thus be introduced at the field level, before the discrete sampling, as the BOLTS model prescribes.

\section{Accuracy of the one-point and two-point statistics}\label{appendix:two-point}

\begin{figure*}
    \centering
    \includegraphics[trim={0cm 1cm 0cm 0.5cm},clip,width=0.86\linewidth]{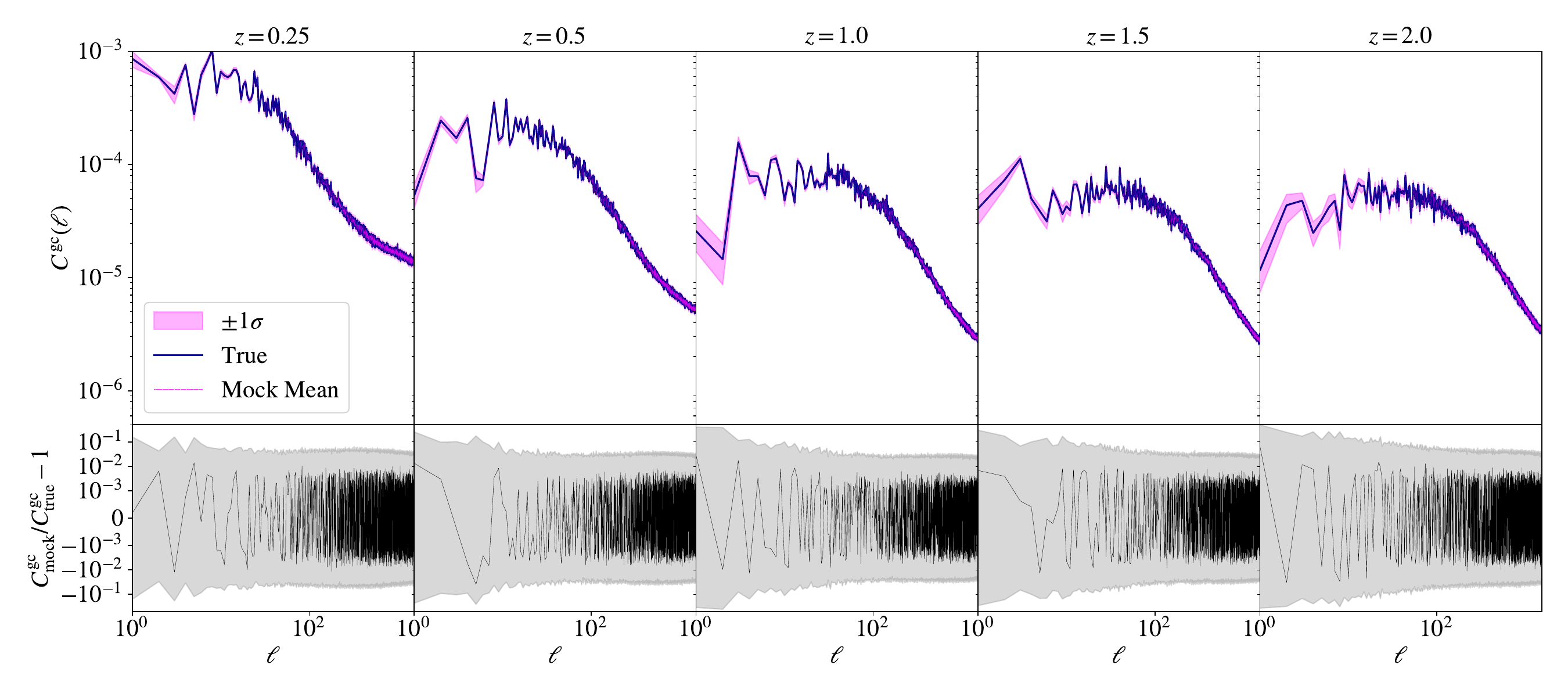}
    \caption{Two-point statistics of the biased density field. Each panel shows the angular cross-spectra $C_{\ell}^{\mathrm{gc}}$ for galaxies with $M_{\mathrm{h}} > 10^{13} \, \mathrm{M}_{\odot}$ at different redshifts. The solid blue line shows the true measurement from the \texttt{FLAMINGO} `L2p8\_m9' hydrodynamical simulation, the magenta line shows the BOLTS mock mean (with the shaded region indicating the standard deviation across 200 realisations). By construction, BOLTS reproduces the two-point statistics of the simulation at all scales. The lower panels show the fractional residuals with respect to the truth. The agreement is similarly good for all other mass bins as well as for the auto-spectra $C_{\ell}^{\mathrm{gg}}$ and $C_{\ell}^{\mathrm{cc}}$, which are not shown here for brevity.}
    \label{fig:2pt_statistics}
\end{figure*}

\begin{figure}
    \centering
    \includegraphics[trim={0 1cm 0 0},clip,width=0.9\linewidth]{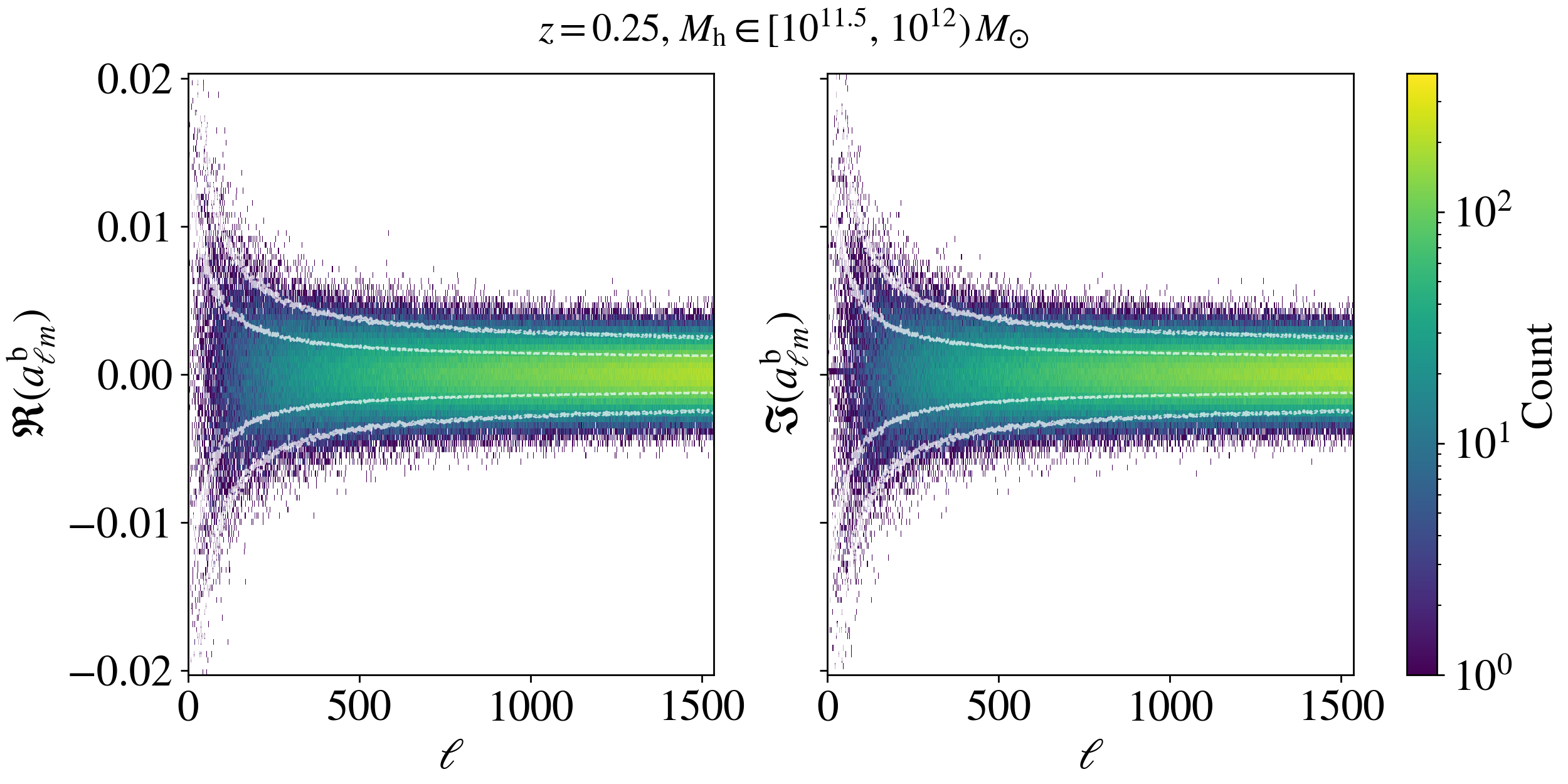}
    \includegraphics[trim={0 1cm 0 0},clip,width=0.9\linewidth]{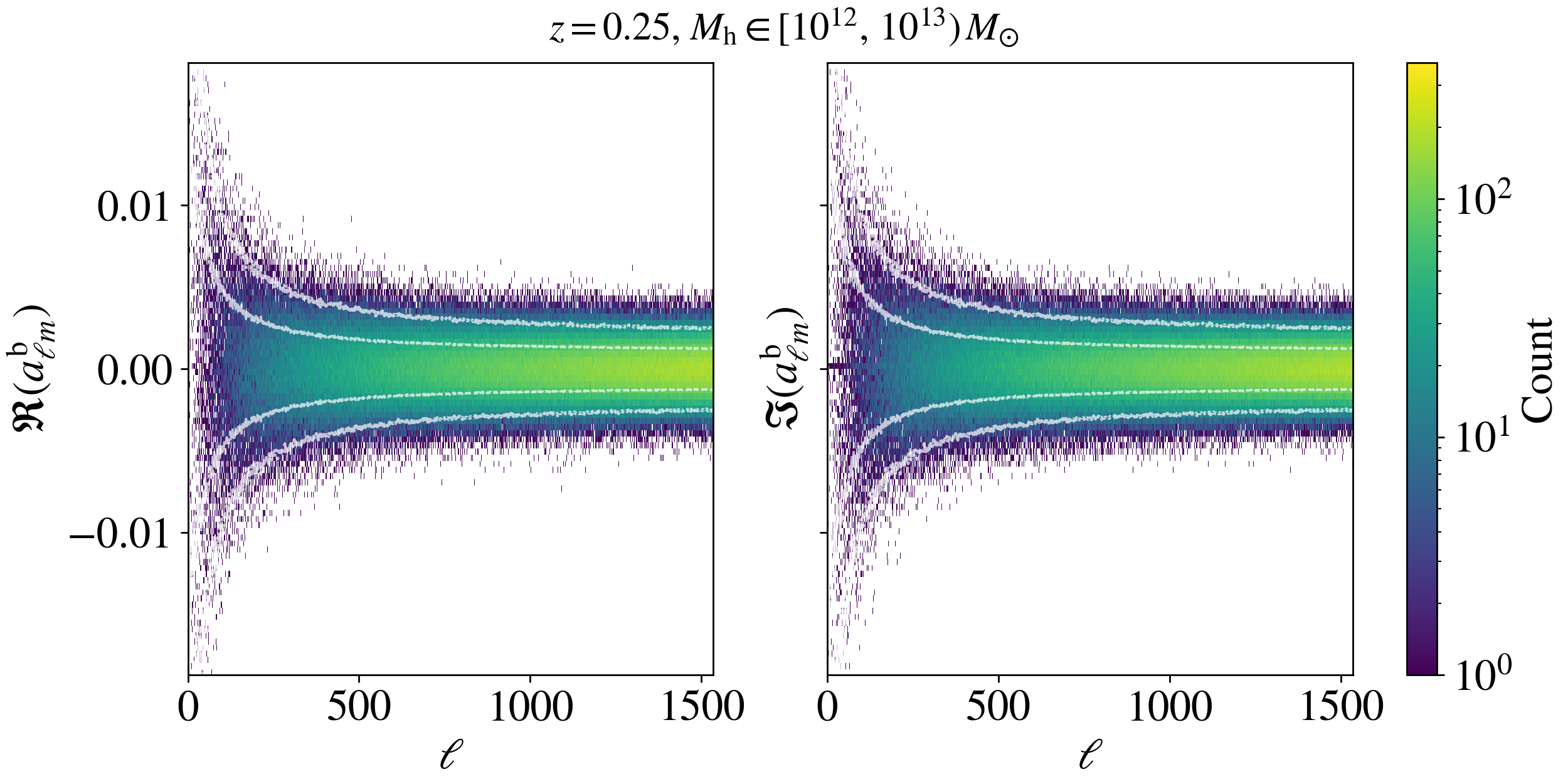}
    \includegraphics[trim={0 0.2cm 0 0},clip,width=0.9\linewidth]{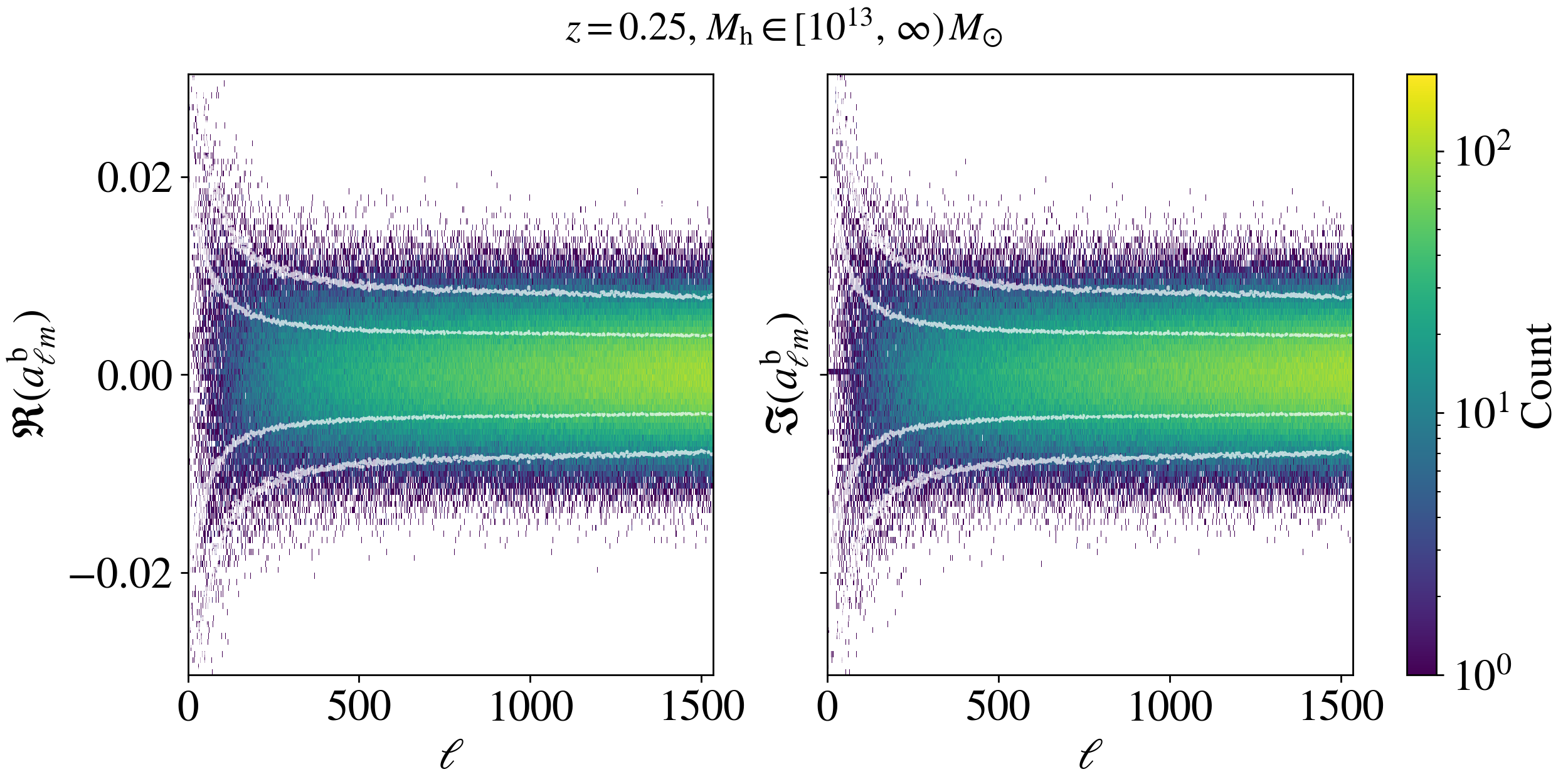}
    \caption{One-point distribution of the spherical harmonic coefficients $a_{\ell m}$ of the galaxy field measured from the \texttt{FLAMINGO} `L2p8\_m9' simulation for a single shell at $z \in [0.25, 0.30)$. The panels correspond to increasing halo mass ranges: $M_{\mathrm{h}} \in [10^{11.5}, 10^{12})$ (top), $[10^{12}, 10^{13})$ (middle), and $> 10^{13} \, \mathrm{M}_{\odot}$ (bottom). The left panels shows the real part of the spherical harmonic coefficients, while the right panels show their imaginary part. The white lines show the 68\% and 95\% quantiles of the distribution of the $a_{\ell m}$ at a given value of $\ell$. While the variance of the distribution decreases with $\ell$, the distribution remains approximately Gaussian at all scales. This is similarly the case across all redshifts and for all halo mass ranges tested.}
    \label{fig:1pt_statistics}
\end{figure}

We compare the one-point and two-point statistics of the synthetic biased fields to those measured from the \texttt{FLAMINGO} simulation. Two-point agreement is guaranteed by construction, since $\beta_{\ell}$ and $\sigma_{\ell}$ are calibrated on $C_{\ell}^{\mathrm{gc}}$ and $C_{\ell}^{\mathrm{gg}}$. Nonetheless,  Figure~\ref{fig:2pt_statistics} checks the internal consistency of the pipeline. At all redshifts and for $M_{\mathrm{h}} > 10^{13}\,\mathrm{M}_{\odot}$, the BOLTS mock mean over 200 realisations agrees with the simulation at the level of the stochastic noise, as it does for the other mass ranges and for $C_{\ell}^{\mathrm{gg}}$.

The one-point statistics provide a non-trivial test: BOLTS assumes that $\gamma_{\ell m}$ is drawn from a complex normal distribution (Equation~\ref{eqn:final_model}). To validate this ansatz, Figure~\ref{fig:1pt_statistics} shows the one-point distributions of the real and imaginary parts of $b_{\ell m}$ at representative angular scales for three halo mass ranges at $z \in [0.25, 0.30)$. For the highest mass bin ($M_{\mathrm{h}} > 10^{13}\,\mathrm{M}_{\odot}$), the Gaussian model with equal variances ($\sigma_{1,\ell}^2 = \sigma_{2,\ell}^2$) provides an excellent fit at all scales shown. At lower masses ($M_{\mathrm{h}} \in [10^{11.5}, 10^{12})$ and $[10^{12}, 10^{13})$), the distributions develop heavier tails at high $\ell$, reflecting the growing importance of shot noise and non-linear effects for low-mass halos, at all redshifts examined. The Gaussian approximation remains adequate for generating synthetic fields that capture the dominant two-point and bispectrum-level signatures.

\begin{figure}
    \centering
    \includegraphics[trim={0cm 0.5cm 0cm 0cm},clip,width=0.85\linewidth]{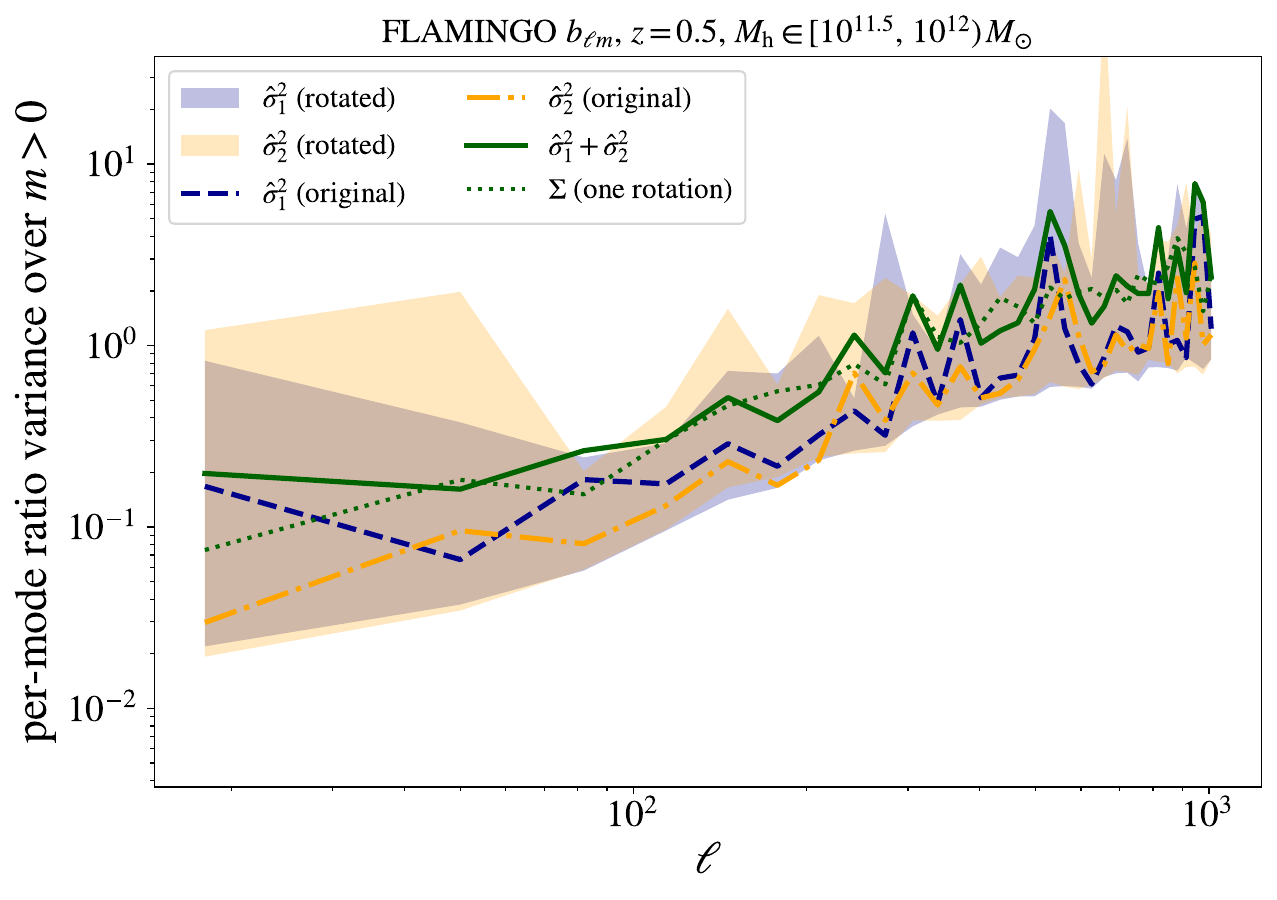}
    \caption{Rotation test of the split of the stochastic bias variance into real and imaginary parts. For the $z \in [0.50, 0.55)$ shell of the `L2p8\_m9' simulation and galaxies with $M_{\mathrm{h}} \in [10^{11.5}, 10^{12})\,\mathrm{M}_{\odot}$, the per-mode bias ratios $b_{\ell m} = a^{\mathrm{g}}_{\ell m} / a^{\mathrm{c}}_{\ell m}$ are formed and the variances of their real and imaginary parts over the $m > 0$ modes of each multipole are measured in bins of $\Delta\ell = 32$: in the original map (dark blue dashed and orange dot-dashed) and after 20 random rotations applied to both fields to give the error intervals (shaded regions of the same colours). The sum of both components is given by the green solid line, while the green dotted line shows a single rotation.
}
    \label{fig:rotation_test}
\end{figure}

Figure~\ref{fig:rotation_test} tests the equal split of the stochastic variance between real and imaginary parts (Section~\ref{derivation}) independently of the BOLTS assumptions. We measure the variances of the per-mode ratios in \texttt{FLAMINGO} and re-measure them after random rotations of the projected fields. The per-mode ratio $a^{\mathrm{g}}_{\ell m}/a^{\mathrm{c}}_{\ell m}$ has Cauchy-like tails \citep[Prop.~3a]{baldi2007characterizations}, so its sample variance over the $2\ell$ modes of a multipole is a heavy-tailed statistic. The individual components, and at $\ell \lesssim 100$ even their sum, scatter from rotation to rotation, while the asymmetry $(\hat\sigma_1^2 - \hat\sigma_2^2)/(\hat\sigma_1^2 + \hat\sigma_2^2)$ averaged over $64 \leq \ell \leq 1000$ is $+0.02$ in the simulation frame and $+0.003 \pm 0.009$ over 20 random rotations. The finite-variance measure $\Re \sum_m \epsilon_{\ell m}^2 \delta^{\mathrm{c}*2}_{\ell m} / \sum_m |\epsilon_{\ell m}|^2 |\delta^{\mathrm{c}}_{\ell m}|^2$, with $\epsilon_{\ell m} = a^{\mathrm{g}}_{\ell m} - \hat\beta_{\ell} a^{\mathrm{c}}_{\ell m}$, which equals the asymmetry for the model of Equation~\ref{eqn:final_model}, is within $\pm 0.6\%$ of zero at $\ell \geq 256$ for all three mass bins at $z = 0.5$ and for $M_{\mathrm{h}} > 10^{13}\,\mathrm{M}_{\odot}$ at $z = 1.5$. Its imaginary part, which measures $\rho$, is likewise consistent with zero.

\section{Model independent measurements of stochasticity and non-locality}\label{appendix:model_independent}

\begin{figure}
    \centering
    \includegraphics[trim={0cm 0.3cm 0cm 0cm},clip,width=0.75\linewidth]{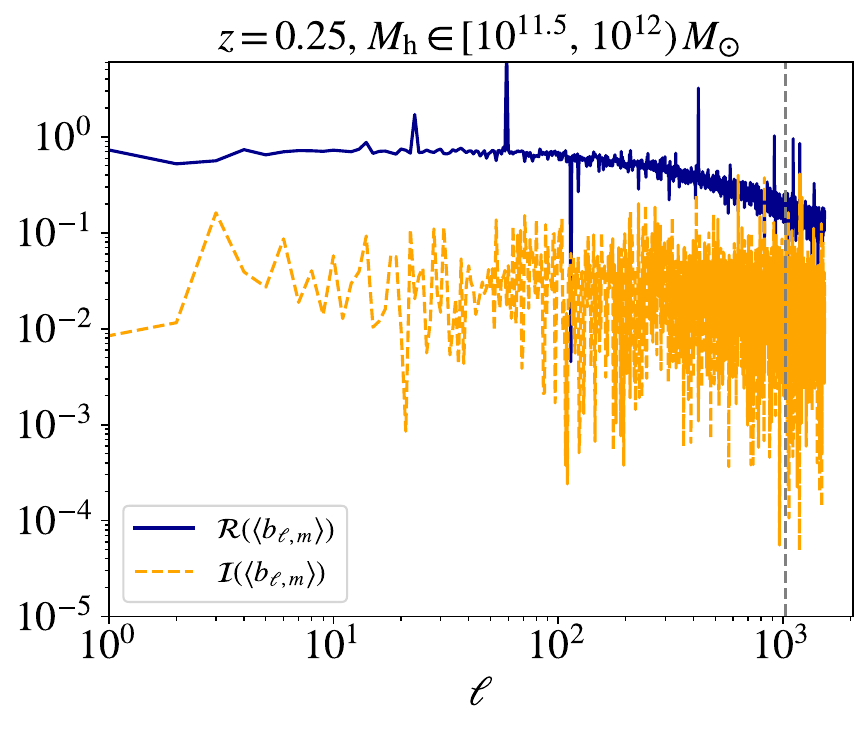}
    \caption{Model-independent measurements of the mean galaxy bias, $\langle b_{\ell m} \rangle_m$, given  $M_{\mathrm{h}} \in [10^{11.5}, 10^{12})$ as a function of angular scale, $\ell$, in the $z \in [0.25, 0.3)$ shell of the of the `L2p8\_m9' simulation (the real part with a solid dark blue line and the imaginary part with a dashed orange line). The vertical grey dashed line indicates the pixel scale. The presence of non-locality is indicated by a scale dependence in $\langle b_{\ell m} \rangle_m$, while the presence of stochasticity is indicated a scatter in $\langle b_{\ell m} \rangle_m$. The scatter is in turn also scale-dependent as the variance changes with $\ell$. The pattern is similar across all redshifts examined and halo mass ranges, with the strongest signatures of non-locality and stochasticity appearing at low masses and low redshifts.}
    \label{fig:model_independent}
\end{figure}

Here we outline model-independent diagnostics of stochasticity and non-locality in the galaxy bias that do not assume the functional form of the BOLTS model.

For a general linear bias model $\delta^{\mathrm{b}}_{\ell m} = b_{\ell m} \, \delta_{\ell m}$, the mean over all $m$ modes at fixed $\ell$ estimates the total bias $b_{\ell} = \langle b_{\ell m} \rangle_m$. The presence of non-locality is then indicated by a scale dependence in $b_{\ell}$, while the presence of stochasticity is indicated by a non-zero total complex variance $\sigma_{\ell}^2 = \langle |b_{\ell m} - b_{\ell}|^2 \rangle_m$. 

Figure~\ref{fig:model_independent} shows these measurements on the spherical harmonic coefficients of the `L2p8\_m9' galaxy fields. They confirm both non-locality and stochasticity across all halo mass ranges and redshifts examined, with the strongest signatures at low masses and low redshifts. The scale dependence of $\beta_{\ell}$ and $\sigma_{\ell}$ is qualitatively consistent with the trends captured by the BOLTS fitting functions, though the per-$\ell$ measurements are noisy due to the limited number of $m$ modes available for averaging at each $\ell$.

\section{Testing field resolution effects}\label{appendix:order_of_ops}

The choice of HEALPix resolution $N_{\mathrm{side}}$ affects the BOLTS calibration in two ways. First, we restrict to $\ell \le 2 N_{\mathrm{side}}$, well below the transform
limit $3 N_{\mathrm{side}} - 1$, so lower resolutions truncate the accessible range of angular scales. Second, the pixelisation itself modifies the effective matter and tracer fields through the HEALPix pixel window function $w_{\ell}$, which suppresses power at scales approaching the pixel size, with $w_{\ell} \to 0$ as $\ell \to 2\,N_{\mathrm{side}}$. The latter effect is corrected at the two-point level by dividing the measured auto- and cross-spectra by $w_{\ell}^{2}$ during the calibration outlined in Section~\ref{calibration}. Crucially, because the deterministic bias $\beta_{\ell} = C_{\ell}^{\mathrm{gm}} / C_{\ell}^{\mathrm{mm}}$ is a ratio of spectra that share the same pixel window, the $w_{\ell}^{2}$ factors cancel identically and $\beta_{\ell}$ is insensitive to the deconvolution. For the stochastic amplitude $\sigma_{\ell}$, which involves the difference $C_{\ell}^{\mathrm{gg}} - 4 \pi / N_{\mathrm{g}}$, the deconvolution is necessary to correctly separate the cosmological signal from shot noise. However, it amplifies noise at $\ell$ close to $\ell_{\mathrm{max}}$ where $w_{\ell} \to 0$.

We verify the robustness of our results by repeating the calibration at $N_{\mathrm{side}} = 256$, $512$, and $1024$ for halos with $M_{\mathrm{h}} > 10^{13}\,\mathrm{M}_{\odot}$ at $z = 0.25$. Note that here we use the shot-noise-subtracted estimator, unlike the raw estimator of the main text. When comparing to the highest-resolution calibration ($N_{\mathrm{side}} = 1024$), the deterministic bias $\beta_{\ell}$ at $N_{\mathrm{side}} = 512$ agrees to better than 2\% for the vast majority ($\gtrsim 95$\%) of individual multipoles over $\ell \le 512$, with binned ($\Delta\ell = 20$) residuals remaining below 0.5\%. At $N_{\mathrm{side}} = 256$, the per-multipole scatter is somewhat larger (roughly 84\% of modes lie within 2\% at $\ell \le 400$) but the smooth trend still agrees at the per cent level, confirming that the measurement is not driven by pixelisation artefacts. The stochastic amplitude $\sigma_{\ell}$ is inherently noisier: for the high-mass bin considered here, $\sigma_{\ell}$ is consistent with zero at the majority of individual multipoles, and where it is non-zero the fractional residuals between resolutions are dominated by noise rather than by a systematic pixelisation effect. We find that the semi-empirical fitting functions (Equations~\ref{eqn:beta_l_fit} and~\ref{eqn:sigma_l_fit}), when fitted independently at each resolution over $\ell \le 0.75\,\ell_{\mathrm{max}}$, yield similar evaluated curves for the $N_{\mathrm{side}} = 512$ and $1024$ calibrations (within a few per cent), though the individual best-fitting parameters can differ due to degeneracies in the parametric form. The $N_{\mathrm{side}} = 256$ fits diverge more noticeably, reflecting the narrower $\ell$ range available for constraining the functional shape. We therefore adopt $N_{\mathrm{side}} = 512$ throughout, a good compromise between angular dynamic range ($\ell_{\mathrm{max}} = 1024$) and computational cost. For field-level forward modelling, the resolution should be matched to that of the survey under consideration.

As an additional consistency check, we compare two orderings of the resolution and calibration steps. In the first procedure, we calibrate $\beta_{\ell}$ and $\sigma_{\ell}$ at the native resolution $N_{\mathrm{side}} = 1024$ and restrict the resulting calibration to $\ell \le 2\,N_{\mathrm{side,low}}$. In the second, we first degrade the input matter and tracer maps to $N_{\mathrm{side}} = 256$ using the \textsc{HealPy} \texttt{ud\_grade} routine and then calibrate the model at the lower resolution. If the BOLTS parameterisation correctly separates scale-dependent bias from pixelisation effects, the two procedures should yield consistent calibrations over the common multipole range. For $\beta_{\ell}$, we find typical (median) agreement at the sub-per-cent level, with 90\% of multipoles agreeing to better than ${\sim}\,2$\% for $\ell \le 300$, rising to ${\sim}\,3$\% at $\ell \le 400$ as the deconvolution of the lower-resolution pixel window amplifies noise near $\ell_{\mathrm{max}}$. The stochastic amplitude $\sigma_{\ell}$ shows larger scatter between the two procedures, particularly at low $\ell$ where $\sigma_{\ell}$ is small and the fractional differences are amplified. This confirms that the calibrated parameters are not significantly coupled to the pixel window, so the model can be applied at resolutions other than the calibration one, provided the multipoles are well resolved at both.


\bsp	
\label{lastpage}
\end{document}